\documentclass[structabstract]{aa} 
\newcommand{\kms}{$\mathrm{km\, s^{-1}\, }$}
\usepackage{natbib}
\usepackage{graphicx}
\usepackage{longtable,lscape}
\usepackage{float}

\usepackage{txfonts}
\begin{document}
   \title{Limits on intermediate-mass black holes in six Galactic globular clusters with integral-field spectroscopy. \thanks{Based on observations collected at the European Organization for Astronomical Research in the Southern Hemisphere, Chile (085.D-0928, 086.D-0573).}}
   
   \titlerunning{Limits on intermediate-mass black holes in six Galactic globular clusters}

   \author{N. L\"utzgendorf
          \inst{1}
          \and
          M. Kissler-Patig\inst{2}
          \and
          K. Gebhardt\inst{3}          
          \and
          H. Baumgardt\inst{4}          
          \and
          E. Noyola\inst{5,6}
		  \and
          P. T. de Zeeuw\inst{1,7}
          \and
          N. Neumayer\inst{1}
          \and
          B. Jalali\inst{8}
          \and
          A. Feldmeier \inst{1}
          }

   \institute{European Southern Observatory (ESO),
              Karl-Schwarzschild-Strasse 2, 85748 Garching, Germany\\
              \email{nluetzge@eso.org}
         \and     
              Gemini Observatory, Northern Operations Center, 670 N. A'ohoku Place
Hilo, Hawaii, 96720, USA
         \and
			 Astronomy Department, University of Texas at Austin, 
			 Austin, TX 78712, USA 
         \and
			 School of Mathematics and Physics, University of Queensland, 
			 Brisbane, QLD 4072, Australia
         \and
             Instituto de Astronomia, Universidad Nacional Autonoma de Mexico (UNAM), 
             A.P. 70-264, 04510 Mexico
         \and
			 University Observatory, Ludwig Maximilians University, 
			 Munich D-81679, Germany	 
         \and
			 Sterrewacht Leiden, Leiden University, 
			 Postbus 9513, 2300 RA Leiden, The Netherlands
         \and
	         I.Physikalisches Institut, Universit\"at zu K\"oln, 
    	     Z\"ulpicher Str. 77, 50937 K\"oln, Germany}

   \date{Received August 30, 2012; accepted December 11, 2012}

 
  \abstract
   {The formation of supermassive black holes at high redshift still remains a puzzle to astronomers. No accretion mechanism can explain the fast growth from a stellar mass black hole to several billion solar masses in less than one Gyr.  The growth of supermassive black holes becomes reasonable only when starting from a massive seed black hole with mass of the order of $10^2 - 10^5 \ M_{\odot}$. Intermediate-mass black holes are therefore an important field of research. Especially the possibility of finding them in the centers of globular clusters has recently drawn attention. Searching for kinematic signatures of a dark mass in the centers of globular clusters provides a unique test for the existence of intermediate-mass black holes and will shed light on the process of black-hole formation and cluster evolution. }
   {We are investigating six galactic globular clusters for the presence of an intermediate-mass black hole at their centers. Based on their kinematic and photometric properties, we selected the globular clusters NGC 1851, NGC 1904 (M79), NGC 5694, NGC 5824, NGC 6093 (M80) and NGC 6266 (M62).}
   {We use integral field spectroscopy in order to obtain the central velocity-dispersion profile of each cluster. In addition we complete these profiles with outer kinematic points from previous measurements for the clusters NGC 1851, NGC 1094, NGC 5824 and NGC 6093. We also compute the cluster photometric center and the surface brightness profile using HST data. After combining these datasets we compare them to analytic Jeans models. We use varying $M/L_V$ profiles for clusters with enough data points in order to reproduce their kinematic profiles in an optimal way. Finally, we vary the mass of the central black hole and test whether the cluster is better fitted with or without an intermediate-mass black hole.}
  {We present the statistical significance, including upper limits, of the black-hole mass for each cluster. NGC 1904 and NGC 6266 provide the highest significance for a black hole. Jeans models in combination with a $M/L_V$ profile obtained from N-body simulations (in the case of NGC~6266) predict a central black hole of $M_{\bullet} = (3 \pm 1) \times 10^3 \, M_{\odot}$ for NGC 1904 and $M_{\bullet} = (2 \pm 1) \times 10^3 \, M_{\odot}$ for NGC 6266. Furthermore, we discuss the possible influence of dark remnants and mass segregation at the center of the cluster on the detection of an IMBH.}
   {}

   \maketitle

   \maketitle
%

\section{Introduction}

The study of globular clusters has recently provided some fascinating new discoveries. Not long ago, globular clusters were assumed to be single population, isotropic stellar systems, formed by a single starburst in the early universe. With new generation instruments and space telescopes, however, this point of view has changed. Multiple stellar populations have been detected in many globular clusters and indicate a complicated formation mechanism \citep[eg.][]{bedin_2004,piotto_2007,piotto_2008}. Also, several clusters show rotation and high radial anisotropy in their outskirts \citep{ibata_2011} and turned out to be more complicated dynamical systems than previously assumed. Further, over the last decade, hints were found that massive central black holes reside at the centers of some globular clusters \citep[e.g.][]{gebhardt_2005,noyola_2008,nora11}, although other studies challenge these findings \citep[e.g.][]{vdMA_2010,strader_2012}.

\begin{table*}\tiny
\caption{Properties of the six globular clusters from the references: NG=\cite{noyola_2006}, H= \cite{harris_1996}, PM=\cite{pryor_1993}, M=\cite{mclaughlin_2005} and L=this work (see Section \ref{phot_center}).}             
\label{tab:prop}      
\centering
\begin{tabular}{l l l l l l l l}
\hline \hline
\noalign{\smallskip}
 Parameter 									& NGC 1851 											& NGC 1904 											& NGC 5694 											& NGC 5824 											& NGC 6093 											& NGC 6266 											& Ref.\\
 \noalign{\smallskip}
\hline
\noalign{\smallskip}
 RA (J2000)  								& $05:14:06.7$										& $05:24:11.0$									 	& $14:39:36.3$									  	& $15:03:58.6$									  	& $16:17:02.4$									  	& $17:01:13.0$									  	& L \\ 
 DEC (J2000) 								& $-40:02:49.3$ 										& $-24:31:27.9$ 	 									& $-26:32:19.6$										& $-33:04:05.3$	 									& $-22:58:32.6$	 									& $-30:06:48.2$	 									& L \\ 
Galactic Longitude $l$ 						& $244.51 °$ 										& $227.23 °$ 										& $331.06 °$											& $332.55 °$ 										& $352.67 °$ 										& $353.57 °$ 										& H \\
Galactic Latitude $b$ 						& $-35.04 °$ 										& $-29.35 °$ 										& $30.36 °$ 											& $22.07 °$											& $19.46 °$ 											& $7.23 °$ 											& H \\
Distance from the Sun $D$		 	& $12.1 \ \mathrm{kpc}$ 								& $12.9 \ \mathrm{kpc}$ 								& $34.7 \ \mathrm{kpc}$ 								& $32.0 \ \mathrm{kpc}$ 								& $10.0 \ \mathrm{kpc}$ 								& $6.9 \ \mathrm{kpc}$ 								& H \\
Core Radius $r_c$							& $2.00''$ 											& $5.6''$ 											& $2.2''$ 											& $1.4''$ 											& $4.5''$ 											& $6.6''$ 											& NG \\
Central Concentration $c$ 					& $2.32$ 											& $1.72$ 											& $1.84$ 											& $2.45$ 											& $1.95$ 											& $1.70$ 											& H  \\
SB slope $\alpha$		 					& $-0.38 \pm 0.11$ 									& $-0.03 \pm 0.07$ 									& $-0.19 \pm 0.11$									& $-0.36 \pm 0.16$									& $-0.16 \pm 0.07$									& $-0.13 \pm 0.08$									& NG  \\
Radial Velocity V$_r$ 						& $320.5 \ \mathrm{km}/\mathrm{s}$ 					& $206.0  \ \mathrm{km}/\mathrm{s}$ 					& $-144.1 \ \mathrm{km}/\mathrm{s}$ 					& $-27.5 \ \mathrm{km}/\mathrm{s}$ 					& $8.2 \ \mathrm{km}/\mathrm{s}$ 					& $-70.1 \ \mathrm{km}/\mathrm{s}$ 					& H \\ 
Velocity Dispersion $\sigma$ 				& $10.40 \ \mathrm{km}/\mathrm{s}$ 					& $4.81 \ \mathrm{km}/\mathrm{s}$ 					& $5.5 \ \mathrm{km}/\mathrm{s}$ 					& $12.8 \ \mathrm{km}/\mathrm{s}$ 					& $13.3 \ \mathrm{km}/\mathrm{s}$ 					& $13.7 \ \mathrm{km}/\mathrm{s}$ 					& PM \\
Metallicity $[\mathrm{Fe}/\mathrm{H}]$ 		& $-1.22 \ \mathrm{dex}$ 							& $-1.57 \ \mathrm{dex}$ 							& $-1.86 \ \mathrm{dex}$ 							& $-1.85 \ \mathrm{dex}$ 							& $-1.75 \ \mathrm{dex}$ 							& $-1.29 \ \mathrm{dex}$ 							& H \\
Total Mass $\log M_{tot}$					& $5.50 \pm 0.04$ 									& $5.32 \pm 0.08$ 									& $5.40 \pm 0.06$									& $5.86 \pm 0.04$									& $5.44 \pm 0.44$ 									& $5.92 \pm 0.07$									& M \\ 
Reddening E(B-V) 							& $0.02 $ 											& $0.01 $ 											& $0.09 $ 											& $0.13 $ 											& $0.18 $ 											& $0.40 $ 											& H \\
Absolute Magnitude $M_{Vt}$ 					& $-8.33 \ \mathrm{mag}$ 							& $-7.86 \ \mathrm{mag}$ 							& $-7.81 \ \mathrm{mag}$ 							& $-8.84 \ \mathrm{mag}$ 							& $-8.23 \ \mathrm{mag}$ 							& $-9.18 \ \mathrm{mag}$ 							& H  \\
\noalign{\smallskip}
\hline 
\end{tabular} 
\end{table*}

Intermediate-mass black holes are defined as covering a mass range of $10^2 - 10^5 \ M_{\odot}$ and have become a promising field of research. With their existence it could be possible to explain the rapid growth of supermassive black holes, which are observed at large redshifts \citep{fan_2006} by acting as seed black holes \citep[eg.][]{ebisuzaki_2001,tanaka_2009}. Further, detections of ultraluminous X-ray sources at off-center positions in distant galaxies \citep[eg.][]{matsumoto_2001,fabiano_2001,soria_2010} suggest intermediate-mass black holes, supposedly originating from accreted and disrupted globular clusters or dwarf galaxies. Another motivation for searching for intermediate-mass black holes is the observed relation between the black-hole mass and the velocity dispersion of its host galaxy \citep[e.g.][]{ferrarese_2000,gebhardt_2000b, gultekin_2009}. Exploring this relation in the lower mass range, where we find intermediate-mass black holes, will give important information about the origin and validation of this correlation.

The formation of intermediate-mass black holes can occur by the direct collapse of very massive first generation stars \citep[Population III stars,][]{madau_2001}, or runaway merging in dense young star clusters \citep{zwart_2004, gurkan_2004, freitag_2006}. This makes globular clusters excellent environments for intermediate-mass black holes. The currently best candidates for hosting an intermediate-mass black hole at their centers are the globular clusters $\omega$ Centauri \citep[NGC 5139,][]{noyola_2008, noyola_2010,vdMA_2010,jalali_2012}, G1 in M31 \citep[][]{gebhardt_2005}, and NGC 6388 \citep{nora11}. All of these very massive globular clusters show kinematic signatures of a central dark mass in their velocity-dispersion profiles. For $\omega$ Centauri, \cite{vdMA_2010} investigated a large dataset of HST proper motions and found less evidence for a central IMBH than proposed by \cite{noyola_2008}. Using a new kinematic center, however, \cite{noyola_2010} and \cite{jalali_2012} confirmed the signature of a dark mass in the center of $\omega$ Centauri and proposed a $\sim 40~000 \, M_{\odot}$ IMBH.

For G1, X-ray and radio emission were detected at its center. They are consistent with a black hole of the same mass as derived by kinematic measurements \citep{pooly_2006, kong_2007,ulvestad_2007}. This result, however, was recently challenged by \cite{miller-jones_2012} who found no radio signature at the center of G1 when repeating the observations. Also \cite{strader_2012} investigated a set of globular clusters in order to test for X-ray and radio signatures of intermediate-mass black holes. All of their fluxes lead to very low upper limits on a possible black-hole mass. Further, \cite{cseh_2010} and \cite{Lu_2011} investigated radio and X-ray emission in the globular clusters NGC 6388 and $\omega$ Centauri, respectively and found upper limits of $\sim 1500 \, M_{\odot}$ for NGC 6388 and  $1100 - 5200  \, M_{\odot}$  for $\omega$ Centauri. However, various assumptions about the gas accretion process, such as clumpyness of the gas distribution, accretion efficiency and the translation of X-ray fluxes to bolometric fluxes to black hole masses, have to be made. Thus, non-detection in X-ray and radio are difficult to interpret in terms of a black hole mass upper limit.

\begin{figure*}
  \centering \includegraphics[width=6.0cm]{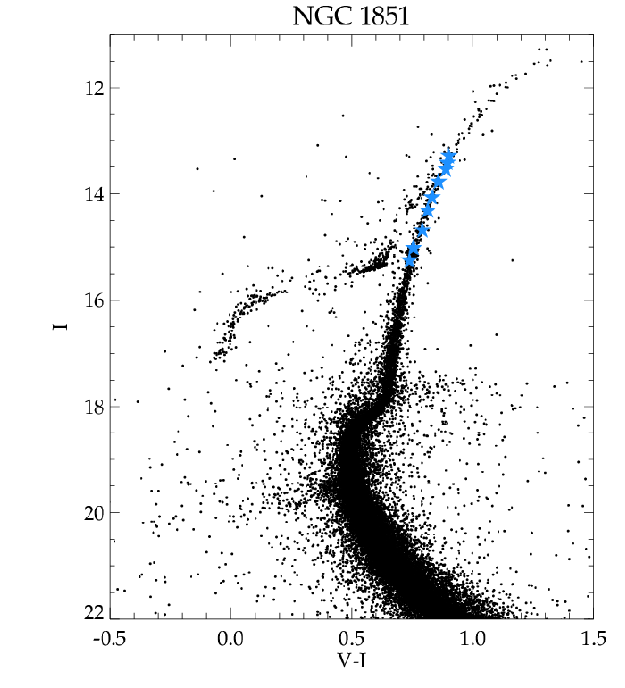}
  \centering \includegraphics[width=6.0cm]{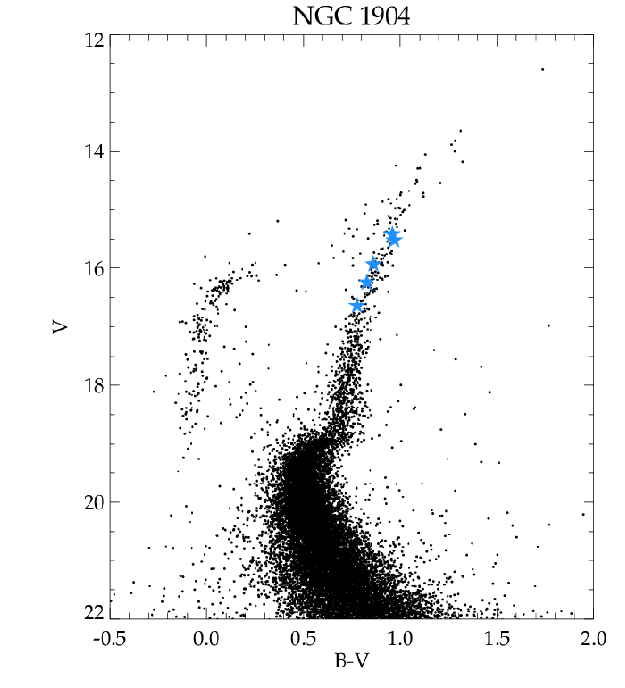}
  \centering \includegraphics[width=6.0cm]{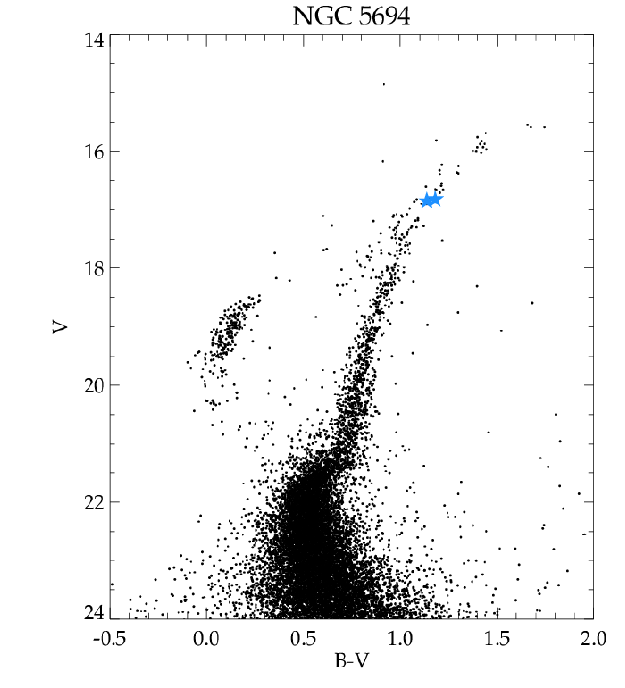}
  \centering \includegraphics[width=6.0cm]{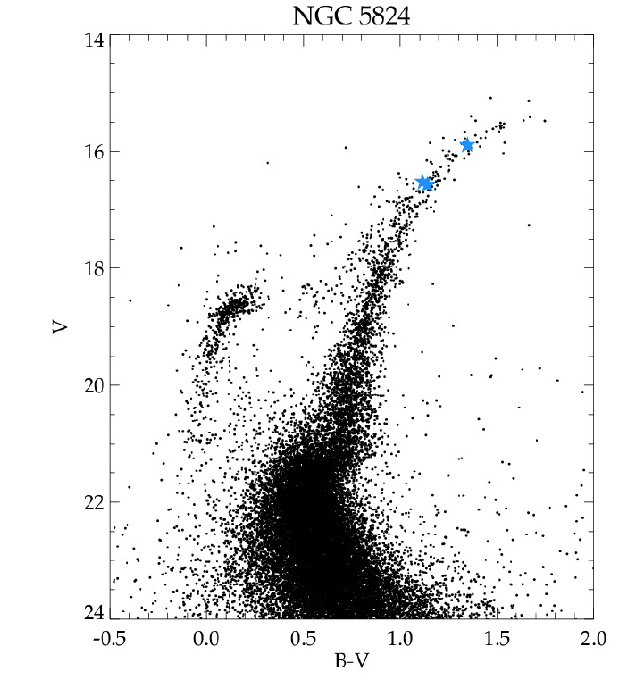}
  \centering \includegraphics[width=6.0cm]{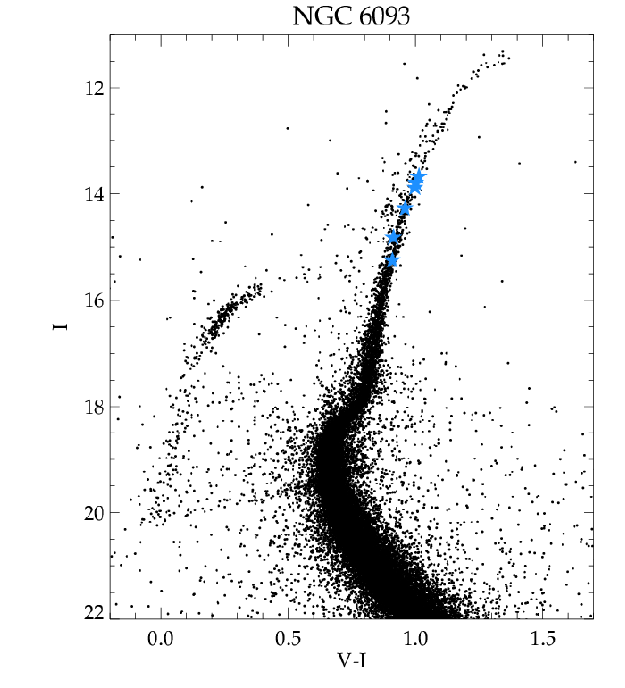}
  \centering \includegraphics[width=6.0cm]{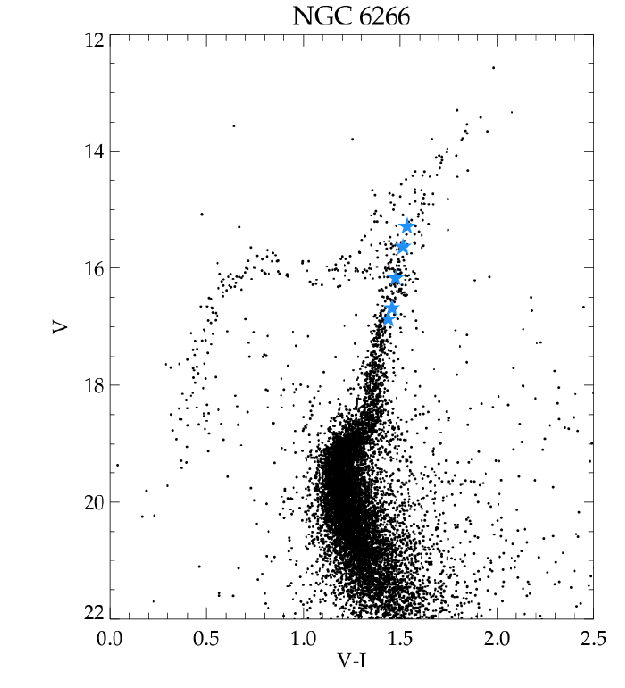}
  \caption{The color-magnitude diagrams of the six clusters of our sample. Overplotted are the stars which we used to create the combined template for the kinematic analysis (blue stars). For NGC 1851 and NGC 6093 the catalogs provided by \citet{anderson_2008} are used.}
  \label{fig:cmd}
\end{figure*}
Kinematic observations of individual stars in the central regions of globular clusters are not trivial. Without adaptive optics or space telescopes the central regions are not resolved due to their enormous stellar densities. Integrated light spectroscopy and proper motions from space telescope measurements (where the observations are not limited by the seeing) are therefore currently the only methods to derive the very central kinematics. 

When selecting a globular cluster candidate for investigating the possible presence of an intermediate-mass black hole we consider the absolute brightness of the cluster (since this is an indicator of its mass), its central velocity dispersion and the inner slope of its surface brightness profile. A massive cluster is much more likely to host a massive black hole than a less massive cluster, according to formation scenarios such as runaway merging, that require many massive stars in a dense environment to form a black hole. Furthermore, it has been suggested that the most massive globular clusters are stripped cores of dwarf galaxies and therefore also more likely to host central black holes. In addition, a high central velocity dispersion indicates a high mass concentration in the center of the globular cluster and, according to the observed $M - \sigma$ correlation, also a high black-hole mass. However, as shown in the case of M15, a rising velocity-dispersion profile is not always a sign for an intermediate-mass black hole \citep[e.g.][]{dull_1997,baumgardt_2003a,baumgardt_2005,bosch_2006}. The dynamical state of that cluster also needs to be considered. A collapsed core, as in the case of M15,  indicates a high level of mass segregation, that can also rise the velocity-dispersion in the center profile without needing to invoke a black hole. 

N-body simulations have shown that a shallow cusp in the surface brightness profile is expected for a cluster hosting an intermediate-mass black hole \citep{baumgardt_2005}. The simulations predict slopes in the surface brightness profiles between $\alpha = -0.1$ to $-0.3$ (where $\alpha$ is the slope of the surface luminosity density $I(r) \varpropto r^{\, \alpha}$). The values of the slope of the six globular clusters in our sample are taken from \cite{noyola_2006} and listed in Table \ref{tab:prop} together with other properties. The Galactic globular clusters NGC 1851, NGC 1904 (M79), NGC 5694, NGC 5824, NGC 6093 (M80) and NGC 6266 (M62) have distances of $\sim 7 - 35$ kpc and belong to the more massive clusters of our Galaxy. We note that not all clusters fulfill all three criteria (high mass, high dispersion and intermediate surface brightness slope). However, it is important to also investigate clusters which are, at this stage, considered unlikely to host IMBHs in order to verify the assumptions and to compare the 'good' with 'bad' candidate systems. Systematic differences might help to identify hosts of IMBHs candidate in the future.

This paper is structured in the following way: Section \ref{sec:phot} describes the photometric datasets, the computation of the color-magnitude diagrams, center determinations and derivation of the surface-brightness profiles. In Section \ref{sec:spec} we describe the spectroscopic data and the reduction and in Section \ref{sec:kin} the computation of the velocity maps as well as the velocity-dispersion profiles. The kinematic data is then combined with the photometric data and compared to dynamical Jeans models in Section \ref{sec:jeans}. In the final Section \ref{sec:con} we summarize our results and discuss the conclusions we derive from the data analysis.


\section{Photometry}\label{sec:phot}

We first analyze the photometric properties of our six clusters. This Section describes the various methods used to measure the cluster center and the light profiles. Since the photometric data are from different observations it is crucial to keep the analysis as homogeneous as possible in order to compare the results of the different clusters with each other.
                               
\subsection{Photometric catalogs} \label{sec_cmd}
 

For our photometric analysis we use data from the Hubble Space Telescope (HST) Archive. For the majority of the clusters (NGC 1904, NGC 5694, NGC 5824, NGC 6266) the data were obtained with the Wide Field and Planetary Camera~2 (WFPC2) within different projects (see Table \ref{tab:obsphot} for more details). When available, we use images in the filters $I$ (F814W) and $V$ (F555W, F606W), since the kinematic data were obtained in a wavelength range close to the $I$ filter. However, only half of the clusters were observed in both filters. For NGC 1904, NGC 5694 and NGC 5824, only observations in the $B$ (F439W) and V filters are available. 

For NGC 6266 we additionally use data obtained with the Advanced Camera for Surveys (ACS) in the Wide-Field Channel  in the filters $B$ and $R$ (F625W). Unfortunately, the high exposure time of $t = 340$ s saturated the brightest stars in the pointing. Therefore, we did not use these images for most of our analysis since we need the brightest stars in order to register them in our IFU pointing. The large field of view of the ACS is nevertheless useful to check our center determination.

\begin{table}\tiny
\caption{HST observations of the clusters.}             
\label{tab:obsphot}      
\centering
\begin{tabular}{cccc}
\hline \hline
\noalign{\smallskip}
Cluster 	&  Date & Pogram ID  & Filters \\
 \noalign{\smallskip}
\hline
\noalign{\smallskip}
NGC 1851 	&2006-05-01 & GO-10775 (PI: Sarajedini) 	& 	F606W, F814W 	 \\
NGC 1904 	&1995-10-14 & GO-6095 (PI: Djorgovski) 	& F439W, F555W \\
NGC 5694 	&1996-01-08 & GO-5902 (PI: Fahlman) 	&F439W, F555W \\
NGC 5824 	&1996-01-12 & GO-5902 (PI: Fahlman) 	& F439W, F555W \\
NGC 6093 	&2006-04-09 & GO-10775 (PI: Sarajedini) 	&   F606W, F814W \\
NGC 6266 	&2000-08-21 & GO-8709 (PI: Ferraro) 	& F555W, F814W \\
			&2004-08-01 & GO-10120 (PI: Anderson) &  F439W, F625W \\
\noalign{\smallskip}
\hline 
\end{tabular} 
\end{table}

All images are analyzed using \textit{daophot II}, \textit{allstar} and \textit{allframe} by \citet{stetson_1987}. The routines are ideally suited for globular clusters since they were developed for photometry in crowded fields. A detailed description of the individual steps can be found in \cite{nora11}. The stars are identified in the image and a point spread function (PSF) is fitted using the routines \textit{find}, \textit{phot} and \textit{psf}. The task \textit{allstars} then groups the neighboring stars to apply the multiple-profile-fitting routine simultaneously. Afterwards, the \textit{find} task is applied again to find, in the star-subtracted image, stars which were not found in the first run. This procedure is performed in all bands independently and the catalogs are combined using the programs \textit{daomaster} and \textit{allframes}. After constructing the final catalog for the long and short exposures, we combine the two datasets and obtain the final catalog which is used for further analysis.

For NGC 1851 and NGC 6093 we use images and catalogs provided by the  ACS Survey of Galactic Globular Clusters \footnote{available at $\, \,$ http://www.astro.ufl.edu/$\sim$ata/public$\_$hstgc/databases.html} \citep{anderson_2008}. The catalogs cover a large spatial area ($\sim 100''$) and provide very accurate photometry and astrometry for the clusters in the bands $I$ and $V$. A high-quality science image (retrieved from the HST archive) which consists of the reduced, calibrated, and combined ACS images in the $V$-band is also available and is used for further analysis such as determining the surface brightness profile. Figure \ref{fig:cmd} shows the color-magnitude diagrams (CMD) of our six clusters overplotted with the template stars used for the kinematic analysis.  

\begin{figure*}
  \centering \includegraphics[width=6.0cm]{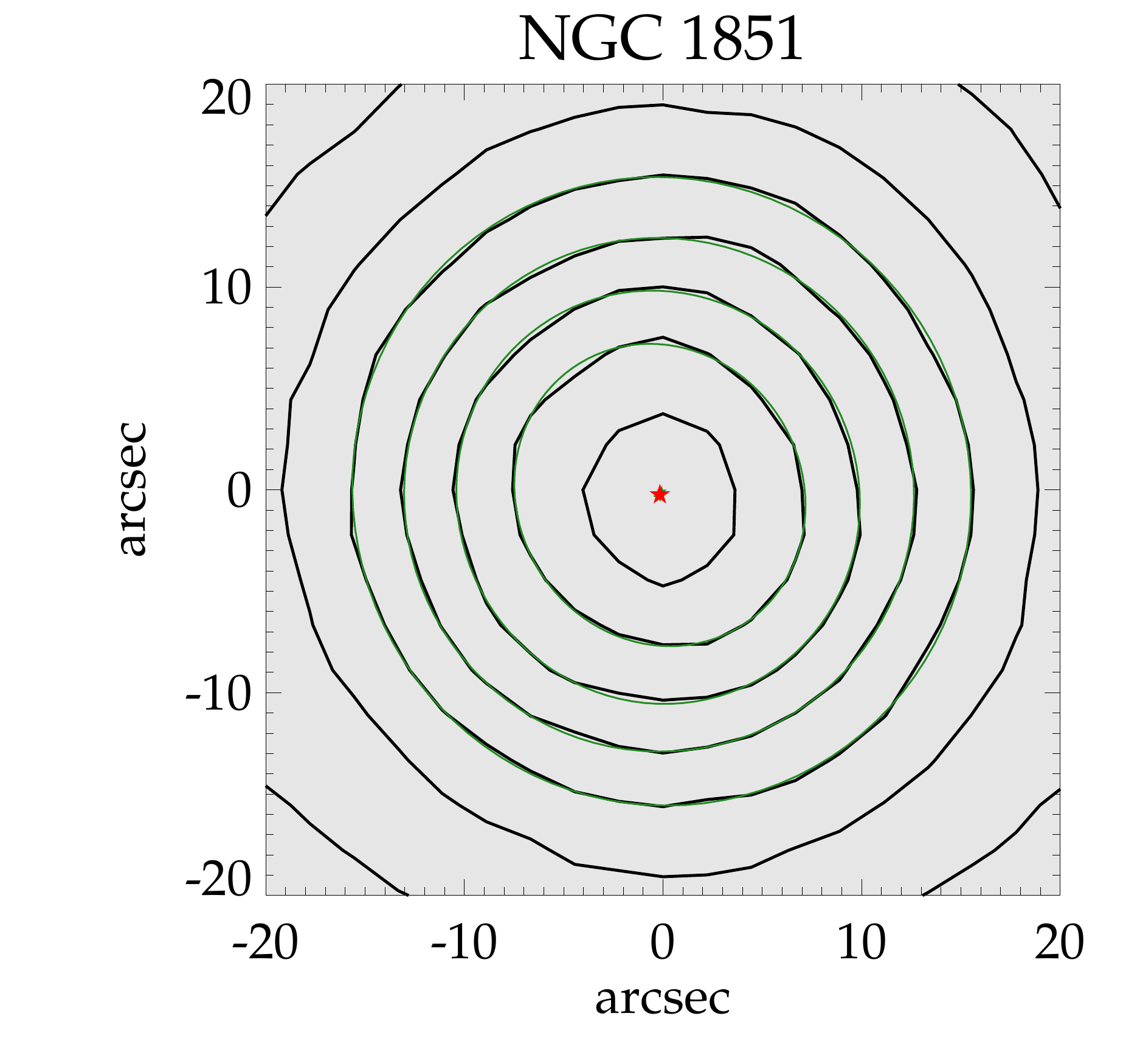}
  \centering \includegraphics[width=6.0cm]{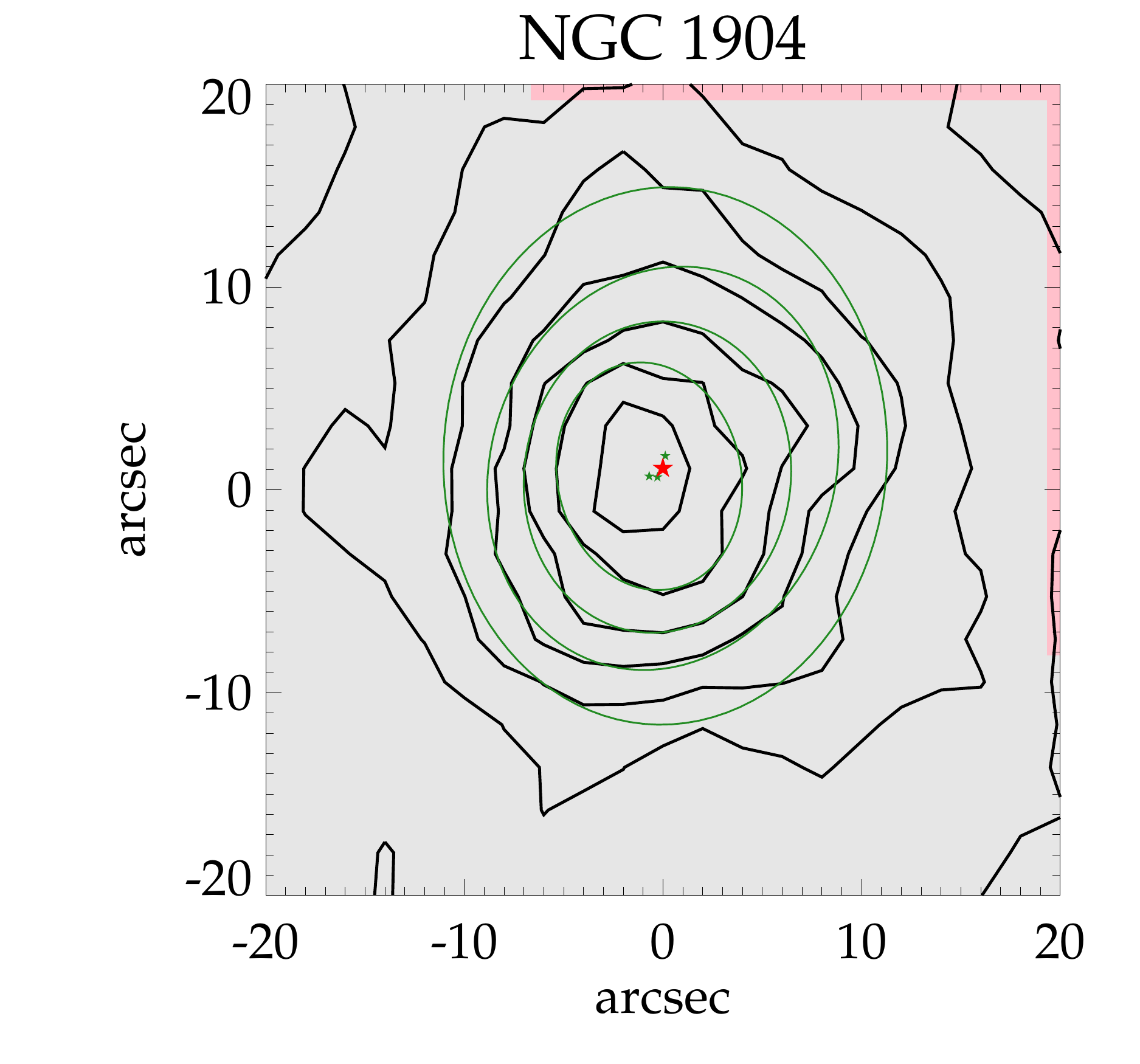}
  \centering \includegraphics[width=6.0cm]{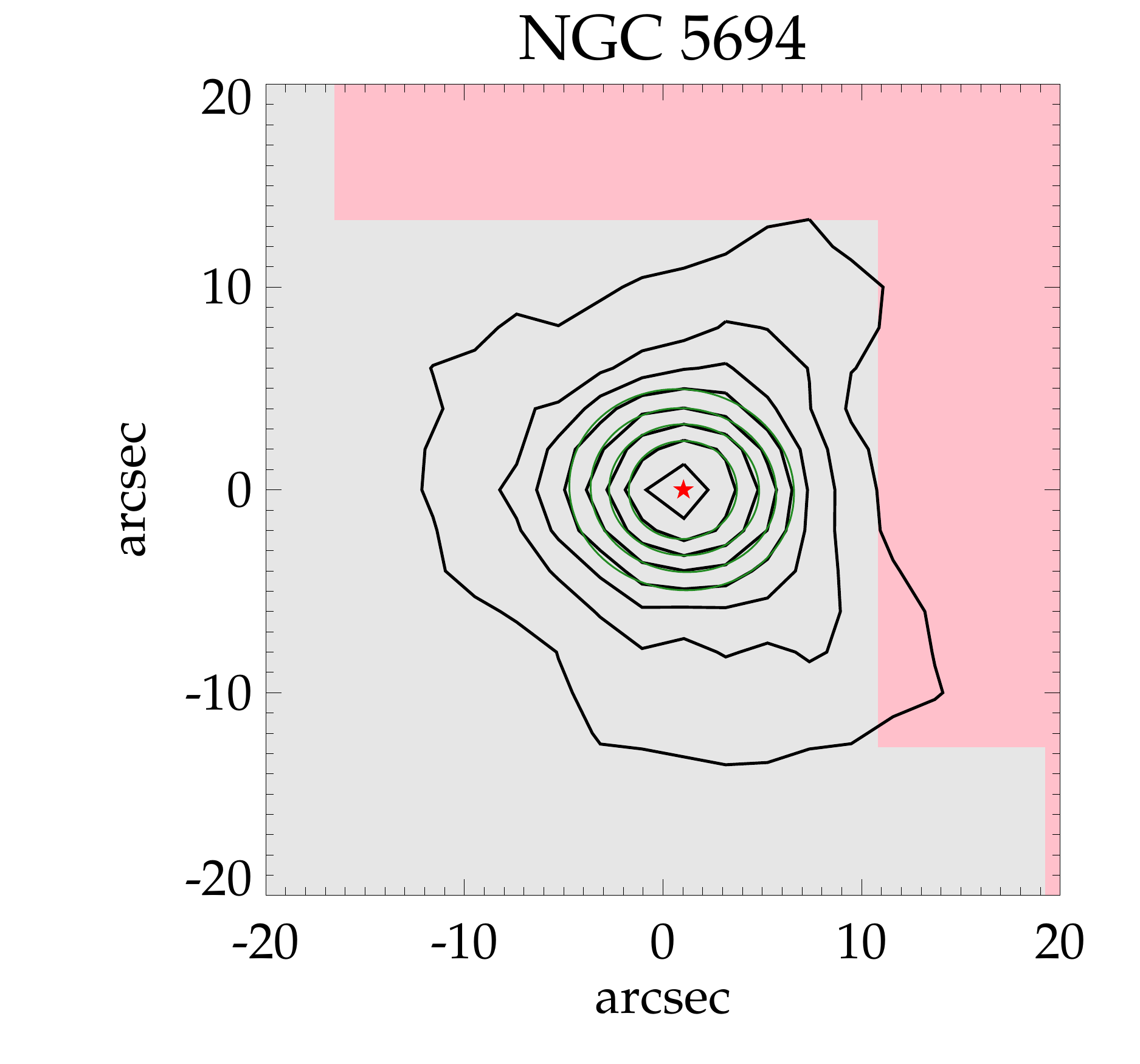}\\
  \centering \includegraphics[width=6.0cm]{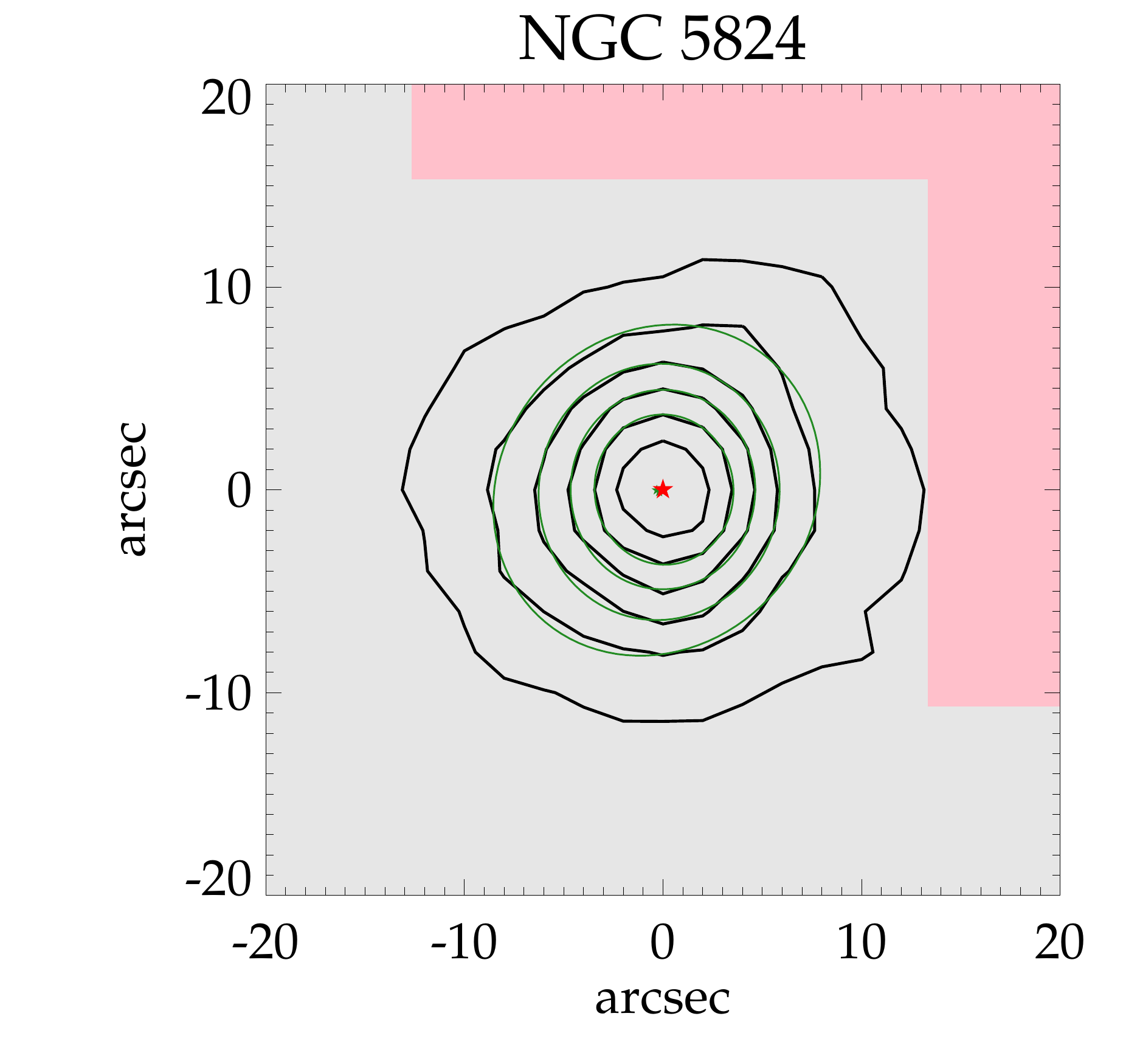}
  \centering \includegraphics[width=6.0cm]{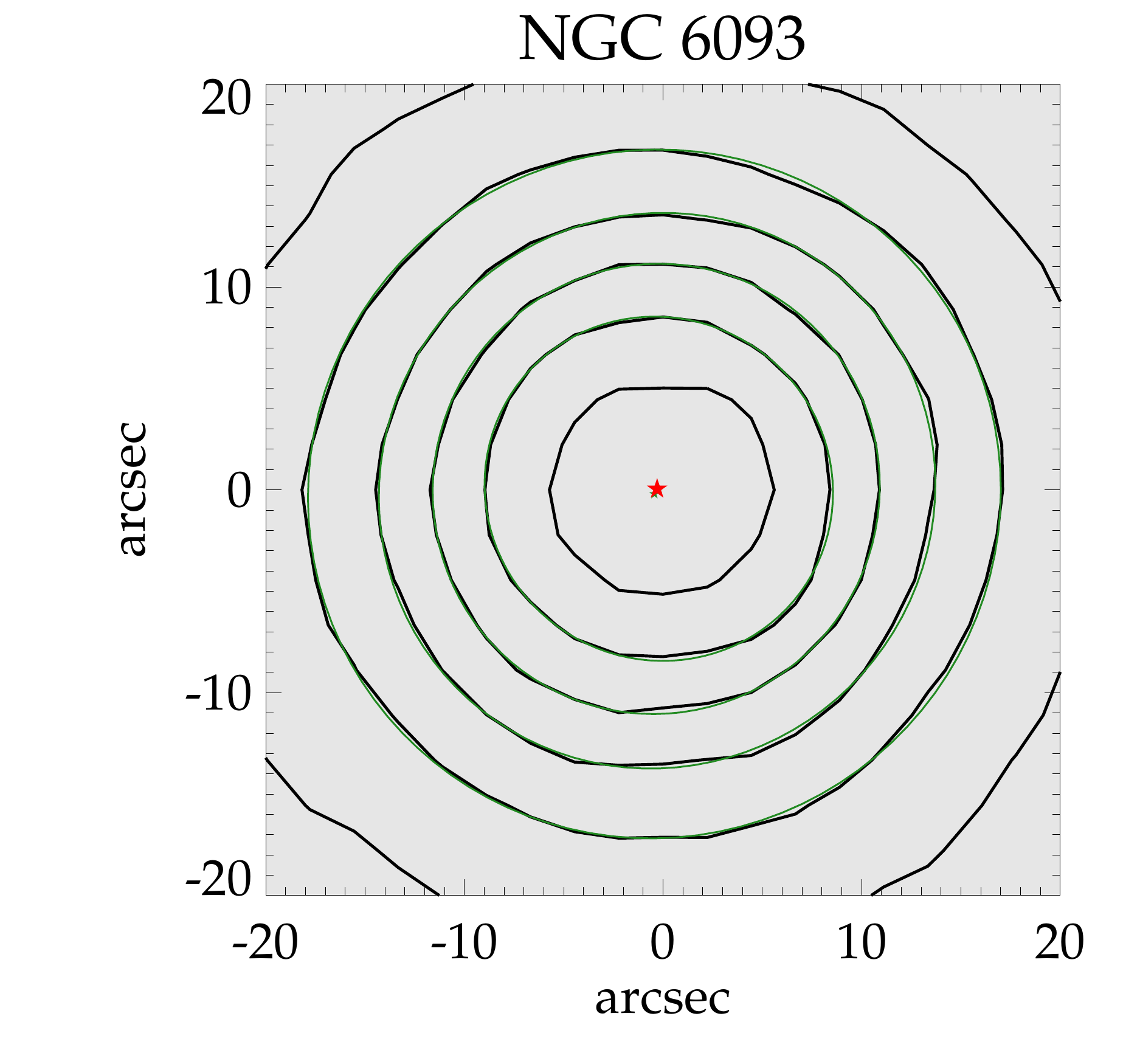}
  \centering \includegraphics[width=6.0cm]{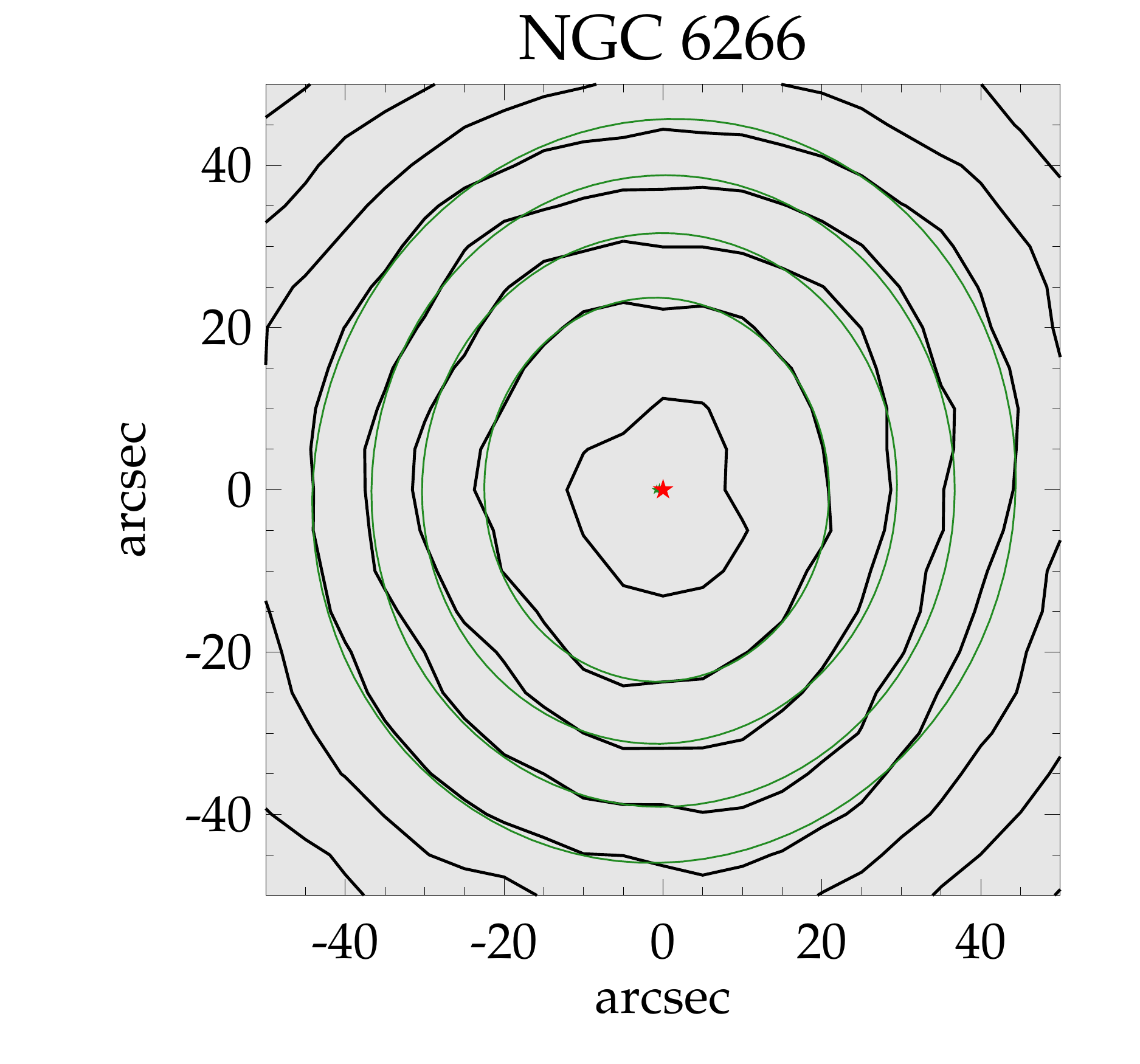}
  \caption{Isodensity contours for the center determination of our sample. The ellipses fitted to four of the contours are shown in green and the resulting center is marked by the red star. The different shades mark the area where the contours can be trusted (gray area) and where they are contaminated by edge effects (pink area).}
  \label{fig:isodens}
\end{figure*}

\subsection{Cluster center determination} \label{phot_center}

The center determination is critical for this work as errors in the center position can cause differences in the velocity-dispersion profile and the surface brightness profile as shown in the case of $\omega$ Centauri \citep{noyola_2008,vdMA_2010}, which is especially tricky due to its large core radius. Despite the small cores of the clusters in out sample, we find some differences in the position of the center with respect to previous measurements.

For the center determination we use the method described in \cite{goldsbury_2010}. This method is based on isodensity contours which are fitted by ellipses to find the center. We overlay the star catalog with a grid of equally spaced points. The grid extends $20''$ in each direction for the WFPC2 images and $50''$ for the ACS images with a spacing of $2''$. On every point a circle is drawn with a radius of $4''$ (WFPC2) or $10''$ (ACS) and the density of stars in this circle is calculated. With this we create an isodensity contour plot of 8 contours. We flag and exclude grid points where the radius exceeds the image dimensions in order to avoid edge effects. Figure \ref{fig:isodens} shows the contour plots from the center determination for our six clusters. The gray shaded areas mark the grid points not affected by the image borders, while the pink shaded areas mark the ones we exclude from the analysis. Finally, we fit ellipses to the four innermost contours leaving out the central region since it is dominated by small number statistics. The center is derived by taking the average value of the different ellipse centers.

For the WFPC2 images, a larger grid and smoothing radius are desired in order to get wide and smooth isodensity contours. Our dataset for these clusters, however, is restricted to the small and asymmetric field of view of the WFPC2 chip. A larger smoothing radius than we use would enlarge the region of excluded points. The values listed above are the best compromise between smoothing and enough contours to fit.

\begin{figure}
  \centering \includegraphics[width=0.5 \textwidth]{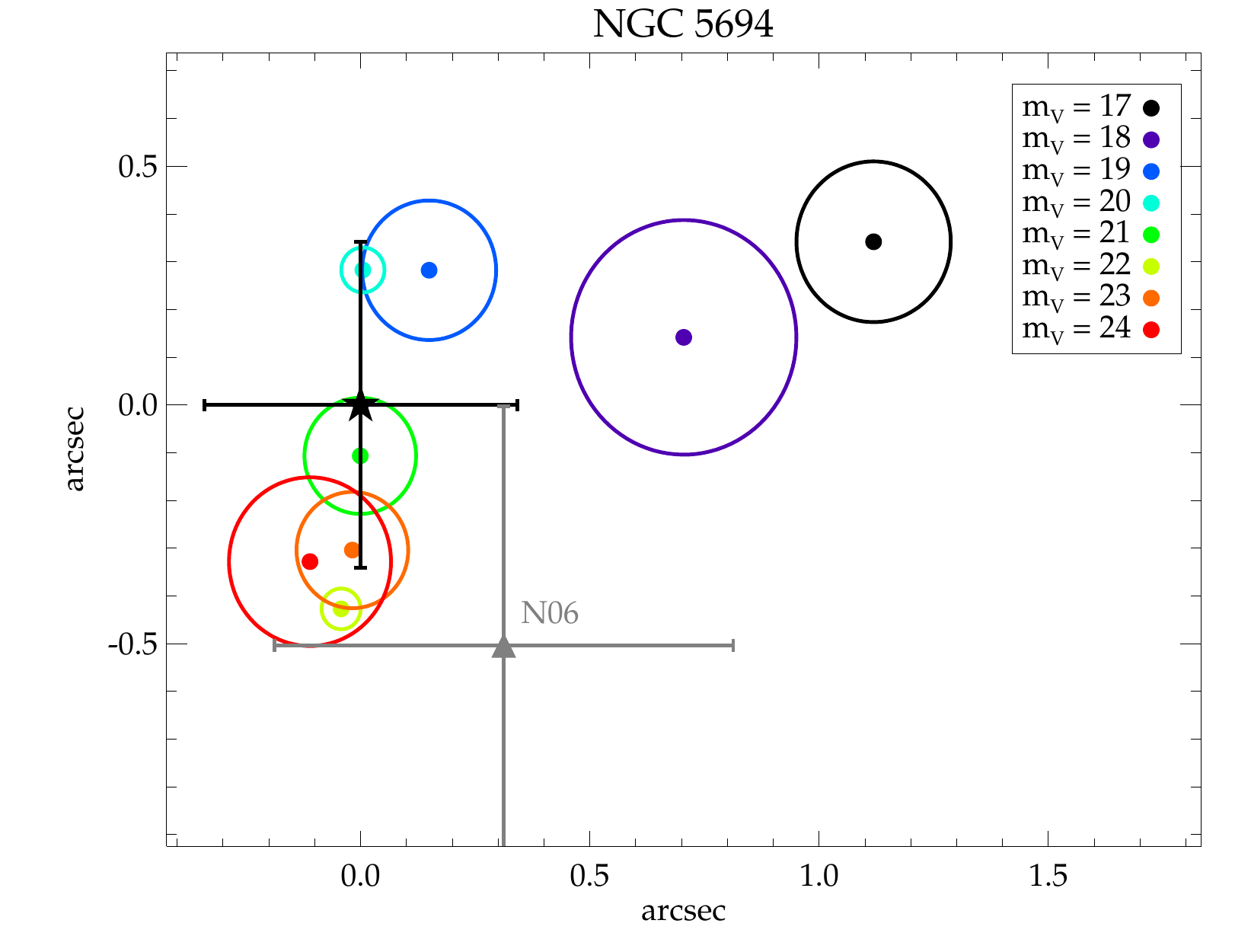}
  \caption{The center positions applying the isodensity method with different magnitude cuts for NGC 5694. The circles represent the error bars of the centers. Overplotted are also the final derived center (black star) and the center determined by \citet{noyola_2006} (gray triangle).}
  \label{fig:magcut}
\end{figure}

We try different magnitude cuts on our catalog in order to investigate the effect on the center determination. We find that this changes the results by far more than the error bars would suggest. Figure \ref{fig:magcut} shows the different centers and their error bars for NGC 5694. The plot shows that the center moves depending on the magnitude cut. We observed this behavior in all of our clusters including the ones where we used ACS images. If the magnitude cut is too low and we include many faint stars then the analysis suffers from incompleteness in the center. This creates holes in the distribution of the faint stars where they are affected by bright stars \citep[as described in][]{goldsbury_2010}. This biases the statistics towards these holes.  On the other side, if the cut is too bright and we only have bright stars, the total statistics is dominated by shotnoise due to the small number of bright stars.

In order to get a stable estimate of the cluster center we use an error weighted mean of the center positions for the cuts $m_V = 19, 20 , 21, 22, 23$ (which show the smallest errors) and determine the error from the scatter. For the final result we adopt the maximum error for both dimensions. Figure \ref{fig:magcut} shows the final position of the center in comparison with the center obtained by \cite{noyola_2006}.

In order to verify our results we tried further methods for the center determination. The first alternative method is using pie wedges instead of isodensity contours on the photometric catalog. For a grid of trial centers, a circle is applied to each center and divided into an even number of pie wedges. For each opposing wedge pair, the cumulative distribution of stars is derived and compared. From the differences between the wedges of one circle, each of the trial centers is assigned a value and a contour plot results. By deriving the minimum of this contour plot we derive the center of symmetry, thus the center of the cluster. This has been done for different numbers of pie wedges and orientations as well as the same magnitude cuts as in the previous method. We find the same behavior of the center position dependent on the magnitude cuts. After applying the same method for finding the average center position out of the different magnitude cut dependent centers, we have a very good agreement between the centers obtained with pie wedges and with isodensity contours. 

\begin{figure*}
  \centering \includegraphics[width=5.5cm]{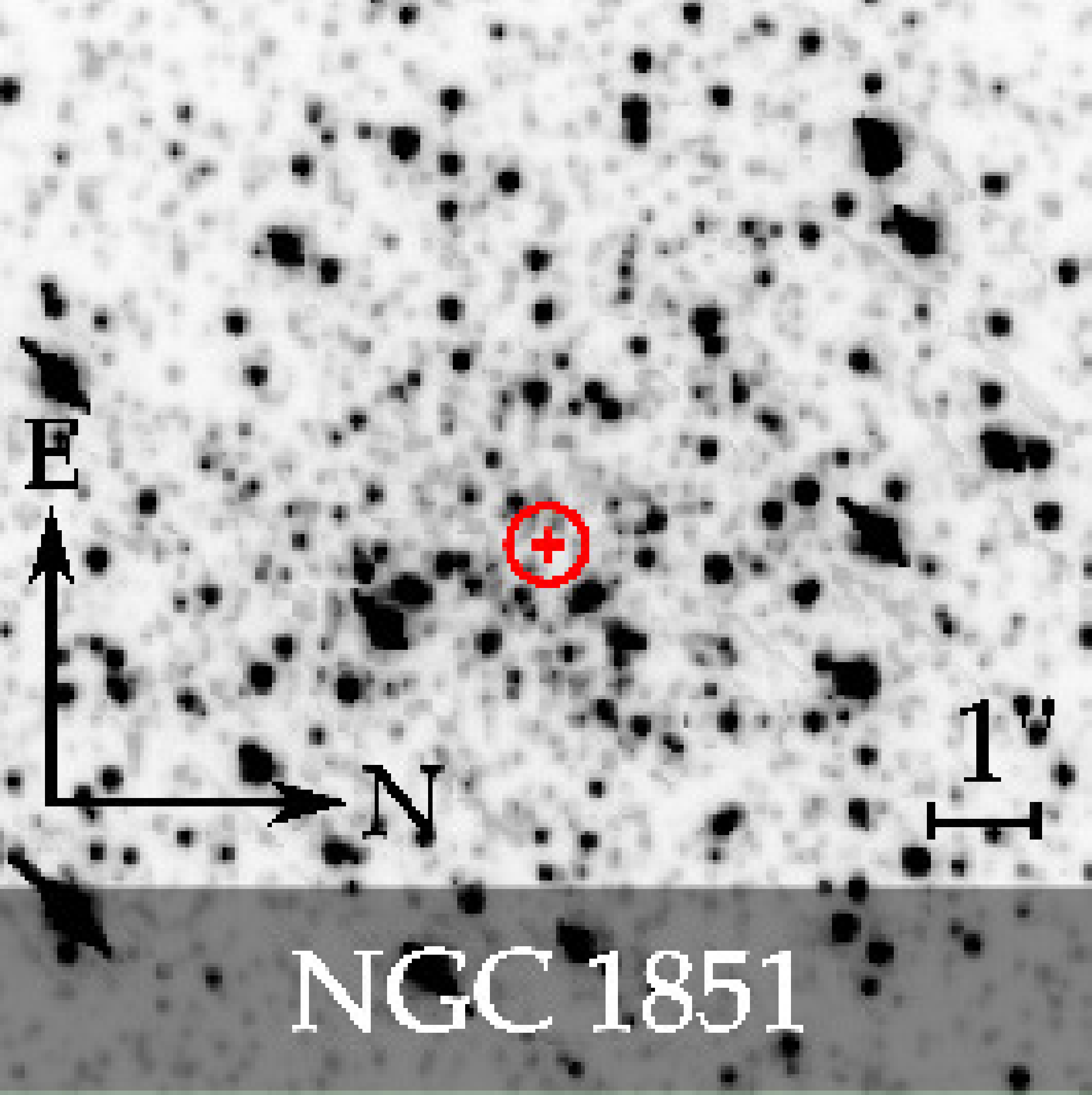}
  \centering \includegraphics[width=5.5cm]{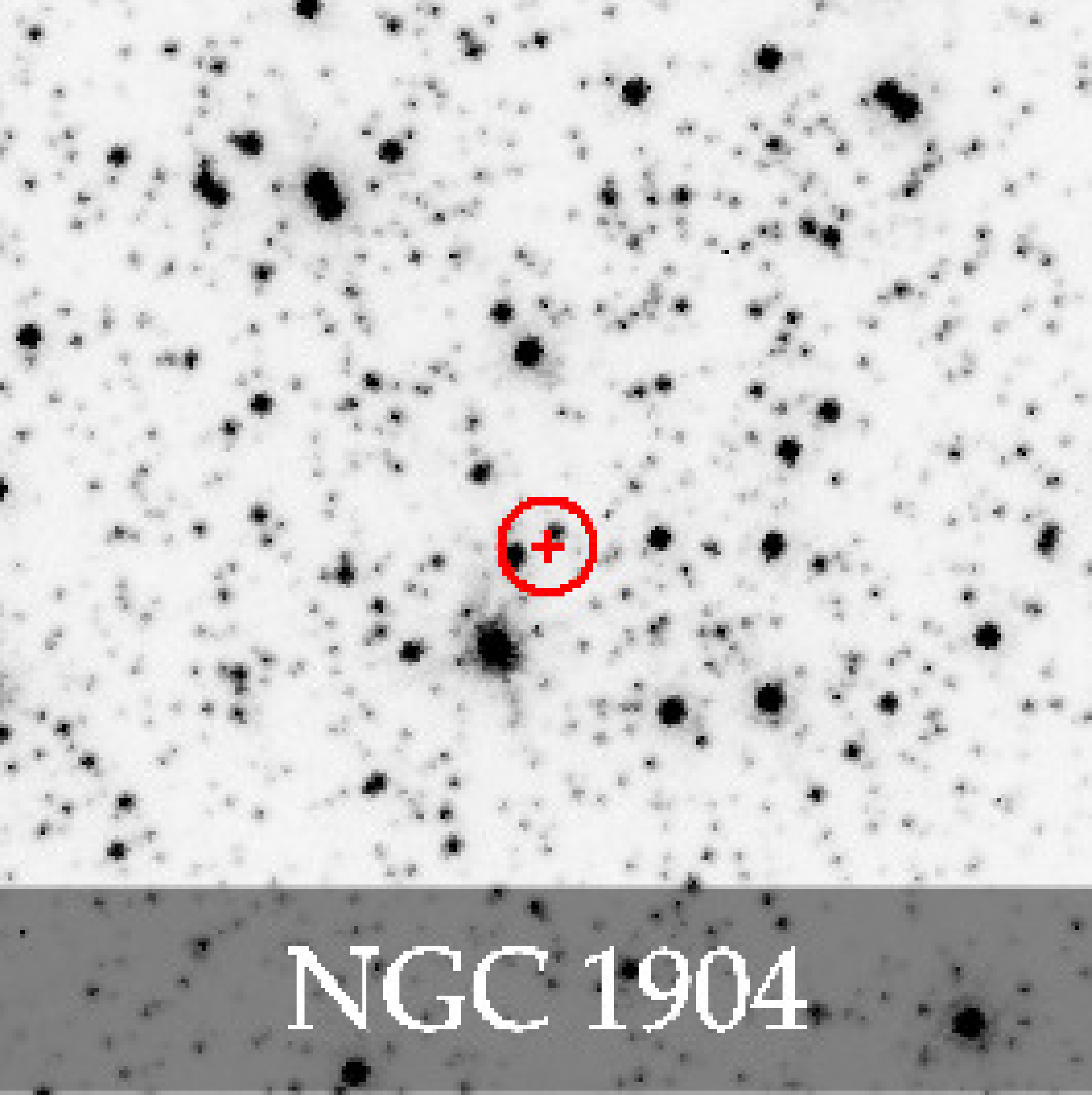}
  \centering \includegraphics[width=5.5cm]{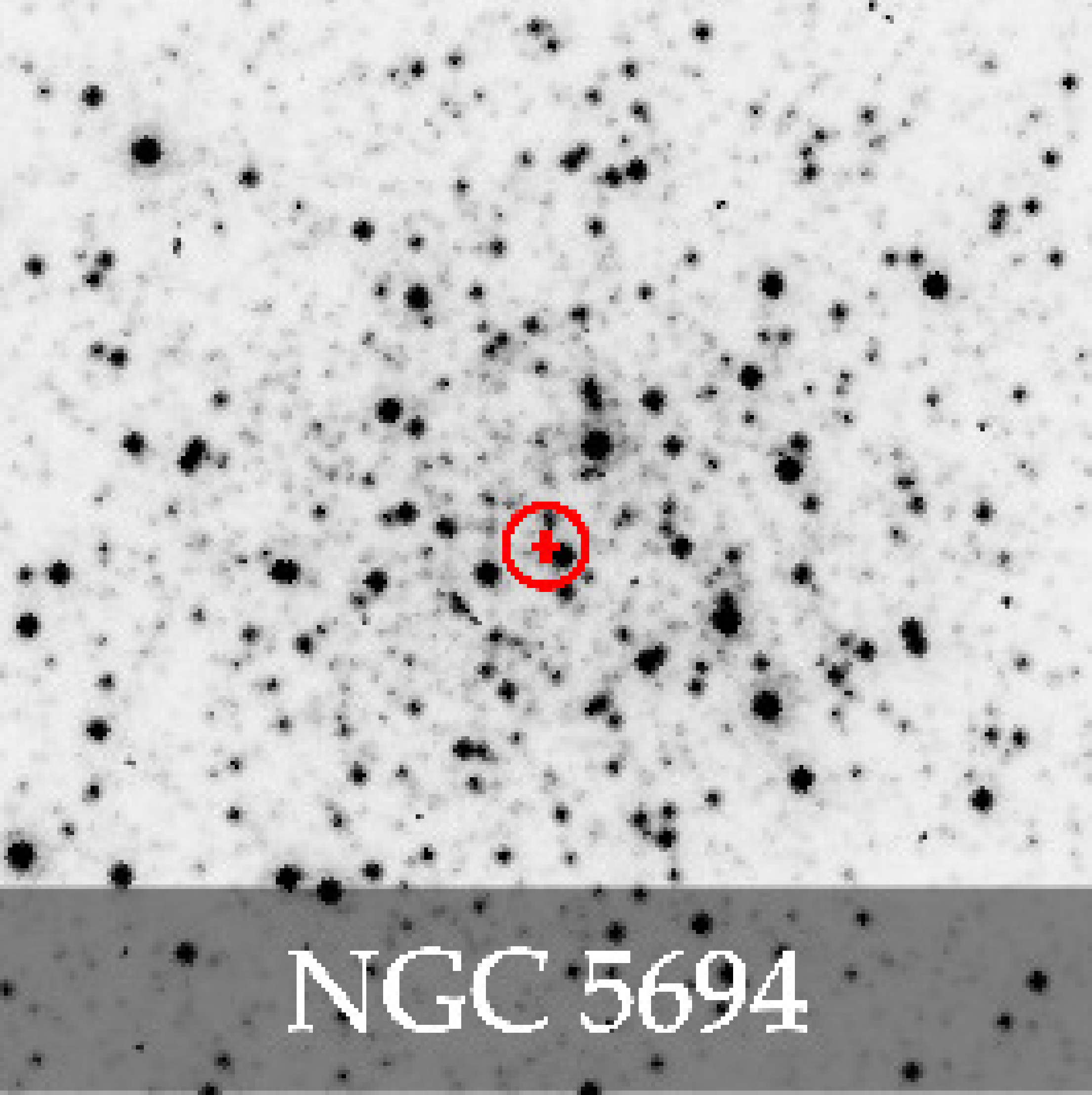}\\
  \centering \includegraphics[width=5.5cm]{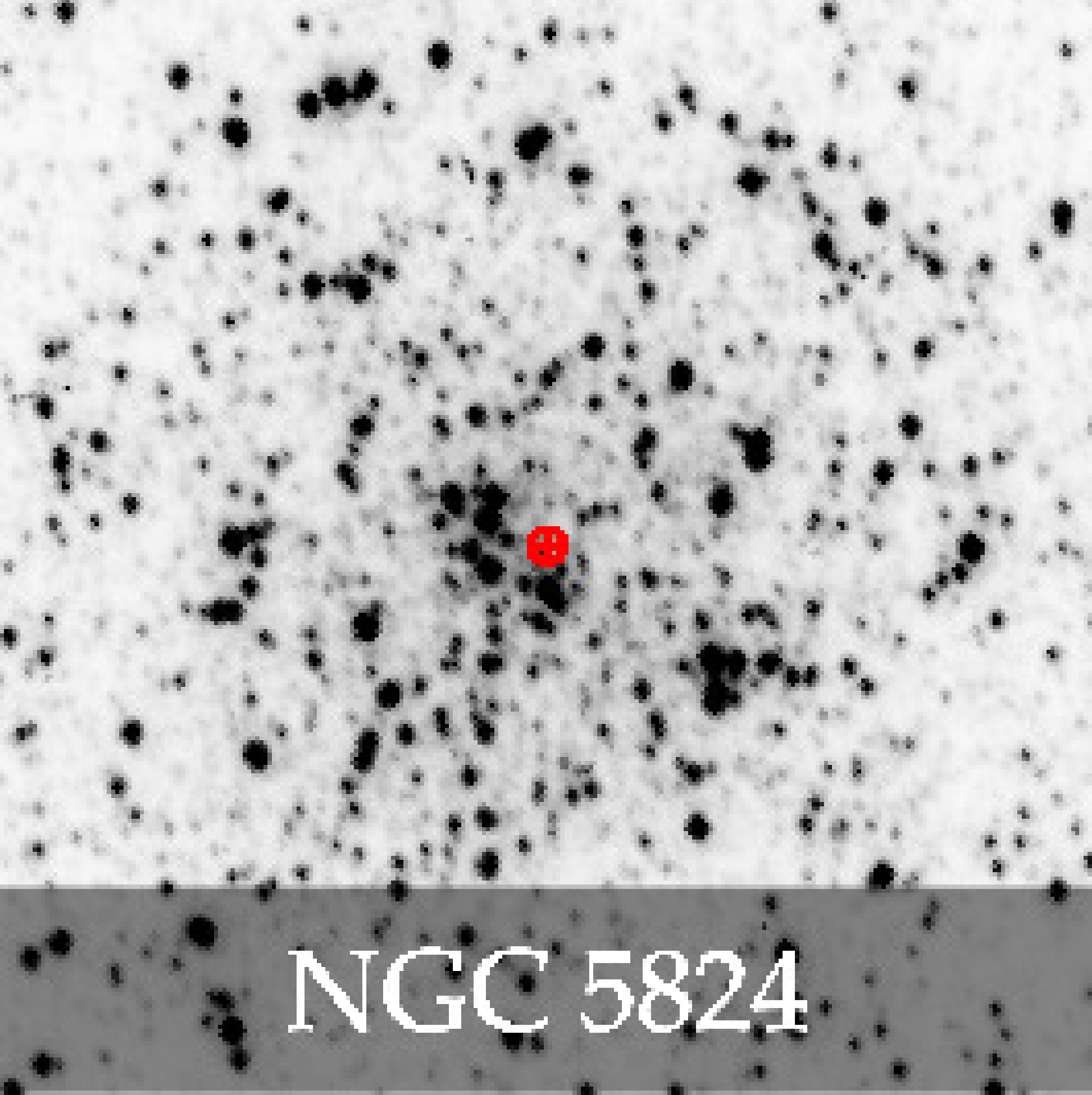}
  \centering \includegraphics[width=5.5cm]{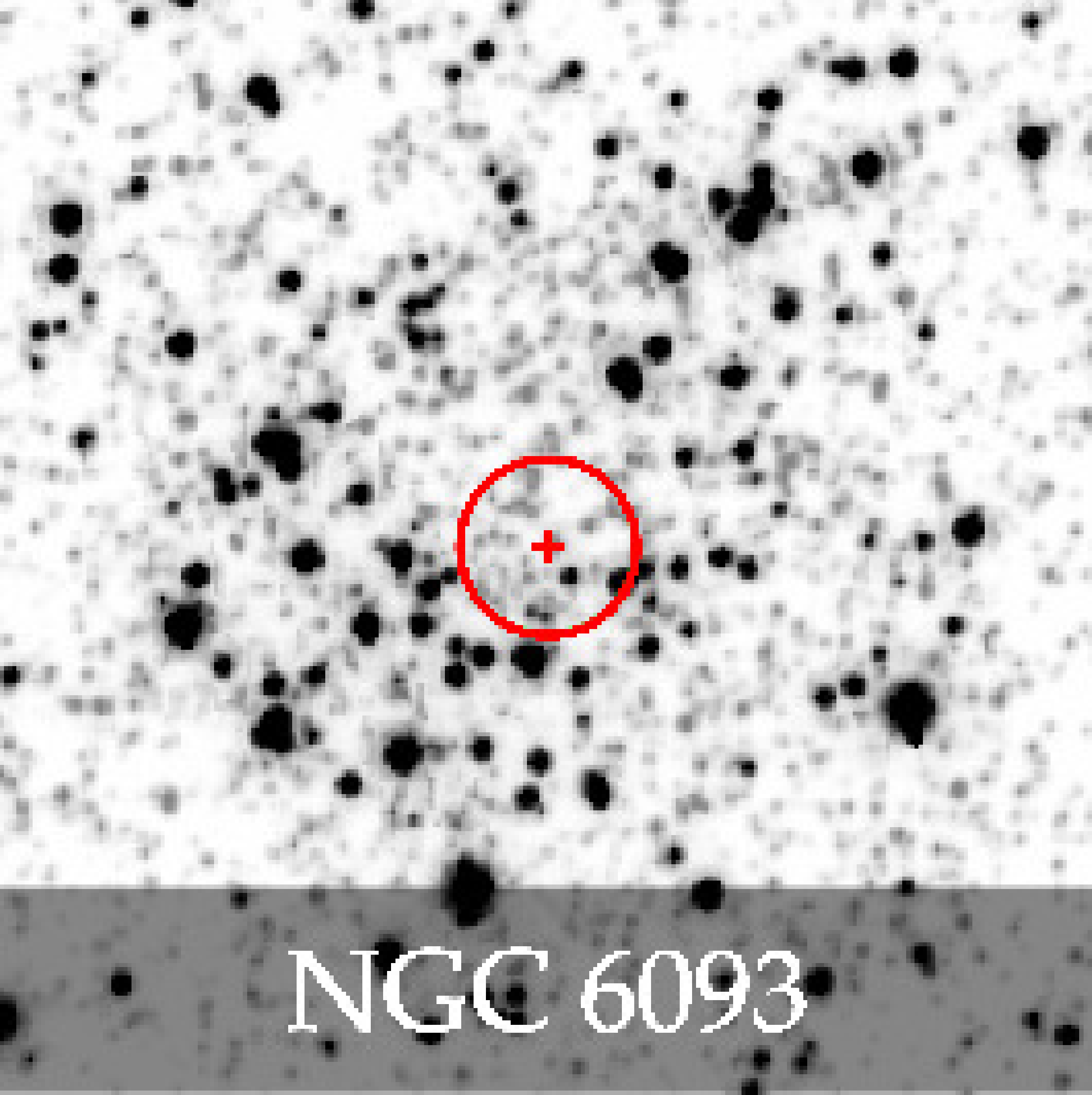}
  \centering \includegraphics[width=5.5cm]{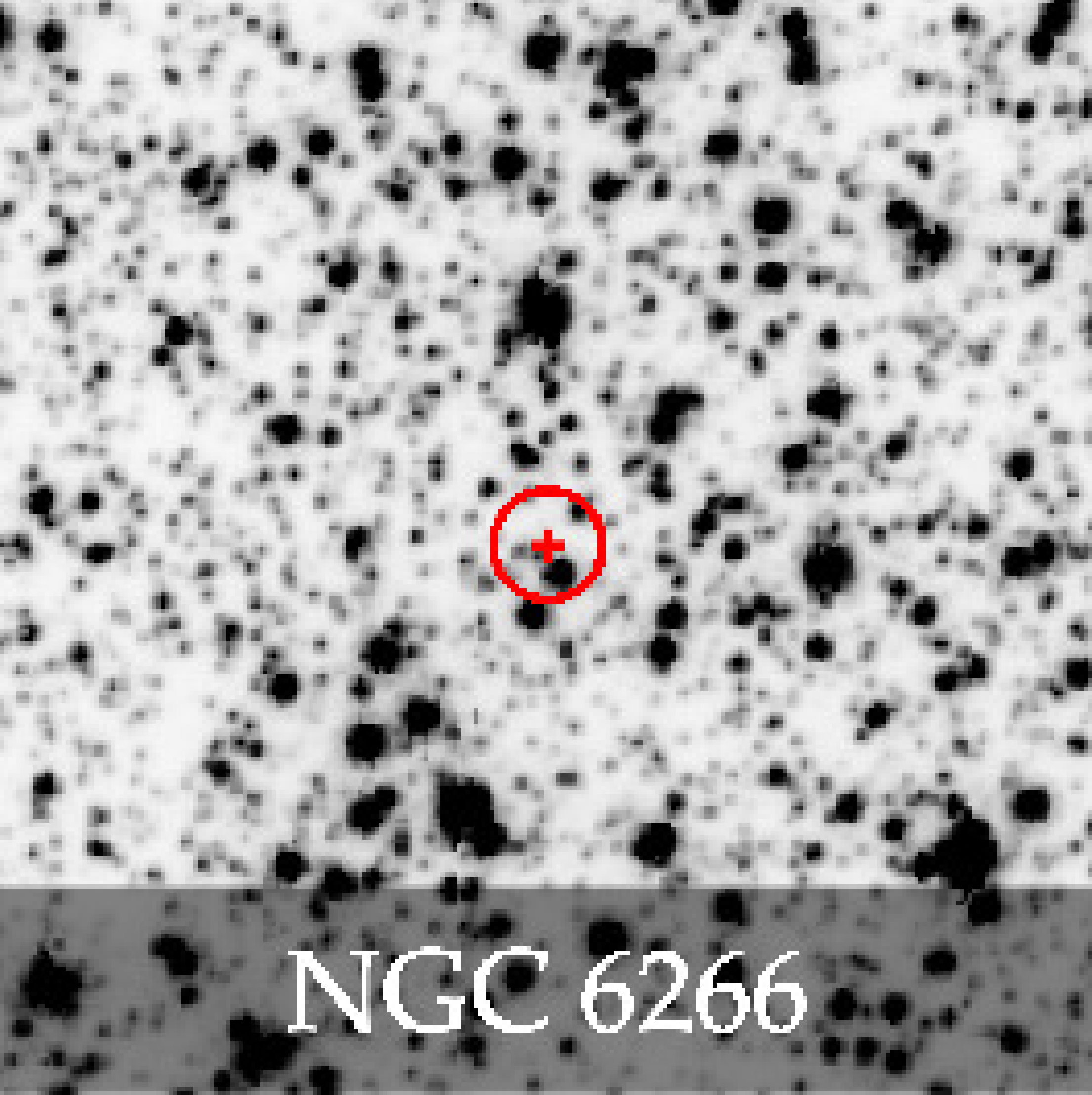}
  \caption{Finding charts for the centers of the six clusters. All images are orientated in the same way as the first one in the left panel. The size of the circle corresponds to the error of the center position.}
  \label{fig:find}
\end{figure*}

As a third test we create isolight-density contours from the image. After masking out the brightest stars, the image is processed like the star catalogue by overlaying a grid with a spacing of $2''$, applying a circle with a radius $4''$, and evaluating the light per pixel evaluated for each grid point. From this, a contour plot is derived and ellipses fitted to find the center of the light distribution. Despite the masking, the contours are still very noisy due to contamination of bright stars. However, the centers derived from this method also agree within the errorbars with our previous centers. We use the isodensity contour center as our final result which is listed in Table \ref{tab:phot}. In addition, the finding charts for each cluster are shown in Figure \ref{fig:find}.

We compare the derived center positions with the values obtained by \citet{noyola_2006}. In Table \ref{tab:phot} we list the distance of the two center positions. For all clusters the values agree well within our uncertainties and the uncertainty of \citet[][$\Delta=0.5''$]{noyola_2006}. The only exception is NGC 6266. For this clusters the two centers lie almost $3 ''$ apart from each other which is out of range for both uncertainties. Due to the unfortunate position of the center of NGC 6266 on the WFPC2 chip (close to the image boarder), the center position of \citet[][]{noyola_2006} suffers from large uncertainties. We repeat the analysis of \citet[][]{noyola_2006} with the new ACS image and find no discrepancy between the two centers. We therefore, conclude that the center derived in this work is more reliable due to the better quality of the ACS image.

\begin{figure*}
  \centering \includegraphics[width=9.0cm]{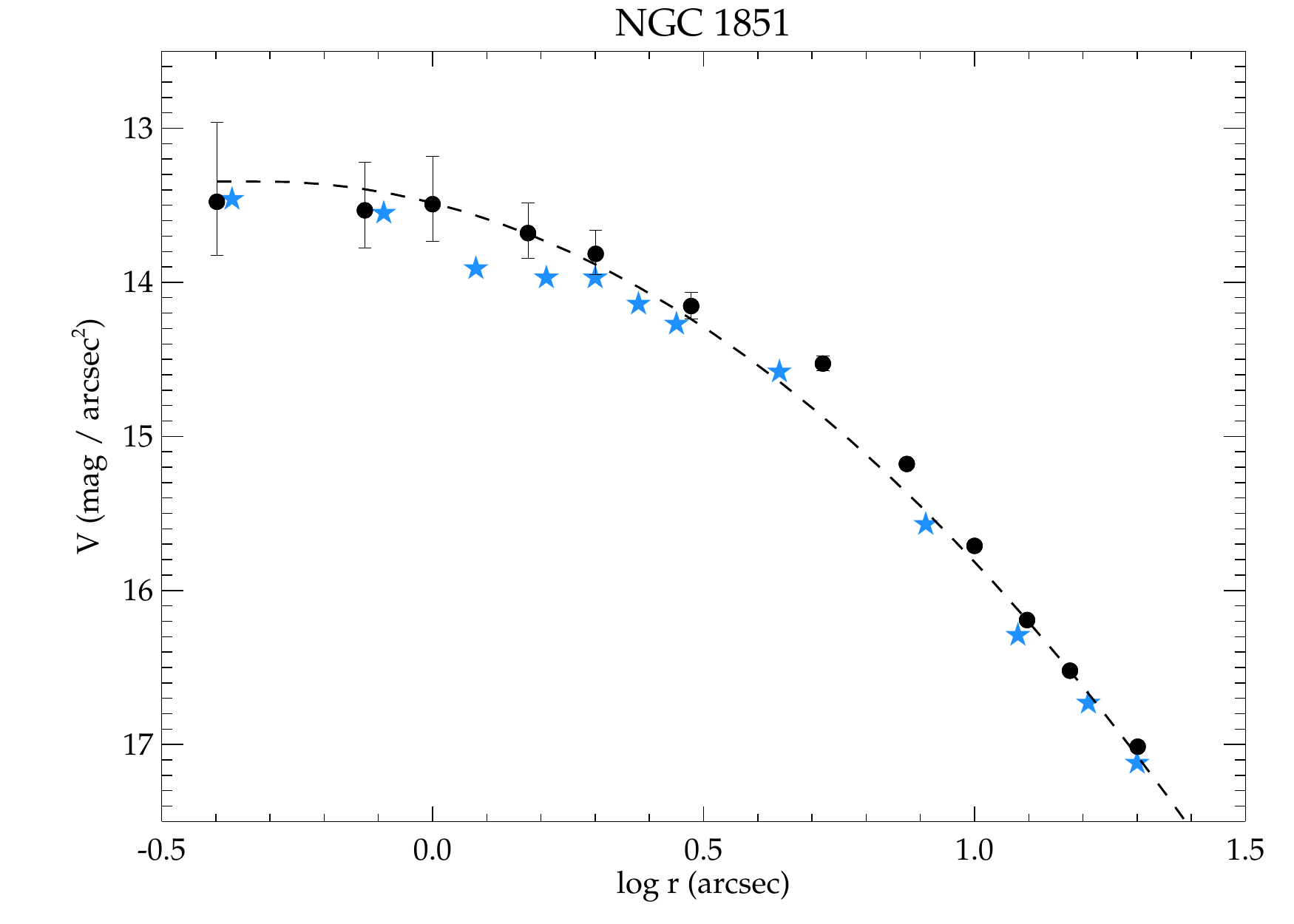}
  \centering \includegraphics[width=9.0cm]{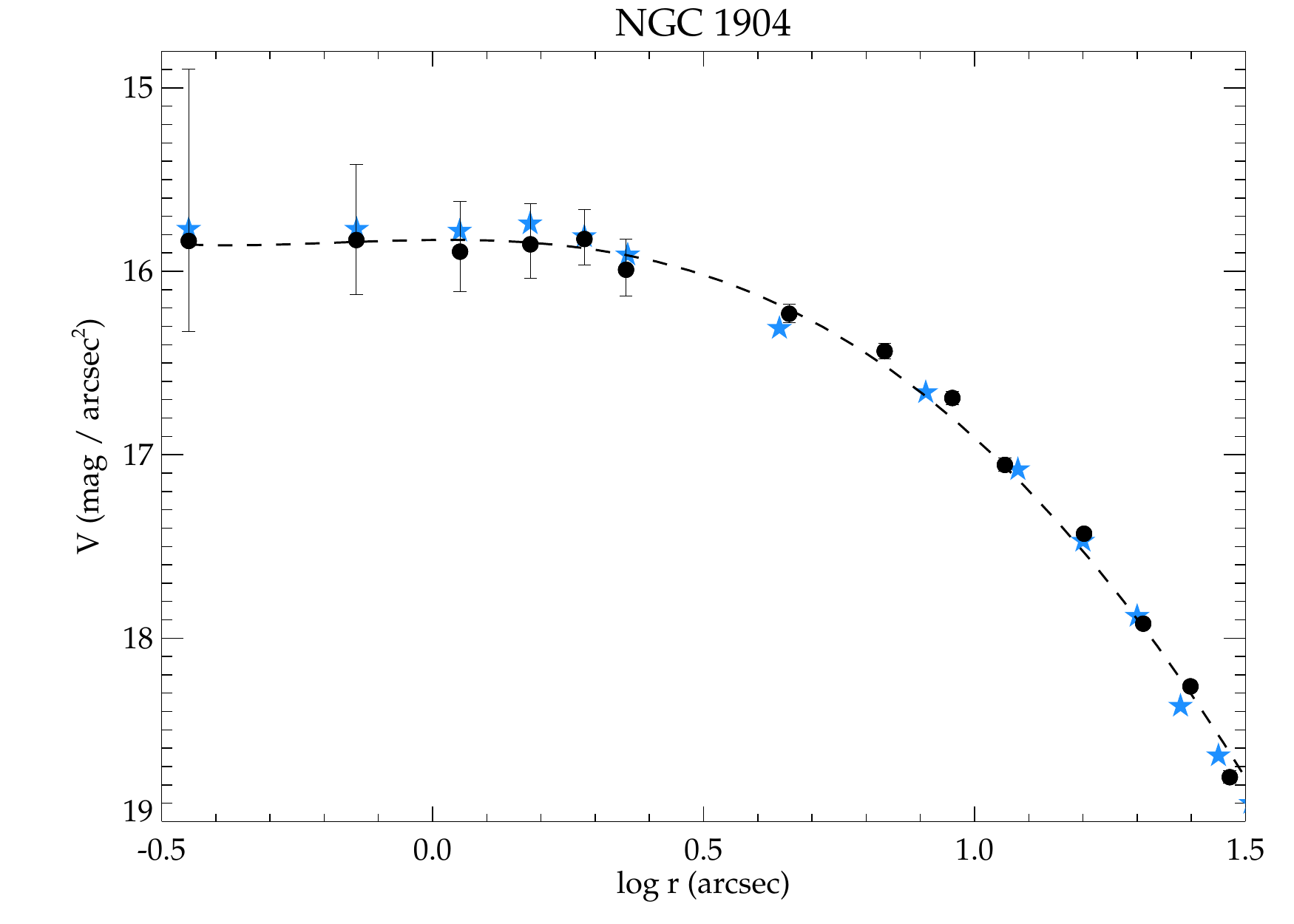}
  \centering \includegraphics[width=9.0cm]{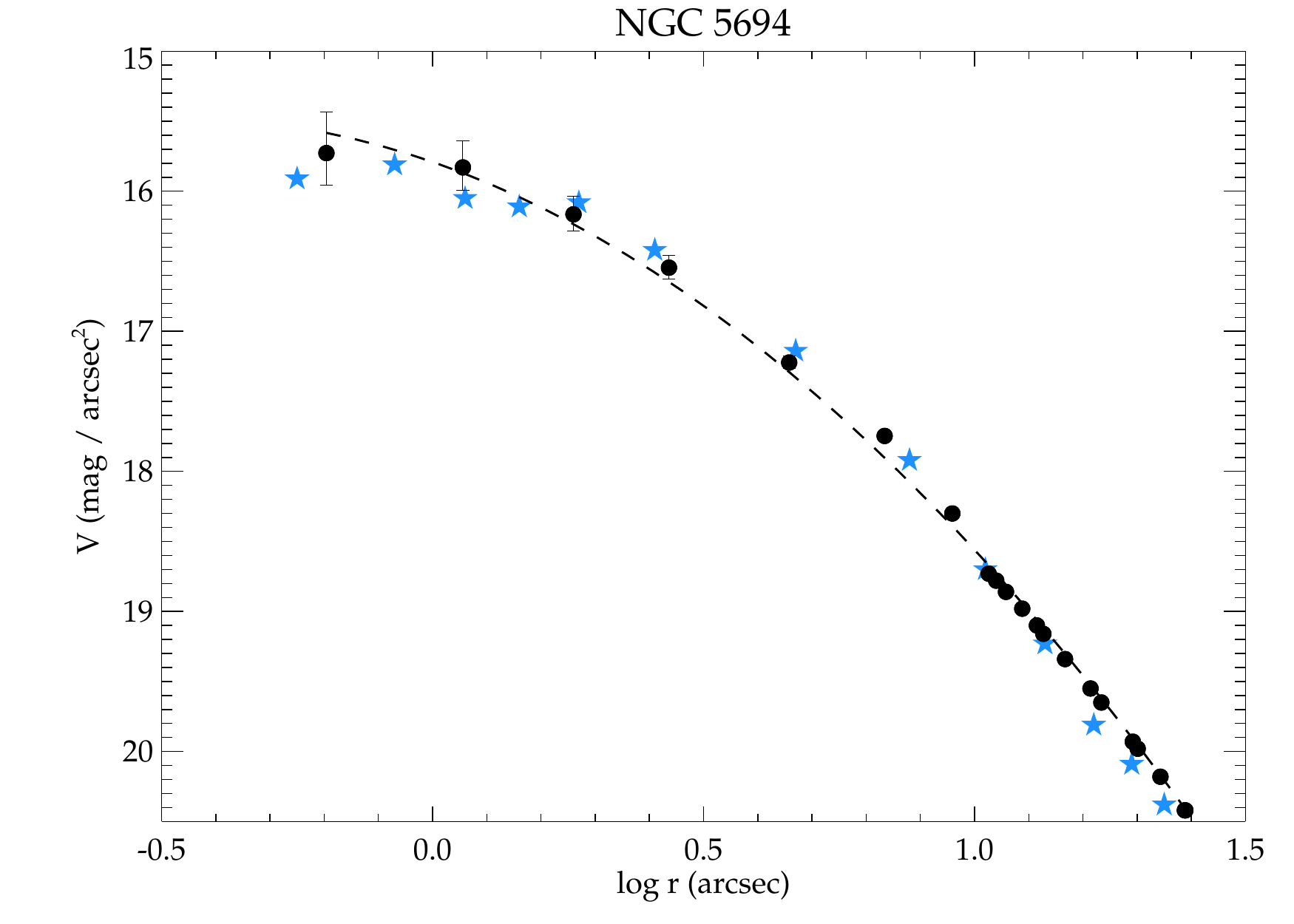}
  \centering \includegraphics[width=9.0cm]{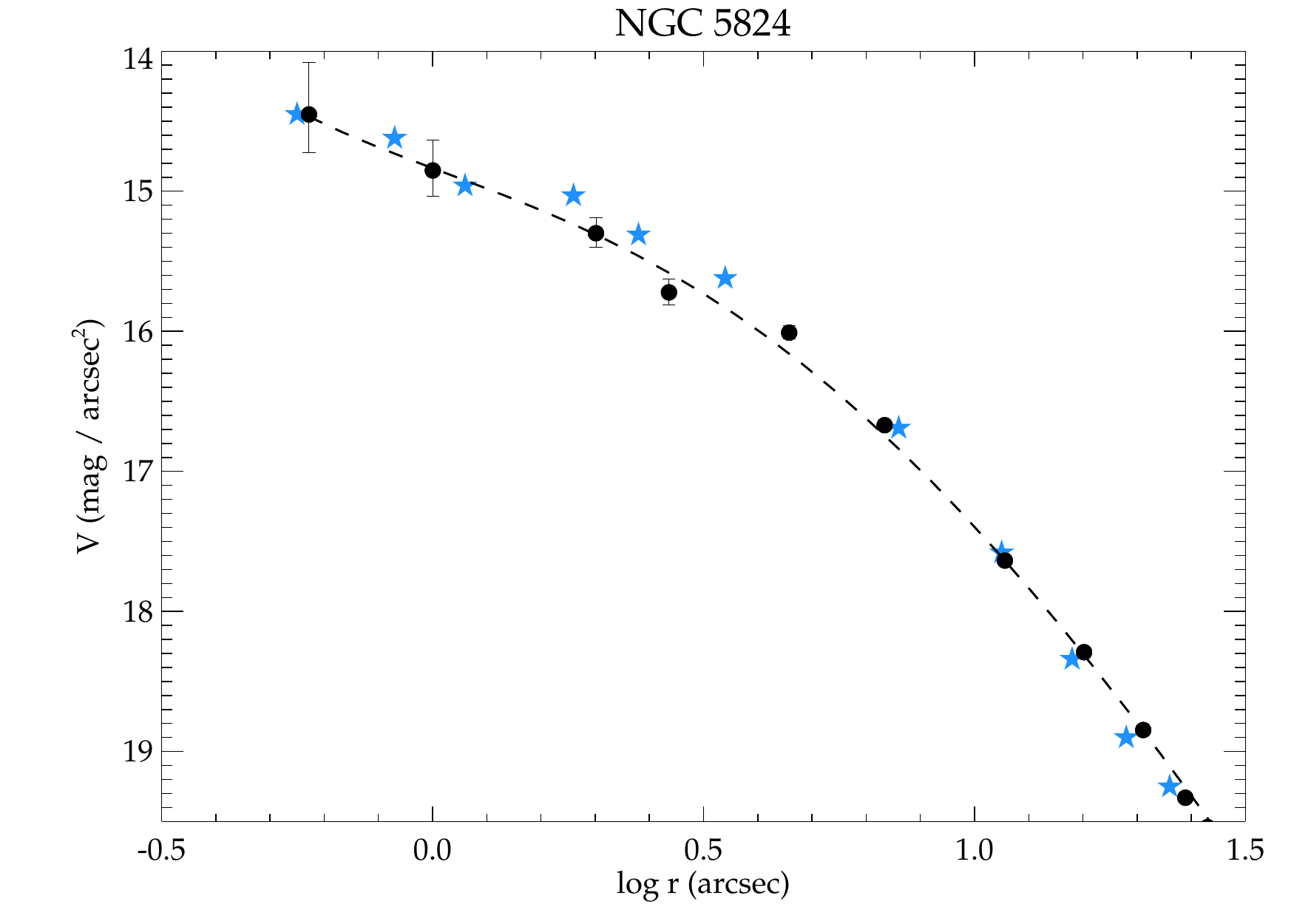}
  \centering \includegraphics[width=9.0cm]{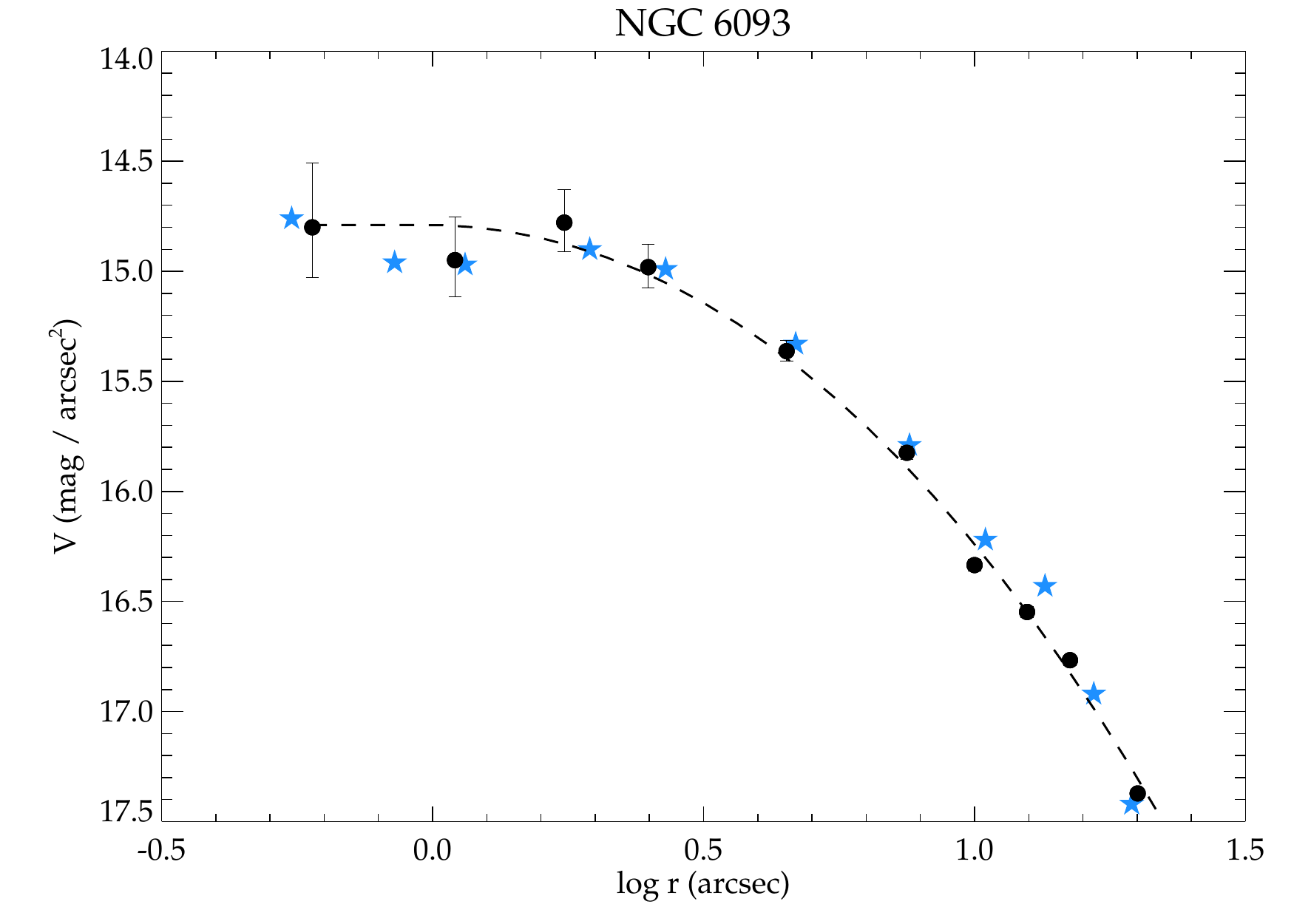}
  \centering \includegraphics[width=9.0cm]{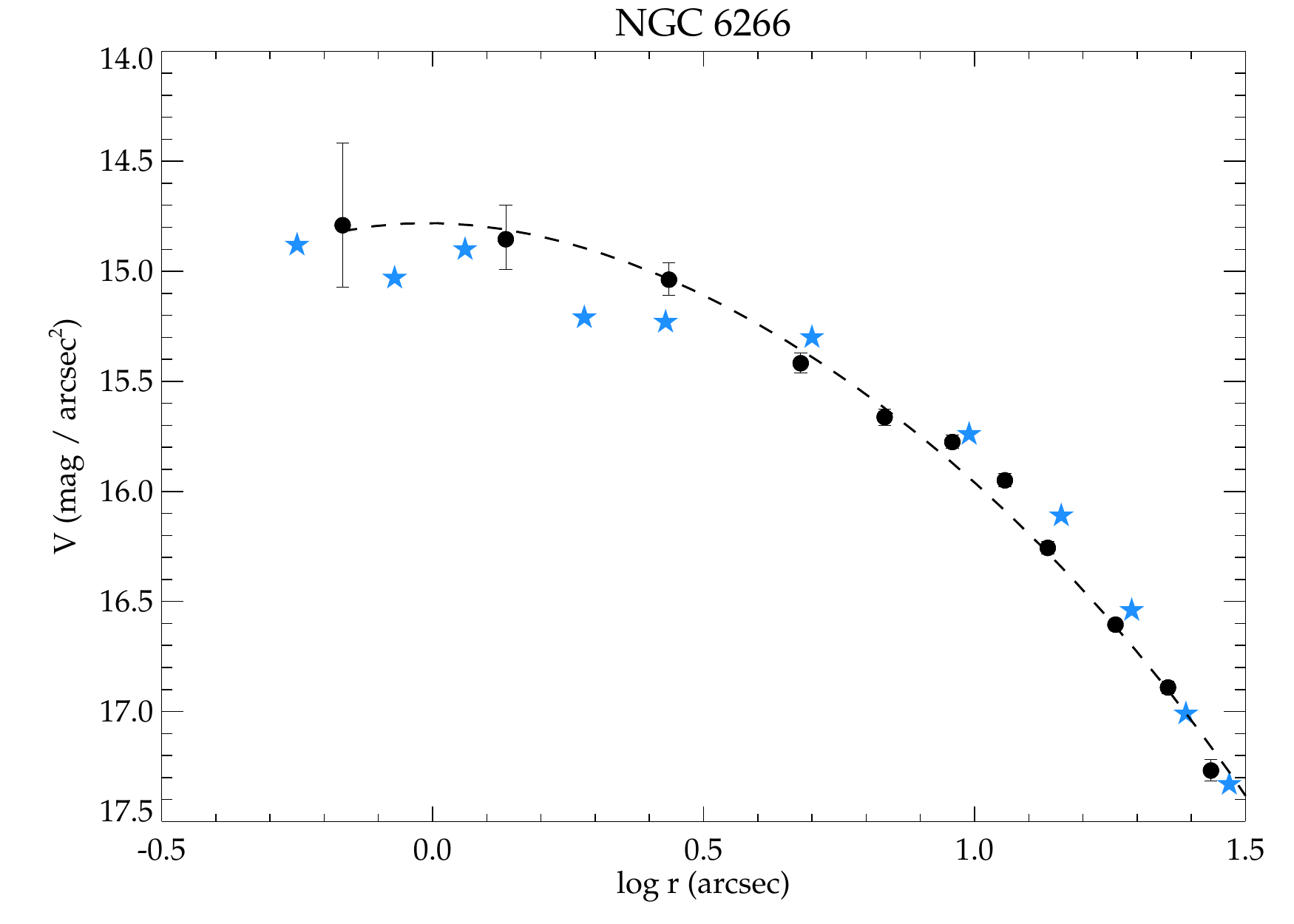}
  \caption{The surface brightness profiles of our six clusters. Black circles are the measured points with our method of combined pixel statistics and star counts, the blue stars show for a comparison the points from the profile obtained by \citet{noyola_2006}. The dashed line shows our smoothed profiles.}
  \label{fig:sb}
\end{figure*}

\subsection{Surface brightness profile} \label{phot_sb}

The next step in the photometric analysis is the computation of the surface brightness profile. This is needed as an input for the dynamical Jeans models and provides important information about the dynamical state of the cluster, e.g. if it is core collapsed or not. Due to the finite number of stars, a surface brightness profile is always affected by shotnoise. Therefore, deriving a smooth profile which can be easily deprojected is not straightforward. 

As described in \citet{luetzgendorf_2012a} we use a combination of pixel statistics and star counts. The pixel statistics are carried out by masking the bright stars in the image and drawing radial bins around the center. In each bin, the counts per pixel are estimated using a robust biweight estimator \citep{beers_1990}. The starcount profile is determined by taking the same bins and calculating the surface brightness density of the remaining bright stars which have been masked out in the previous method. The final profile is the combination of the two methods, scaled to the photometric points of \cite{trager_1995}.  The errors of the profile are derived from Poisson statistics of all stars in each bin. For radii larger than the extent of our images we complete the profile with the points from \cite{trager_1995}. For NGC 1904 this method was not applicable, due to its low central stellar density. When splitting the area into individual stars and light, low number statistics cause large uncertainties. The surface brightness profile for this cluster was therefore derived by pixel statistics only. 

Figure \ref{fig:sb} shows the surface brightness profiles for all the clusters compared with the unsmoothed surface brightness profiles of \citet[][blue stars]{noyola_2006}. The large difference between our SB profile and the one of \citet[][]{noyola_2006} most probably occurs from the difference in the centers. However, except of two points, the profiles still agree within their errorbars. The profiles are not completely smooth. This is natural since a cluster with a finite number of stars never has a smooth profile. However, when deprojecting the profile, little uneven parts in the shape of the profile are enhanced and may create strong features in the resulting density profile. This then further affects the derived dynamical models. In order to avoid this we smooth our profiles before deprojecting them using Tschebychev polynomials \citep[][]{noyola_2006}. The smoothed profiles are also shown in Figure \ref{fig:sb} (dashed lines). From these profiles we derive the total luminosity in the $V$-band of each cluster using Multi-Gaussian-Expansions \citep{emsellem_1994}. The final photometric parameters are listed in Table \ref{tab:phot}.

In order to test how much the uncertainties in the center position affect the SB profile, we run Monte Carlo simulations on the center position and construct a new light profile on every generated center position. We find that the uncertainties produced from a center are within the errorbars of the SB profile and, as shown in Monte Carlo simulations in \cite{luetzgendorf_2012a}, will not effect the measurements on the black-hole mass.


\addtocounter{table}{1}

\begin{table*}
\caption{Photometric properties of the clusters in our sample. The columns are: (1) name of the cluster (2,3) position of the center on the image in pixel, (4) uncertainties in pixel, (5, 6) position of the center on the image in RA and DEC, (7) uncertainties in arcseconds, (8) reference image used for the center determination, (9) distance to the center derived in \citet{noyola_2006}, and (10) total luminosity of the cluster.}             
\label{tab:phot}      
\centering
\begin{tabular}{l r r r r r r r r r}
\hline \hline
\noalign{\smallskip}
Cluster 	&$x_{c}$  (pix)	& $y_{c}$  (pix)& $\Delta$ (pix) & RA$_{c}$  (J2000)		& DEC$_{c}$ (J2000)		&$\Delta ('')$& Reference Image & NG06 ($''$)	&  $\log L_{tot} (L_{\odot})$ \\
 \noalign{\smallskip}
\hline
\noalign{\smallskip}
NGC 1851	& $2919$	& $3011$& $7$	&  $05:14:06.7$ 	&  $-40:02:49.3$		&$0.3$& \verb|ACS SURVEY|		& $0.1$ & $5.25$ \\
NGC 1904	& $262$	& $289$ 	& $9$	&  $05:24:11.0$ 	&  $-24:31:27.9$ 	&$0.5$& \verb|u2vo0402t_c0f.fits|	& $0.0$	& $4.95$ \\
NGC 5694	&  $453$	&  $367$	& $8$	&   $14:39:36.3$	&   $-26:32:19.6$	&$0.4$&  \verb|u2y70105t_c0f.fits|	& $0.7$ & $5.10$ \\	
NGC 5824	& $365$	& $313$	& $3$	&   $15:03:58.6$	&   $-33:04:05.3$	&$0.2$& \verb|u2y70205t_c0f.fits|	& $0.4$ & $5.39$ \\
NGC 6093	& $3007$	& $2943$	& $16$	&  $16:17:02.4$ 	&  $-22:58:32.6$		&$0.7$& \verb|ACS SURVEY|		& $1.0$	& $5.18$ \\			
NGC 6266& $1764$	&$3424$	& $6$	&  $17:01:13.0$	& $-30:06:48.2$ 		&$0.3$& \verb|j92103050_drz.fits| 	& $2.8$	& $5.57$ \\
\noalign{\smallskip}
\hline 
\end{tabular} 
\end{table*}


\section{Spectroscopy}\label{sec:spec}

                        
The inner kinematic data were obtained with the GIRAFFE spectrograph of the FLAMES \citep[Fiber Large Array Multi Element Spectrograph,][]{pasquini_2002} instrument at the Very Large Telescope (VLT) using the ARGUS mode (Large Integral Field Unit). NGC 6266 was observed in March 2010 and the other five clusters were observed in February 2011. The ARGUS unit was set to the 1 : 1.67 magnification scale (pixel size: $0.52 ''$, $14 \times 22$ pixel array). Depending on the core radius and surface brightness the pointings were arranged symmetrically around the center with exposure times of either 600 s or 900 s. The orientation and placement of the individual pointings can be seen in Figure \ref{fig:vel}, where the final velocity maps are shown. For detailed information about the spectroscopic observations we refer to Table \ref{tab:obs}.

\begin{figure}
  \centering \includegraphics[width=0.5 \textwidth]{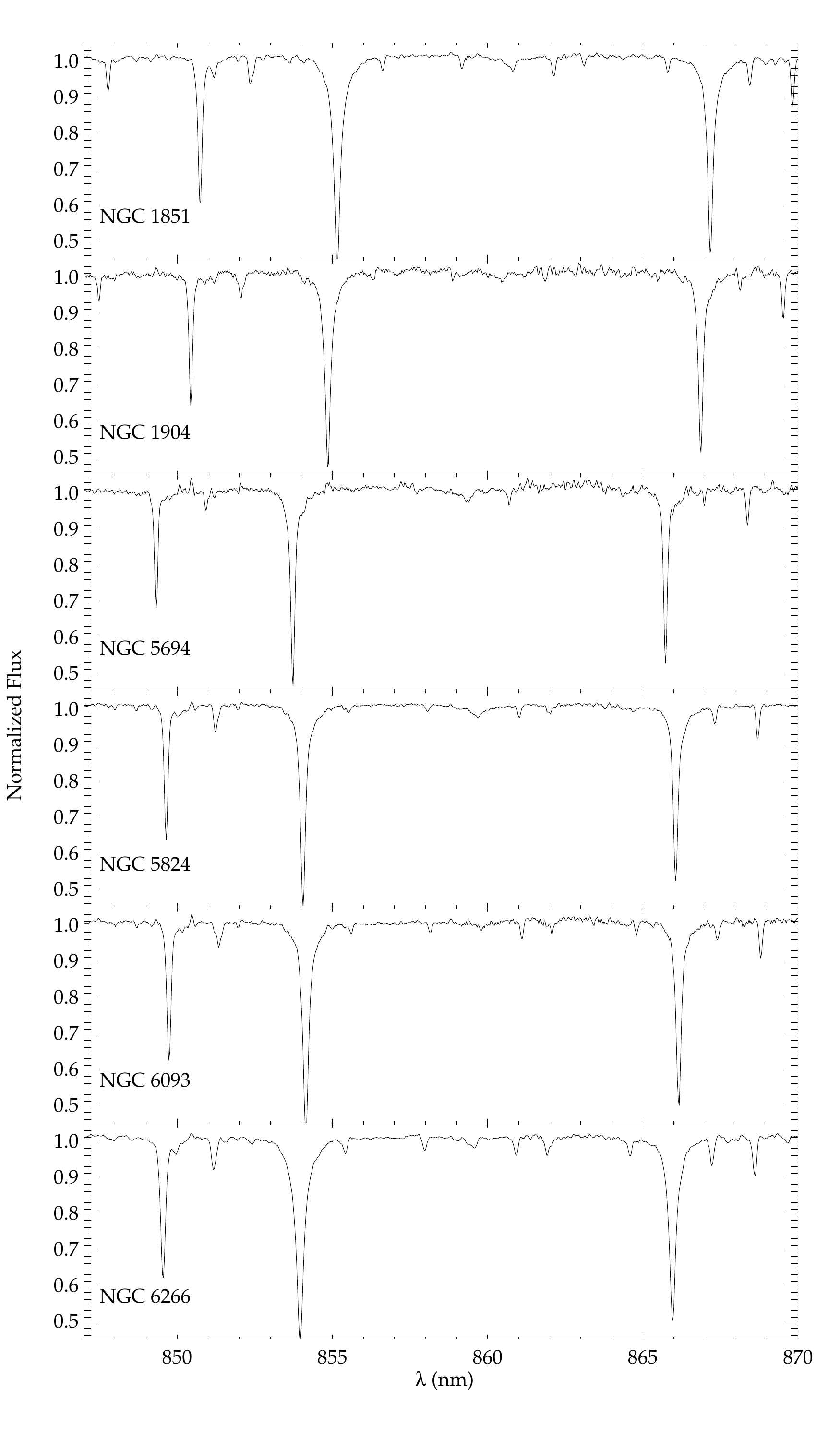}
  \caption{Combined spectra showing the Ca-triplet region, of the second radial bin of each cluster. The spectra are representative for the ones from which we derive the kinematic measurements. }
  \label{fig:spec}
\end{figure}

The observations were taken in the wavelength range of the Calcium Triplet ($\sim 850 \, \mathrm{nm}$), a prominent absorption feature ideally suited for kinematic measurements (example spectra are shown in Figure \ref{fig:spec}). With the low spectral resolution mode set-up LR8 ($820-940 \, \mathrm{nm}, \, \mathrm{R} = 10400$) and sufficient S/N ($\sim 100$) we can measure robust velocity dispersions down to 8 \kms with an accuracy of of 1-2 \kms using a nonparametric fitting algorithm. Below that value, we are sensitive to template mismatch and continuum placement. We report three velocity dispersions that are below 8 \kms, where we have performed simulations in order to understand the uncertainties. While the uncertainties are measured robustly, bias in the values can be large comparatively. We do not attempt to estimate the bias for these low values since they will have little consequence for the enclosed mass profile given the other well-measured points.

The reduction of all clusters was done homogeneously using the GIRAFFE pipeline programmed by the European Southern Observatory (ESO) in combination with additional programs. The pipeline with its five recipes (\textit{gimasterbias, gimasterdark, gimasterflat, giwavecalibration, giscience}) is described in more detail in \cite{nora11}. Before applying this pipeline, however, we use the program LA-Cosmic developed by \cite{Lacos} to remove the cosmic rays from our raw spectra. The final routine \textit{giscience} produces a reduced science frame as well as the extracted and rebinned spectra frame from the input observations. A reconstructed image of the respective field of view of the ARGUS observations is also produced by the pipeline. This is later used to match the observations with the HST image.

In order to subtract the sky from the spectra using the 14 skyfibres from the ARGUS observations, we use the IDL program \textit{argus\_ specsub}\footnote{The IDL version can be found at http://nora.luetzgendorf.de} which uses the algorithm of a routine by Mike Irwin \citep{battaglia_2008}. The IDL version of the program does almost the same as the original program except the final cross correlation. For a detailed description of the routine see \citet{battaglia_2008} and \citet{luetzgendorf_2012a}. Due to cross talk between bright stars in object fibres and adjacent sky fibres, some of the sky fibres were contaminated and not usable for the analysis and therefore had to be excluded. In order to avoid bright stars dominating the averaged spectra when they are combined, we apply a normalization to the spectra by fitting a spline to the continuum and divide the spectra by it.

\begin{table}
\caption{Spectroscopic observations of the clusters. The columns list the cluster name, the Program ID of the individual observations, the exposure time of the ARGUS observation ($t_{exp}$), the numbers of exposures and the position angle of the ARGUS array (measured from North to East, the long axis aligned with North-South) for different pointings.}             
\label{tab:obs}      
\centering
\begin{tabular}{cccccc}
\hline \hline
\noalign{\smallskip}
Cluster &  Pogram ID & $t_{exp}$ &  Exposures  & Position Angle &  \\
        &            & [s] &  & [degree] &  \\
 \noalign{\smallskip}
\hline
\noalign{\smallskip}
NGC 1851 	& 086.D-0573 	& 600  	& 24 	& 0 		&\\
NGC 1904 	& 086.D-0573 	& 600  	& 12 	& 0 		&\\
 			&  				& 900  	& 12 	& 0 		&\\
NGC 5694 	& 086.D-0573 	& 900 	& 3 		& 0 		&\\
NGC 5824 	& 086.D-0573 	& 600 	& 3 		& 0 		&\\
 			& 				& 900 	& 6 		& 0 		&\\
NGC 6093 	& 086.D-0573 	& 600  	& 12 	& 0 		&\\
NGC 6266 	& 085.D-0928 	& 600 	& 9  	& 0  	& \\
         	&             	& 600 	& 9  	& 45 	& \\
         	&             	& 600 	& 9  	& 90 	& \\
         	&             	& 600 	& 12  	& 135 	& \\
\noalign{\smallskip}
\hline 
\end{tabular} 
\end{table}


\section{Kinematics} \label{sec:kin}

This section describes how the velocity map and the velocity-dispersion profile are derived and matched with the photometric data. The kinematic profile is later used to fit analytic Jeans models and set constraints on the mass of a possible black hole.

\subsection{Registration of images} \label{subsec:recon}

Before deriving the velocity for each spaxel, the single pointings need to be combined according to their relative shifts. This can be done by using the header information of each pointing and using their relative shifts to each other. This works well for pointings with the same position angle since the relative shifts are quite accurate. For rotated pointings and for absolute positions on the sky, the header information found in the reconstructed ARGUS pointings is not accurate enough. Due to the fact that ARGUS is not exactly rotated around its center, different rotation angles are offset from each other. For this reason, we have to reconstruct each ARGUS pointing on top of the HST image. This is done by two-dimensional cross correlation. The HST image is convolved with a Gaussian according to the seeing of the VLT observations and iteratively cross-correlated with each of the ARGUS pointings. With this method we obtain an accurate position ($\sim 0.5''$) of the pointings on the HST images and thus, their absolute positions and relative shifts to each other. When available we used $I$-band images to do the cross correlation. For the ACS images no calibrated and combined $I$-band image was available. We therefore used the catalog to produce an artificial $I$-band image using the IRAF routine \textit{mkobjects}.

\subsection{Velocity map} \label{subsec:velmap}

The velocity map is computed by arranging the pointings according to their positions, overlaying a grid with a spaxel size of $0.52 ''$, and combining the spectra (e.g. in overlapping regions or multiple exposures) in each spaxel of the grid. The velocity of each spaxel is measured on the combined spectra using the penalized pixel-fitting (pPXF) program developed by \cite{cappellari_2004}. 

In order to test individual spaxels for shot noise, we apply the same routines as described in \cite{nora11}. At every position of a star in the catalog, a two dimensional Gaussian is modeled with a standard deviation set to the seeing of the ground based observations ($\mathrm{FWHM} \sim 0.8''$) and scaled to the total flux of the star. We measure the absolute amount and fraction of light that each star contributes to the surrounding spaxels. For each spaxel we then have the following information: a) how many stars contribute to the light of that spaxel, and b) which fraction of the total light is contributed by each star. As in the case of NGC 2808 \citep{luetzgendorf_2012a}, we mask out spaxels for which either fewer than 10 stars contribute to the light, or a single star contributes more than 60~\% of the light. This leaves us with $2/3$ of the total number of spectra. We also test the effect of the mapping by comparing the velocity-dispersion profiles of the mapped and the unmapped spectra. The deviations are all within the uncertainties of the measurements and therefore not critical. However, in order to include any contribution by systematic errors we apply the mapping which also partly results in smoother profiles.

The template for measuring the velocities and velocity dispersions was taken from the same observations of each cluster in the following way: First, our shot noise routine calculates for each spaxel in the pointing how many stars contribute to the light of each spaxel. This is useful to mask spaxels which are dominated by a single star and to find the spaxels which could be used as templates. We select our template candidates by sorting out all spaxels where a single star contributes by more than 80 \%. These are identified in the color magnitude diagram and excluded if they are possible foreground stars. For each of the clusters we select 6 - 10 template candidates, depending on how many stars are available. The candidates are then divided into equal numbers of faint and bright stars. We bring all templates to the same velocity scale by measuring the relative velocities and shifting them and we combine the faint and the bright candidates to two master templates. We perform the kinematic measurements with the faint and the bright master template. For most of the clusters we find that the combined faint template fits the data better than the bright template, except for NGC 1904 and NGC 5694 where the bright template fits better. We also fit the spectra using all individual templates where the weight on each template is a free parameter. This test results in high weights for the fainter templates. For all further analyses we use the best fitting master template (faint: NGC 1851, NGC 5824, NGC 6093, NGC 6266, bright: NGC 1904, NGC 5694). The positions of the stars used to create the master template for each cluster in the CMD are displayed in Figure \ref{fig:cmd} (blue star symbols).

\begin{figure*}
  \centering \includegraphics[width=0.85\textwidth]{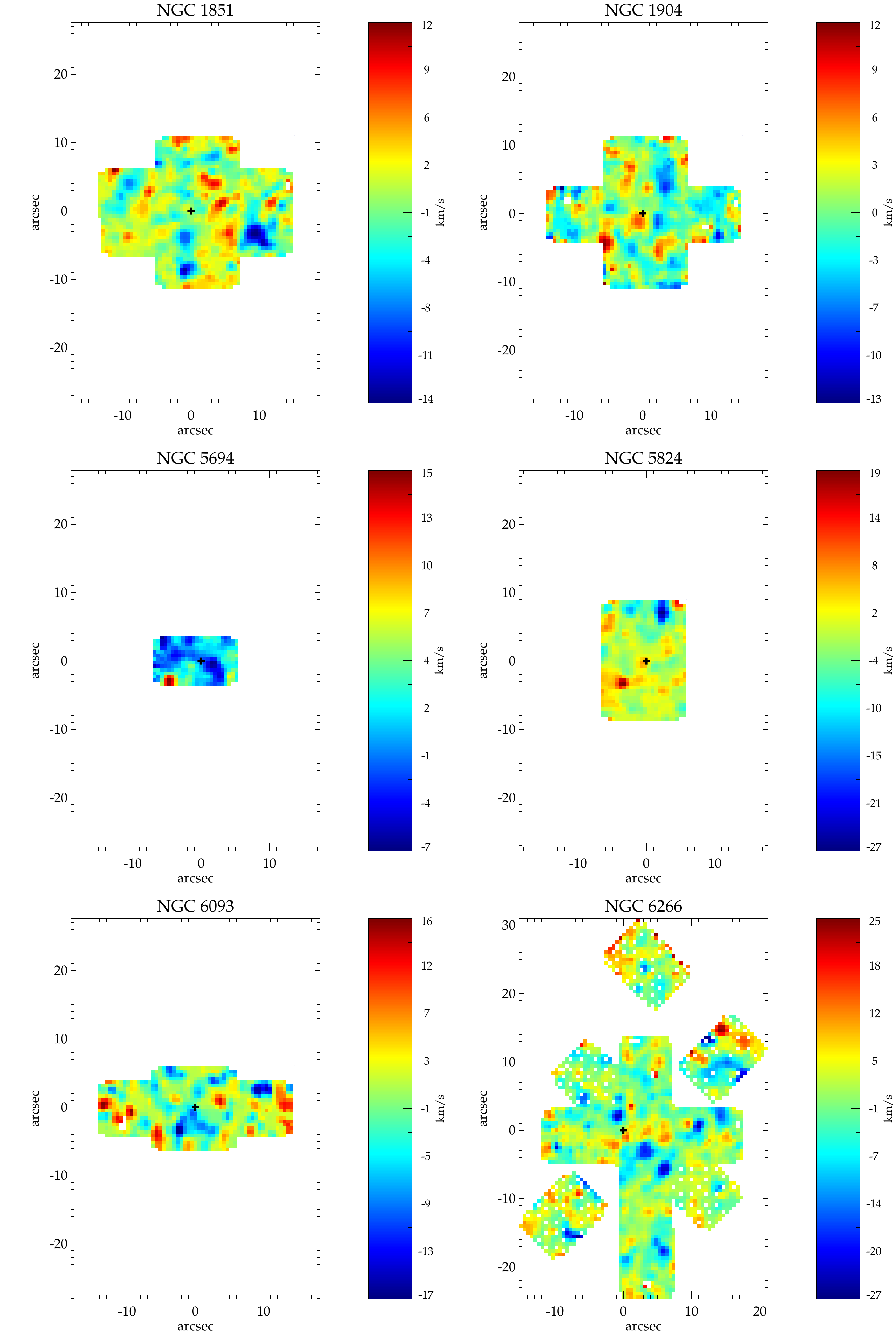}
  \caption{The final velocity maps of the clusters in our sample. The velocity scale has been adjusted to the mean velocity of each cluster. The black cross marks the center of the cluster. The orientation is the same as in Figure \ref{fig:find}}
  \label{fig:vel}
\end{figure*}

The final velocity maps of all clusters are shown in Figure \ref{fig:vel}. The velocities are scaled to the reference frame of each cluster and the cluster centers are marked with the black crosses.

\subsection{Inner velocity-dispersion profile}

In order to fit our data with dynamical models and to estimate possible black-hole masses, we need a velocity dispersion profile. We obtain this profile by combining the spectra in radial bins around the center using a sigma clipping algorithm to remove any remaining cosmic rays. From the combined spectra we measure the velocity dispersion with the same template as for the velocity map using a non-parametric line of sight velocity distribution (LOSVD) fitting \citep{gebhardt_2000a,pinkney_2003}. In this method a LOSVD histogram is smoothed and convolved with the template spectrum and fitted to the observations using $\chi^2$ statistics. In addition, the routine is able to fit for a continuum offset and an equivalent width. Several starting values such as the bin width of the LOSVD and the smoothing factor are tested by running Monte-Carlo simulations until the parameters are found which produce the smallest biases. This method turned out to be more stable at this resolution than pPXF which produces different results depending on whether one fits two (V, $\sigma$) or four (V, $\sigma, h_3, h_4$ ) Gauss-Hermite moments. We find that the pPXF results are consistent within the uncertainties with the non-parametric fits when fitting four moments. We also try different combinations of bins in order to find the smoothest profile for each cluster. 

The errors are estimated in two ways: The first one originates from the kinematic measurements and is derived through Monte-Carlo simulations of the observed spectra for 100 different realizations. The scatter in the results is then determined using a robust biweight estimator. The second error is calculated from the shotnoise in each bin as described in \citet{luetzgendorf_2012a}. For each bin we run Monte Carlo simulations using the information from our shot noise routine described in Section \ref{subsec:velmap}. With this information we know how many stars contribute by what amount to each spaxel, and how many spaxels are added up in each bin. We simulate a spaxel by assigning random velocities, drawn from a Gaussian distribution ($\sigma = 10$ \kms), to each star contributing to a spaxel, shifting the template spectra by this velocity and averaging the spectra after weighting them by their luminosity. The resulting spaxels are normalized, combined and the kinematics measured. The shotnoise error for each bin is derived from the spread in the dispersion measurements of the fake spaxel after 1000 realizations. The final value results from a quadratic summation of both errors assuming that they are independent of each other. Table \ref{tab:moments} lists the derived kinematic profiles for all clusters together with additional outer kinematics as described  in the next section.

In addition to the velocity dispersion profile, we estimate the radial velocity of the cluster in a heliocentric reference frame and the effective velocity dispersion $\sigma_e$. The radial velocity is measured by combining all spectra in the pointing and measuring the total velocity relative to the velocity of the template. After correcting this value for the motion of the template and the heliocentric velocity, the absolute radial velocity of all clusters is derived. The values are listed in Table \ref{tab:kin} together with the other results from the spectroscopic analysis. As in \citet{luetzgendorf_2012a} we determine the effective velocity dispersion using equation (1) in \cite{nora11} by adopting the half-mass radius as the effective radius. $\sigma_e$ is also listed them in Table \ref{tab:kin}.


\subsection{Outer kinematics}

In addition to the inner velocity-dispersion points we use outer kinematic points, if available,  to constrain the mass-to-light ratio of the clusters. Outer kinematics are available for NGC 1805, NGC 1904, NGC 6093 and NGC 5824. For NGC 1851 and NGC 1904 they are taken from the dataset obtained by \cite{scarpa_2011} using the FLAMES multi-object spectrograph. The dataset was obtained in order to test Newtonian gravity in globular clusters and therefore extends to large radii ($\sim 30$ pc). The observations were carried out from November 2007 to March 2008 by using the HR9B setup of the FLAMES spectrograph which covers the wavelength range of the magnesium triplet ($5143$ \AA $\ < \lambda < 5346$ \AA) at resolution $R = 25~900$. For further description of the data and data reduction we refer to \cite{scarpa_2011}. 

The data set contains 199 stars for NGC 1851 and 173 stars for NGC 1904 with positions and radial velocities. We measure the outer velocity-dispersion profile by binning the data into bins containing $\sim$ 50 stars each and evaluating a velocity dispersion using a maximum likelihood method introduced by \cite{pryor_1993}. The final profiles are shown in Figure \ref{fig:m1851} and \ref{fig:m1904} together with the inner kinematic points.

For NGC 5824, we use data obtained with the Rutgers Fabry-Perot on the Blanco 4-m telescope at Cerro Tololo Inter-American Observatory (CTIO) observed in March 1994. The Fabry-Perot velocities come from a very similar set of observations and reductions as presented in \cite{gebhardt_1997}. These spectra are centered on a small region around the H$\alpha$ absorption line, with absolute velocity calibration derived from comparison with published radial velocities.  Due to the fact that the data set only contains $\sim 100$ stars, only one kinematic point can be extracted from the sample. This data point, however, is still useful and included in the velocity dispersion profile of NGC 5824 (Figure \ref{fig:m5824}).

The outer kinematics for NGC 6093 were taken with a different Fabry-Perot instrument. As described in \cite{gebhardt_1997} the data were taken in May 1995 using an imaging Fabry-Perot spectrophotometer with the Sub-arcsecond Imaging Spectrograph (SIS) mounted at the Canada-France-Hawaii Telescope (CFHT). Similar to the observations of NGC 5824, a series of exposures stepped by $0.33$ \AA ~were taken across the H$\alpha$ line. For each star and each exposure, a photometric analysis was carried out and a spectrum across the H$\alpha$ line was derived. From this, the kinematics of $\sim875$ stars were obtained. We  combine these velocities in four bins with $\sim200$ stars each and derive the outer velocity-dispersion profile for NGC 6093.

\addtocounter{table}{1}

\subsection{Dynamical distance to NGC~6266}\label{subsec:dyndist}

For NGC~6266 no outer kinematics are found in the literature but recent measurements of proper motions obtained by \cite{mcnamara_2012} can be used to verify the profile and determine the dynamical distance of NGC~6266. At first we compute the velocity-dispersion profile of the proper motions using our derived center position. The ideal way would be to apply the same bins as for our radial measurements to have a direct comparison. Unfortunately, there are only 5 stars in the central bin in the proper motions dataset. The maximum likelihood method from \cite{pryor_1993} which we apply to calculate the velocity dispersions is not reliable for bins with less than 20 stars, but necessary to account for the different errorbars of the velocity measurements. For this reason it is not possible to obtain a data point for the innermost bin from the proper motions. The data is binned in 5 bins containing $\sim 30 - 100$ stars and compared to the velocity-dispersion profile of the from the radial measurements. The proper motions profile is then scaled to the radial profile by applying $\chi^2$ statistics. The final distance is then computed by:

\begin{equation}
\label{eq:hi}
d(pc) = 21.1 \frac{\sum\limits_{i=0}^N (\sigma_{r,i}/\delta \sigma_{p,i})^2}{\sum\limits_{i=0}^N (\sigma_{r,i} \sigma_{p,i})/\delta \sigma_{p,i}^2}
\end{equation} 

Where $\sigma_{r,i}$ and $\sigma_{p,i}$ are the datapoints of the interpolated N radial velocity-dispersion profile and the proper motions profile, respectively, and $\delta\sigma_{p,i}$ are the uncertainties of the proper motions. We derive a final distance of $d = (6.5 \pm 0.3)$ kpc. This is in good agreement with the value obtained by \citet[][$d=(7.2 \pm 0.6)$ kpc]{mcnamara_2012} as well as other published values: $7.9 \pm 1.1$ kpc \citep{illingworth_1976}, $7.0$ kpc \citep{ferraro_1990}, $6.6 \pm 0.5$ \citep{beccari_2006}, $6.7 \pm 1.0$ kpc \citep{contreras_2010}, and $6.9 \pm 0.7$ \citep{harris_1996}. Figure \ref{fig:m6266} shows the final proper motion and radial velocity-dispersion profile compared to Jeans models.

\section{Dynamical Models} \label{sec:jeans}

After extracting photometric and kinematic data from the observations, the data are fed into dynamical models in order to find the best fitting parameters that describe the globular cluster. This Section describes the analytical Jeans models which we use for fitting the data and lists the results from the best fits. 

\begin{figure*}
  \centering \includegraphics[width=\textwidth]{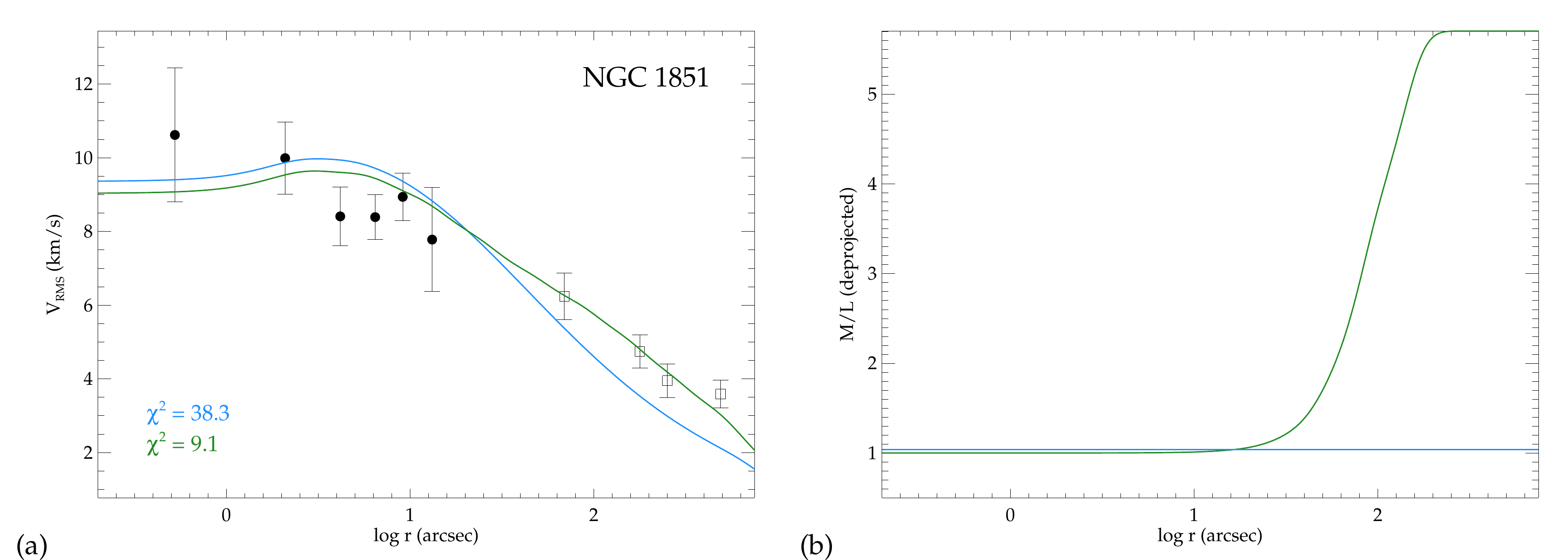}
  \centering \includegraphics[width=\textwidth]{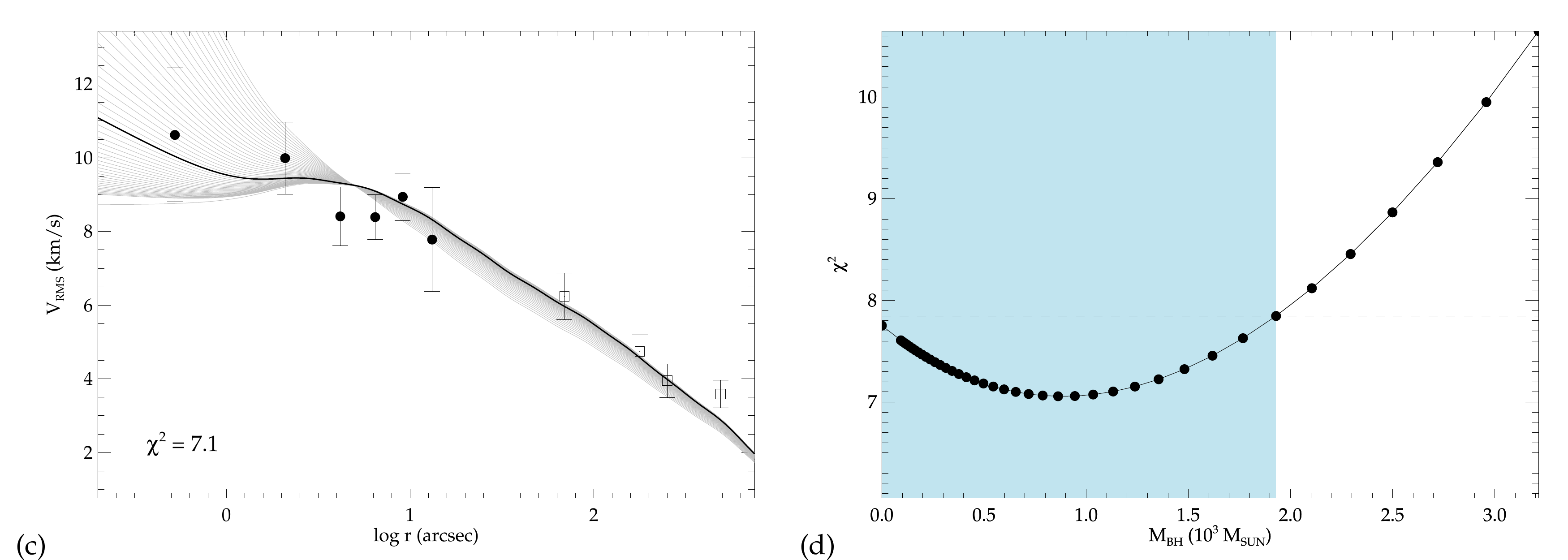}
  \caption{Jeans models for the globular cluster NGC 1851 compared to the datapoints obtained in this work (black bullets) and the outer data points obtained by \citet[][open squares]{scarpa_2011}. Panel a) shows the comparison of a model with and without fitting a $M/L_V$ profile. The corresponding $M/L_V$ profiles are shown in panel b). The panels c) and d) show various Jeans models with the varying $M/L_V$ profile and with different black-hole masses and their $\chi^2$ values. The best fit model is plotted as a black solid line and the blue shaded area in panel d) indicates the $1 \sigma$ error on the black-hole mass.}
  \label{fig:m1851}
\end{figure*}

\subsection{Isotropic Jeans models}

For our analysis we use spherical Jeans models as described by \citet{cappellari_2002,cappellari_2008}. The IDL routines provided by M. Cappellari\footnote{Available at http://www-astro.physics.ox.ac.uk/$\sim$mxc/idl} enable the modeling of the surface brightness profile and fitting the observed velocity data and the mass-to-light ratio at the same time. Using the surface-brightness profile as an input, the routine parametrizes it using a multi-Gaussian expansion (MGE) method developed by \cite{emsellem_1994}. The advantage of this method is that the deprojection can be done analytically since a deprojection of a Gaussian surface brightness results again in a Gaussian luminosity density. 

With the deprojected luminosity profile the code uses the spherical Jeans equation to calculate a velocity dispersion profile. This model profile is then scaled to the observed data using $\chi^2$ statistics. From the scaling factor the global $M/L_V$ is derived. In \cite{nora11} we introduced a radial $M/L_V$ profile \citep[similar to the method described in][]{williams_2009} in order to fit clusters with high mass segregation and therefore highly varying $M/L_V$. This can be done in two ways: 1) using a predefined $M/L_V$ profile or 2) fitting the $M/L_V$ profile to the data. For the latter, a $M/L_V$ profile is computed by multiplying each of the Gaussian components of the luminosity profile with a different factor in order to derive a density profile. The ratio of the density profile to the luminosity profile then gives the $M/L_V$ profile. 

The first method is implemented by parametrizing the input $M/L_V$ profile with an MGE fit. This is done by multiplying the individual gaussians of the MGE-luminosity profile with a set of factors in order to derive the MGE-mass-density profile. From this, an MGE-$M/L_V$ profile is derived by dividing the mass-density profile by the luminosity profile, and compared to the input $M/L_V$ profile. We vary the multiplication factors in order to derive the best fit to the input $M/L_V$ profile. The final model-velocity-dispersion profile is then calculated from the best-fit mass-density profile. 

Fitting an $M/L_V$ profile directly to the kinematic data is done by finding the optimum multiplication factors for the luminosity-Gaussians which minimize the $\chi^2$ of the resulting kinematic fit. The disadvantage of this procedure is that one can only use as many Gaussians as data points are available (otherwise the fit would be degenerate). For a small number of data points this method is therefore difficult to apply since a small number of Gaussians produces large variations in all profiles. A way around this would be a combination of both methods in two steps. In the first step the $M/L_V$ profile is fitted to the data with a limited number of Gaussians and in the second step the noisy $M/L_V$ profile is smoothed and fed again into the routine as a given profile, where a higher number of Gaussians can be used.

\begin{figure*}
  \centering \includegraphics[width=\textwidth]{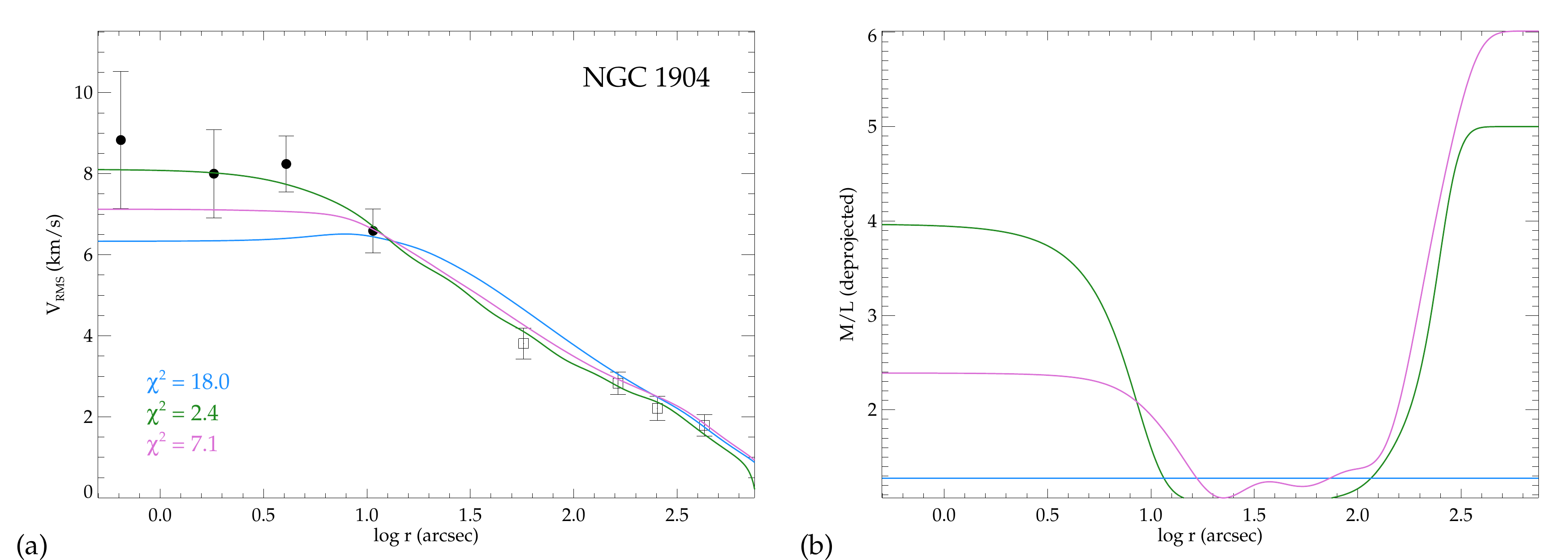}
  \centering \includegraphics[width=\textwidth]{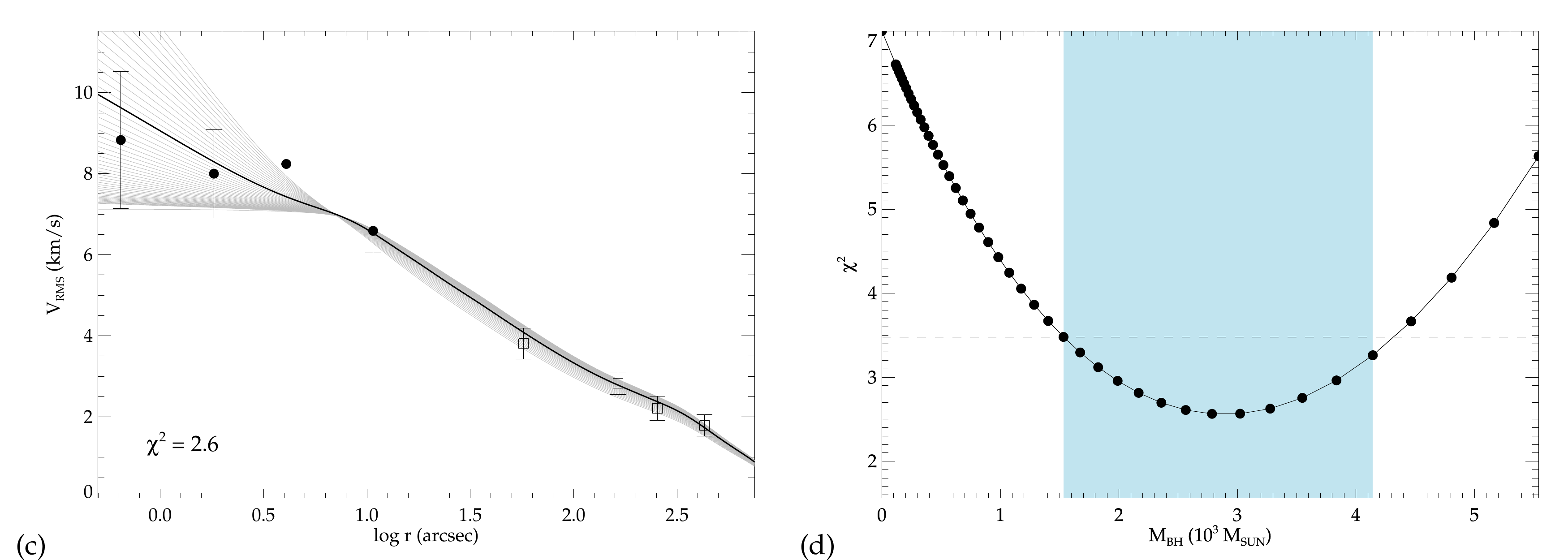}
  \caption{Jeans models for the globular cluster NGC 1904. All four panels and symbols are the same as in Figure \ref{fig:m1851}. The purple line in Panel b) shows the intermediate $M/L_V$ profile which is used also when fitting the black hole mass in Panel c) and d). }
  \label{fig:m1904}
\end{figure*}

The total $M/L_V$ value is derived from the total luminosity $L_{tot}$ and the total mass $M_{tot}$. The uncertainties of the total $M/L_V$ value are derived from a combination of the uncertainties of the mass, obtained from the $\chi^2$ fit to the kinematic data and the luminosity, calculated from the average errors on the distance (5 \%) and the reddening (10 \%) as taken from \cite{harris_1996}. In the following sections we discuss the modeling of the clusters individually. Every cluster is different. Every object is composed of different dataset combinations and therefore needs different treatment in terms of modeling.

Spherical symmetry is in general a valid assumption for globular clusters. Especially in our sample all clusters have observed axial ratios larger or equal to $b/a = 0.95$ \citep{white_1987}. However, ignoring rotation in the outer regions of the cluster would lead to an overestimation of the $M/L_V$ in the outskirts. For this reason we also test for axisymmetric models using the one-dimensional surface brightness profile and the axial ratios as listed in \citet{white_1987}. We find that even for the cluster with the highest flattening (NGC~ 1851, $b/a = 0.95$) the difference in the Jeans models in the outer regions does not exceed $0.1$ \kms. This corresponds to a difference in the maximum $M/L_V$ value in the outskirts of $\sim 0.7$ causing a difference in the total $M/L_V$ of $0.1$ and a difference in the total mass of $\Delta \log M_{tot} = 0.03$. All deviations caused by the flattened models are still within the errorbars of the values measured with the spherical models. Consequently, it is not necessary to use axisymmetric models for these cluster for our purposes.

It is important to state that we do not directly include rotation in our Jeans models. However, since we measure the second moment in the data and models, we in fact do include the contribution of the rotation to the dynamics. Thus, we expect no difference when directly including the rotation. We treat the data and models consistently by integrating over full annulus so any rotation, if present, will be included in the second moment and thus the mass profile.


\subsection{NGC 1851}

The combined kinematic data of NGC 1851 covers a large radial range of the cluster. At first sight, the two datasets of the outer and inner kinematics seem to agree well with each other. However, when comparing the data to a Jeans model with constant $M/L_V$ (Figure \ref{fig:m1851}.a) the datasets seem to be shifted with respect to each other. Figure \ref{fig:m1851}.a also shows a model where the $M/L_V$ profile is fitted to the data (green line). The resulting  $M/L_V$ profile is shown in panel b) of the same Figure. The profile stays flat for a large range of radii and rises by a factor of five in the outskirts ($r > 3$ pc). The shape of the profile is expected since low-mass stars (with high $M/L$) move to the outskirts of the cluster due to mass segregation. The amount with which the $M/L_V$ changes seems high but compared to the $M/L_V$ profile derived from N-body simulations for NGC~6266 (Section \ref{subsec:6266}), still reasonable. When applying these fits to combined datasets one has to keep in mind that there could be still a systematic offset between the data. This could change the shape of the fitted $M/L_V$ profile as explained in the following Section. 

We also test models with different black-hole masses (Figure \ref{fig:m1851}.c and d). For this we use the green model from Figure \ref{fig:m1851}.a and include the potential of a central black hole. The $\chi^2$ values of each model are shown in Figure \ref{fig:m1851}.d as a function of the black-hole mass. It is shown that despite the fact that the best fit requires a black hole of about $1000 \, M_{\odot}$, the data is still consistent with a no-black-hole model within the $1 \sigma$ errors (blue shaded area in Figure \ref{fig:m1851}.d).

\subsection{NGC 1904}

NGC 1904 is an interesting case. Similar to NGC 1851 the large radial coverage of kinematic data points allows to study the dynamic stage of that cluster very well. Figure \ref{fig:m1904} demonstrates that a simple isotropic Jeans model with a constant $M/L_V$ over the entire cluster does not reproduce the data very well (blue line in Figure \ref{fig:m1904}.a). The first question which arises immediately is whether either of the data sets suffers from a systematic shift between inner and outer measurements which causes this mismatch between the points. Since the points from \cite{scarpa_2011} were obtained from individual stars in the outer regions of the cluster, which usually is a safe method, the potential errors are most probable with the data points obtained with the IFUs. 

\begin{figure*}
  \centering \includegraphics[width=\textwidth]{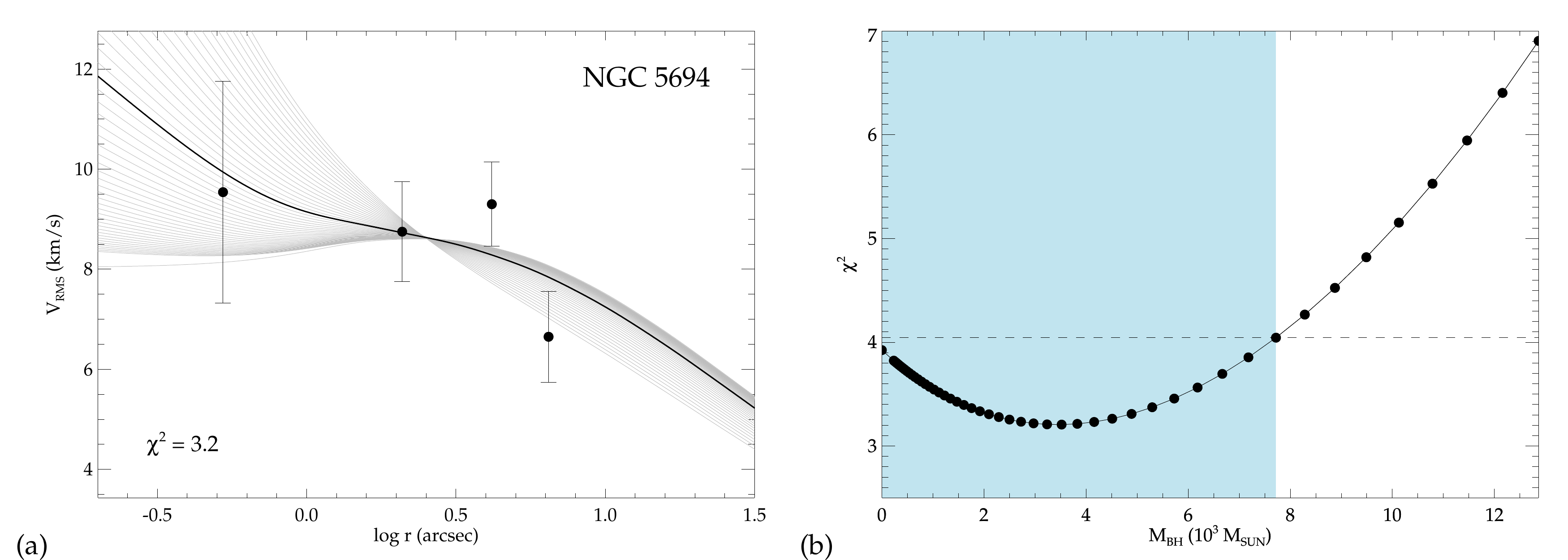}
  \caption{Dynamical models for the globular cluster NGC 5694. a) a grid of models with different black-hole masses. b) $\chi^2$ as a function of black-hole mass. The blue shaded area marks the 1 $\sigma$ error.}
  \label{fig:m5694}
\end{figure*}

In order to test for systematic offsets in our measurements we use different templates and different methods to extract the kinematics (as already explained in the previous section). Using the faint master template results in even higher velocity dispersions ($\sim 1-2 $ \kms) and does not solve the problem. We also extract the brightest stars in the pointing by applying the shotnoise routines and construct a velocity histogram using these stars. By measuring the dispersion of this histogram we get an estimate of the velocity dispersion of the brightest stars in the cluster independent from the line fitting method. We obtain a total velocity dispersion of $\sim 6$ \kms. By testing different luminosity bins of the stars we find that the velocity dispersion drops the fainter the stars are. This arises from the contamination of the faint stars form cluster background cluster light and therefore a bias towards the cluster mean velocities. For this reason the actual velocity dispersion is probably higher than the 6 \kms measured from the bright stars, that are also contaminated by background light by a small percentage. This leads us to the conclusion that the measurement of the integrated light with $\sigma \sim 6 - 9$ \kms is most likely correct correct.

Modeling the final profile of NGC 1904 is not trivial. In order to fit both the inner and the outer data points we fit the $M/L_V$ profile to the data. Figure \ref{fig:m1904}.a shows the result of this fit (green line) and Figure \ref{fig:m1904}.b the corresponding $M/L_V$ profile. In order to reproduce the inner kinematics, the $M/L_V$ profile needs to rise strongly towards the center, to drop further out and to rise again in the outskirts in order to fit the flat outer profile. The strong rise towards the center could be explained with a high concentration of remnants at the center of the cluster. However, this would imply that NGC 1904 is in a core collapse state, which is not supported by its flat surface brightness profile.

The strong rise towards the center is a sign for core collapse, where heavy remnants sink to the center of the cluster. However, for NGC 1904 this scenario is rather unlikely due to its large core and flat surface brightness profile (see Figure \ref{fig:sb}). For this reason the fitted $M/L_V$ profile seems to be not a good descriptions of the clusters properties. We therefore use an intermediate $M/L_V$ profile (Figure \ref{fig:m1904}.a and b, purple line) which does not rise as strong as the previous profile. The shape of the intermediate $M/L_V$ profile is in good agreement with what we find in N-body simulations.

\begin{figure*}
  \centering \includegraphics[width=\textwidth]{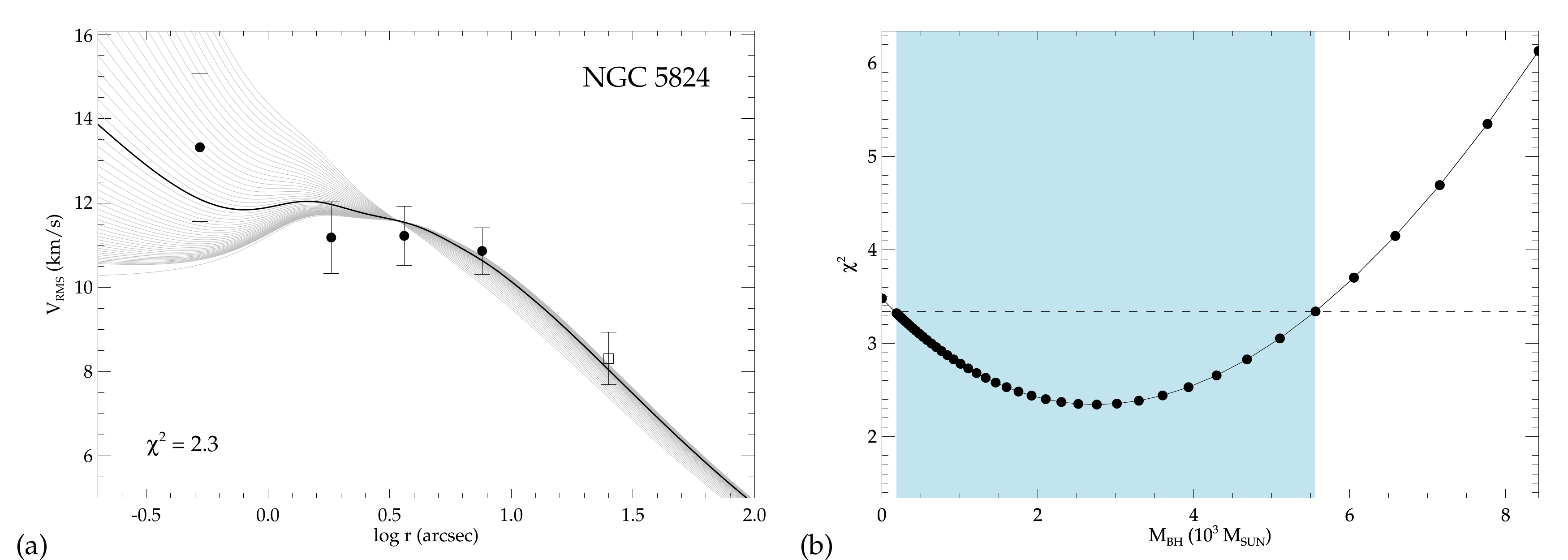}
  \caption{Kinematic profile of NGC 5824 in comparison with different Jeans models.The open square marks the one outer datapoint obtained from Fabry-Perot measurement. The rest of the symbols and panels are the same as in Figure \ref{fig:m5694}.}
  \label{fig:m5824}
\end{figure*}

Using the intermediate $M/L_V$ profile we fit the kinematic profile by adding a black hole in the center of the cluster. In Figure \ref{fig:m1904}.c we show different Jeans models computed for different black-hole masses (grey lines) and constant $M/L_V$. By performing $\chi^2$ statistics (Figure \ref{fig:m1904}.d) we find the model which fits the data best (black line). This model predicts a black hole of  $M_{\bullet} = (3 \pm 1) \times 10^3 M_{\odot}$.  
   

\begin{figure*}
  \centering \includegraphics[width=\textwidth]{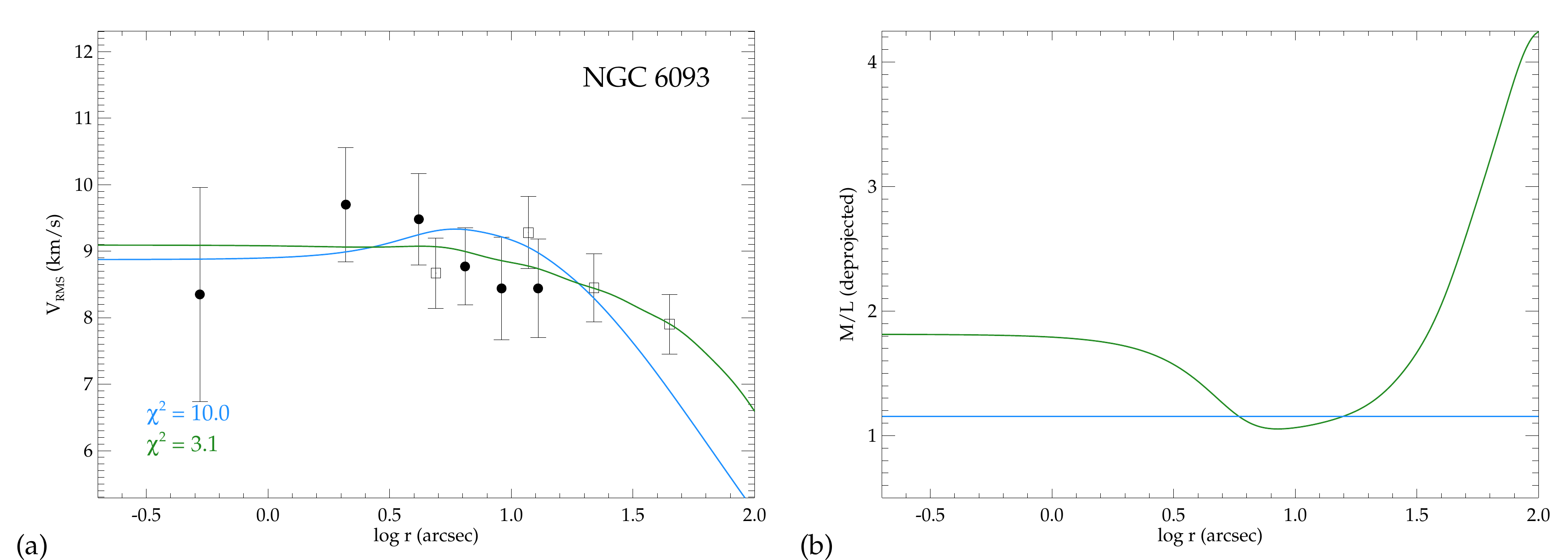}
  \centering \includegraphics[width=\textwidth]{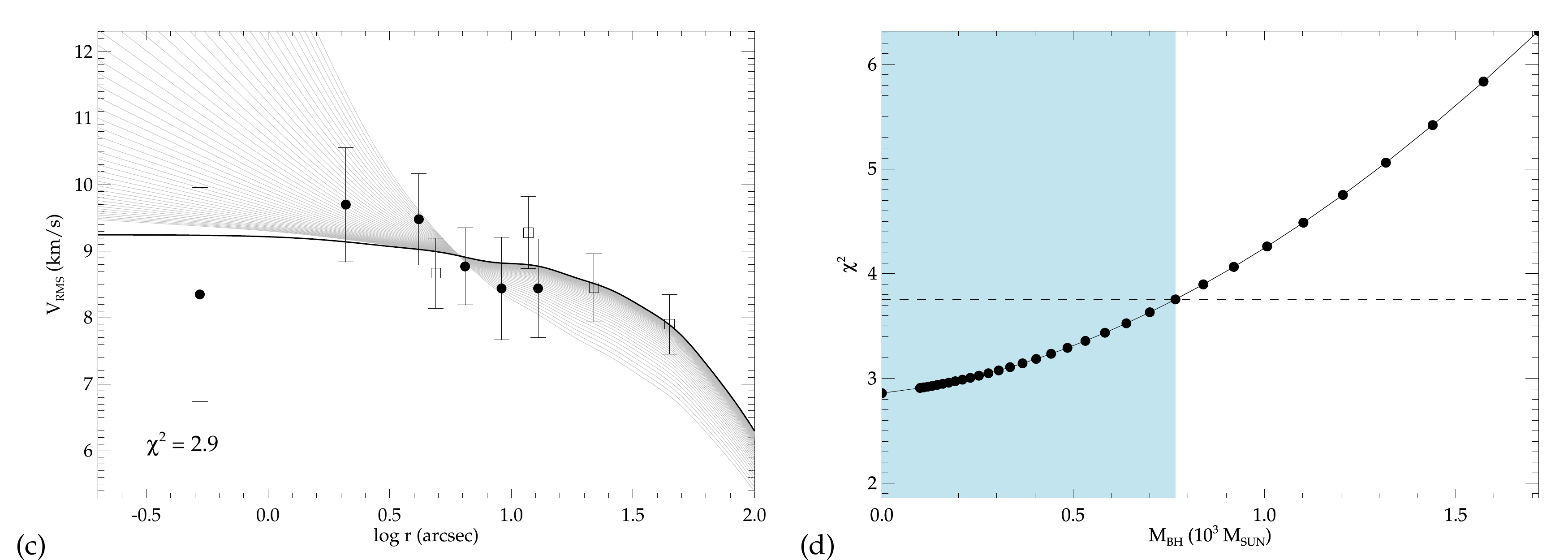}
  \caption{Jeans models for the globular cluster NGC 6093 compared to the datapoints obtained in this work (black bullets) and the outer data points obtained by with the CFHT Fabry-Perot instrument (open squares). Panel a) shows the comparison of a model with (green) and without (blue) fitting a $M/L_V$ profile. The corresponding $M/L_V$ profiles are shown in panel b). The panels c) and d) show various Jeans models with different black-hole masses and their $\chi^2$ values. The best fit model is plotted as a black solid line.}
  \label{fig:m6093}
\end{figure*}

\subsection{NGC 5694}

The limited field of view of the ARGUS observations of NGC 5694 (see Figure \ref{fig:vel}) makes it difficult to unfold the kinematic properties of this cluster. Also, due to its low surface brightness the signal-to-noise of these observations is almost half the signal-to-noise of the other clusters and the uncertainties increase accordingly. Figure \ref{fig:m5694} shows the final datapoints as well as a grid of Jeans models with different black-hole masses. Due to the small coverage of the kinematic profile it is not possible to fit a $M/L_V$ profile to the data and a constant $M/L_V$ is assumed. For this reason the only free parameter we vary is the mass of the central black hole. 

The best fit is provided by a model with a black-hole mass of $M_{\bullet} \sim 4  \times 10^3 \ M_{\odot}$, the model with no black hole however, is not excluded. Unfortunately the errorbars are large and the datapoints too few. The cluster can be considered as a good candidate for hosting an intermediate-mass black hole at its center, but with the current data coverage and quality it is challenging to put strong constraints on the mass of a possible central black hole. For that reason we treat the result of the fit for NGC~5694 with care.

\subsection{NGC 5824}

For NGC 5824 we obtain one outer datapoint from the Fabry-Perot data. This adds up to five kinematic points out to a radius of $25 ''$ ( $\sim 3 $ pc).  The observed profile is in very good agreement with a Jeans model with constant $M/L_V$ (Figure \ref{fig:m5824}.a), and does not require a $M/L_V$ variation. 

We also test for Jeans models with different IMBH masses for NGC 5824. In Figure \ref{fig:m5824}.a we plot the grid of models on top of the data and study their $\chi^2$ values. We find a slightly better fit for a model with a black hole of $M_{\bullet} \sim 2000 \ M_{\odot}$. However, due to the large uncertainty of the innermost point the data is still consistent with a no-IMBH model. 

\subsection{NGC 6093}

For NGC 6093 the outer and inner kinematics seem to agree well within the errorbars. Considering the last data point of the Fabry-Perot data, however, a variable $M/L_V$ profile is needed also for this cluster in order to fit the observational data in an optimal way. Figure \ref{fig:m6093}.a and b show two Jeans models, one with constant (blue) and one with varying (green) $M/L_V$ on top of the data. Note that the $M/L_V$ profile shown in panel b is only reliable out to the last data point where the model was fitted. The outer drop of the profile for radii greater than $200 ''$ is therefore meaningless. 

The comparison with models of different black-hole masses in Figure \ref{fig:m6093}.c and d shows that no black hole is needed in order to fit the kinematic data of NGC 6093 and the profile is flat in the center. 

\begin{figure*}
\centering \includegraphics[width=\textwidth]{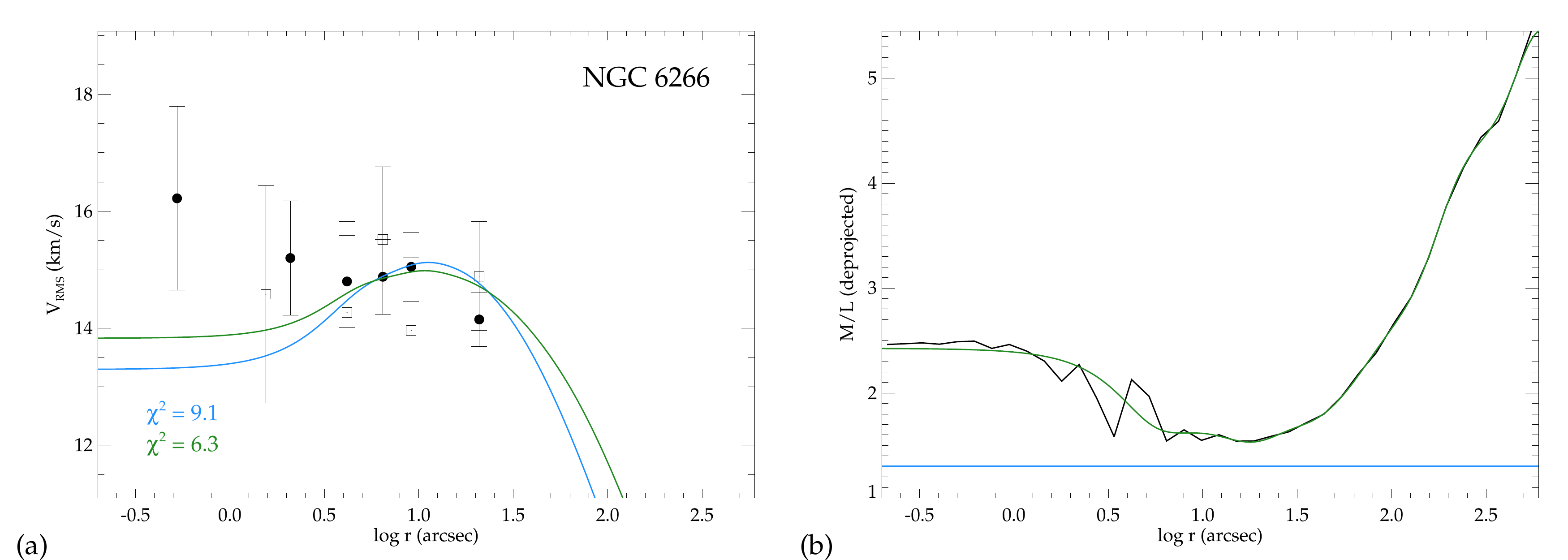}
\centering \includegraphics[width=\textwidth]{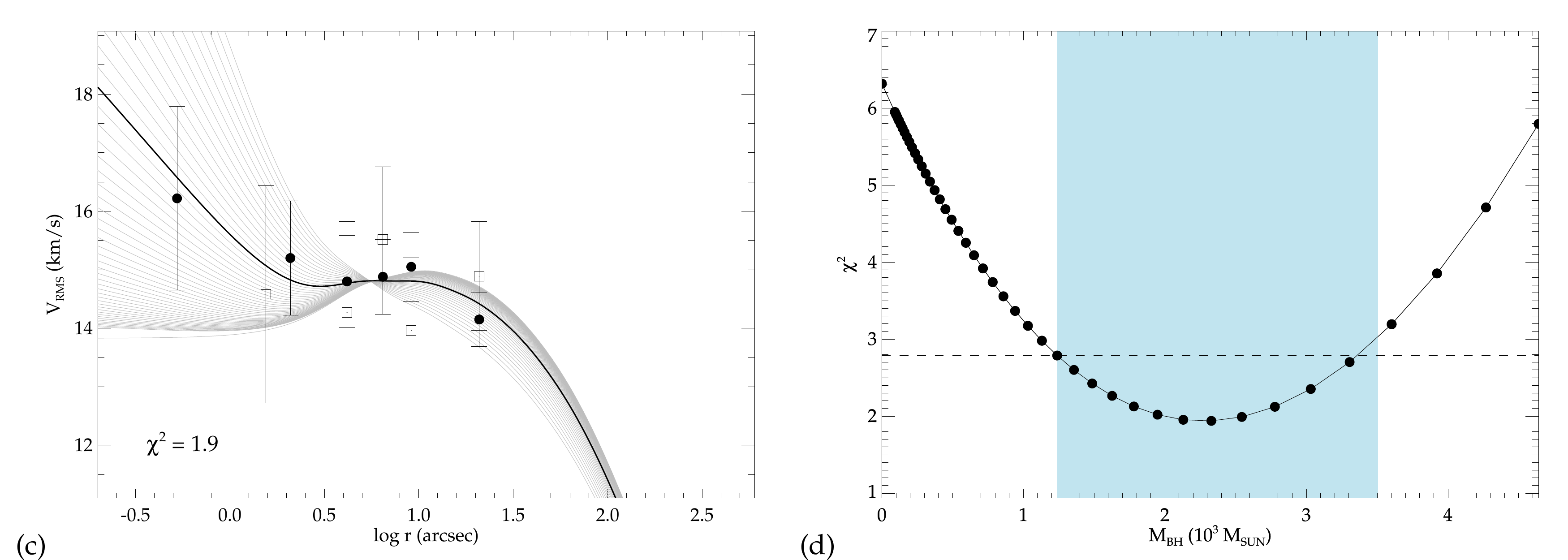}
  \caption{Jeans models, data, and $M/L_V$ profile of NGC 6266. All panels and colors are the same as in the previous figures. The $M/L_V$ profile (black line in panel b) is taken from N-body simulations and used as an input for the Jeans models. The green solid line in panel b) shows the parametrization of this profile. The $1 \sigma$ area of the black-hole mass is visualized as blue shaded area in panel d) as in the previous figures.}
  \label{fig:m6266}
\end{figure*}

\subsection{NGC 6266} \label{subsec:6266}

The velocity-dispersion profile of NGC 6266 shows the most distinct rise in the center of all the clusters. \cite{mcnamara_2012} performed N-body simulations in order to reproduce the kinematic data of NGC 6266 obtained from HST proper motions. We match the proper motions with our radial velocity-dispersion profile and find good agreement in the general shape of the two datasets. For that reason we can use their derived N-body $M/L_V$ profile as an input for our kinematic models. 

The N-body simulations were carried out using the Graphical Processing Unit (GPU)-enabled version of the collisional N-body code NBODY6 \citep{aarseth_1999,Nitadori_2012} on the GPU cluster of the University of Queensland. All simulations started with $N = 50~000$ stars distributed according to a \cite{king_1962} model using different initial concentrations $c$ and half-mass radii $r_h$ and masses following a \cite{kroupa_2001} mass function. The metallicity was set to $[Fe/H] = -1.18$ as given by \cite{harris_1996} and stellar evolution was included according to the routines of \cite{hurley_2000} during the cluster evolution time of 12 Gyr. With different retention fractions for white dwarfs ($100\%$), neutron stars ($10\%$) and black holes ($10\%$), the final population of dark remnants in the cluster consists mainly of white dwarfs which make $30\%$ of the total cluster mass. In general, most of the stellar-mass black holes got kicked out during the cluster lifetime and only $1 - 2$ black holes and about a dozen neutron stars remain.

Figure \ref{fig:m6266} shows this $M/L_V$ profile and the corresponding Jeans model. The $M/L_V$ rises strongly in the outer regions of the cluster which results from mass segregation of low-mass stars which migrate towards larger radii. However, due to the lack of outer kinematics for NGC 6266 we cannot validate this profile with great confidence since the rise of the $M/L_V$ value occurs at radii larger than covered by our kinematic data points. 

As for the previous clusters we also test Jeans models with different black-hole masses for NGC 6266. In Figure \ref{fig:m6266}.a the different models (grey lines) and the best fit (solid line) to the data are visualized. Due to the strong rise of the profile in the inner regions the data requires a model with a central black hole of the mass $M_{\bullet} = (2.5 \pm 1.5 ) \times 10^3 \ M_{\odot}$. Figure \ref{fig:m6266}.c shows the $\chi^2$ as a function of black-hole mass and the $1 \sigma$ error of the best fit (blue shaded area).


\begin{table*}
\caption{Kinematic results of the six clusters. Listed are the measured central velocity dispersion from the IFU data ($\sigma_{tot}$), the effective velocity dispersion measured from the kinematic profile ($\sigma_e$), the total $M/L_V$, the total mass $M_{tot}$, and the limits on the mass of the possible intermediate-mass black hole ($M_{\bullet} $). The result marked with an asterisk should be treated with care. The index V indicates when a model with varying $M/L_V$ profile was applied instead of one with a constant value for $M/L_V$. }             
\label{tab:kin}      
\centering
\begin{tabular}{ccccccc}
\hline \hline
\noalign{\smallskip}
Cluster &  V$_{r}$ & $\sigma_{tot}$ & $\sigma_e$ &  $M/L $ & $\log M_{tot} $ & $M_{\bullet} $ \\
        &  [\kms] & [\kms] & [\kms] &  $[M_{\odot}/L_{\odot}]$ & $[M_{\odot}]$ & $[M_{\odot}]$ \\
 \noalign{\smallskip}
\hline
\noalign{\smallskip}
NGC 1851 & $\phantom{-}322.7 \pm 4.3$	&$\phantom{0}8.4 \pm 0.8$ 	&$\phantom{0}9.3 \pm 0.5$ 	& $2.1^V \pm 0.2$  & $5.57^V \pm 0.03$ & $ < 2^V \times 10^3$ \\
NGC 1904 & $\phantom{-}205.2 \pm 2.5$	&$\phantom{0}7.0 \pm 0.4$ 	&$\phantom{0}8.0 \pm 0.5$ 	& $1.6^V \pm 0.2$  & $5.16^V \pm 0.03$& $(3 \pm 1)^V \times 10^3 $ \\
NGC 5694 & $-140.4 \pm 1.4$ 	&$8.5 \pm 0.4$ &$8.8 \pm 0.6$  & $2.1 \pm 0.4$  & $5.41 \pm 0.05$ & $< 8 \times 10^3 \ast $\\
NGC 5824 & $\phantom{0}-25.0 \pm 0.7 $ 	&$11.1 \pm 0.5$ 	&$11.2 \pm 0.4$	& $1.8 \pm 0.2$  & $5.65 \pm 0.03$ & $< 6 \times 10^3$\\
NGC 6093 & $\phantom{-0}12.1 \pm 0.5 $ 	&$\phantom{0}9.4 \pm 0.4$ 	&$\phantom{0}9.3 \pm 0.3$ 	& $2.3^V \pm 0.3$  & $5.53^V \pm 0.02$ & $< 800^V $\\
NGC 6266 & $\phantom{0}-75.0 \pm 1.0$ 	&$15.0 \pm 2.8$ 	&$15.5 \pm 0.5$ 	& $2.5 \pm 0.4$  & $5.97 \pm 0.01$ & $(2 \pm 1)^V \times 10^3$\\
\noalign{\smallskip}
\hline 
\end{tabular} 
\end{table*}


\section{Summary and Conclusions} \label{sec:con}
                                   

We study the kinematic and photometric properties of the six Galactic globular clusters NGC1851, NGC 1904, NGC 5694, NGC 5824, NGC 6093 and NGC 6266. We use spectroscopic observations obtained with the large integral field unit ARGUS of the FLAMES instrument at the VLT as well as high resolution HST images to obtain the photometric properties. From these images we determine the light-weighted center of each cluster using an isodensity-contour method with different magnitude cuts. We find that the derived centers are in good agreement with \citet{noyola_2006} except for NGC 6266 where the centers lie more than 2$''$ apart from each other. We derive surface brightness profiles through  a combination of star counts and integrated light and determine the central slope of the light profile. 

The kinematic profiles are derived by fitting non-parametric LOSVDs to the spectra combined in radial bins. For NGC 1851 and NGC 1904  outer kinematics were also available from datasets obtained by \cite{scarpa_2011}. Outer kinematics for NGC 5824 and NGC 6093 are taken from Fabry-Perot observations \citep{gebhardt_1997}. From the individual velocities of the stars an outer velocity-dispersion profile is derived using a maximum likelihood method. For NGC 6266 we compare our inner kinematic profile with proper motion data obtained by \cite{mcnamara_2012} and find very good agreement between these two datasets. We feed the surface brightness profile into analytical Jeans models and search for the best fitting model by varying the mass of the central black hole as well as the $M/L_V$ values. 

Table \ref{tab:kin} lists the final results of the Jeans modeling. For the total masses and $M/L_V$ values we use the results of the models which fit the data best, i.e. had the lowest $\chi^2$ values. We indicate the values derived from models with a varying $M/L_V$ profile with the index V. Comparing $M_{tot}$ and $M/L_V$ to the values presented in \cite{mclaughlin_2005}, we find in general a good agreement of the total masses of the clusters. The only exception is NGC 5824 where our derived mass of $\log M = 5.65 \pm 0.03$ is lower than their value of   $\log M = 5.81 \pm 0.04$. This originates in our value for the total luminosity which is already lower than the value of \cite{mclaughlin_2005} while for the other clusters the total luminosities are in good agreement with the values of \cite{mclaughlin_2005}. The overall $M/L_V$  ratios that we derive agree with the predictions of single-stellar population models following standard stellar mass functions at ages of 10 - 12 Gyr as shown in Figure 7 in \cite{Mieske_2008} and \cite{mclaughlin_2005} and also agree with the predicted dynamical $M/L_V$  values from \cite{mclaughlin_2005}, partly due to the large uncertainties of the \cite{mclaughlin_2005} values.  It is not surprising that our dynamical $M/L_V$ values are similar to the stellar population models since our clusters are massive and the mass loss through low-mass stars ejections, which would cause a decrease of the initial $M/L_V$ value  \citep[e.g.][]{baumgardt_2003,Kruijssen_2008,kruijssen_2009}, is not yet efficient enough to lower the $M/L_V$ value by more than 0.1 or 0.2.

For all of the clusters we also compared the data to models with different black hole masses. The final results of these fits are also listed in Table \ref{tab:kin}. For three clusters we only give an upper limit on the black-hole mass since the datapoints are consistent with a no-black-hole model within the $1\sigma$ errorbars. For NGC~1904, NGC~5694 and NGC~6266 the data require a model with a significant black-hole mass. In the case of NGC 1904 we perform tests in order to check whether the high black-hole mass could originate from a systematic offset between inner and outer kinematic profile. We find no evidence for a systematic shift of the inner kinematic profile. Assuming the outer kinematics, which are measured from individual stars, are correct, we conclude a black-hole mass of $(3 \pm 1) \times 10^3 \, M_{\odot}$ for NGC~1904. The result of NGC 5694 has to be treated carefully because of the low radial coverage and needs to be confirmed with more data. For that reason we mark the result of this cluster with an asterisk in Table \ref{tab:kin}.

Also NGC 1851 might require more detailed analysis. NGC 1851 could be the merging product of two globular clusters as the presence of self-enrichment signatures in both populations of the cluster and its photometric properties suggest \citep{carretta_2010,gratton_2012,olszewski_2009}. Also, the recently discovered cold stream in the vicinity of NGC 1851 \citep{sollima_2012} hints towards an accretion scenario where NGC 1851 is the remnant of a nuclear cluster located in a dwarf galaxy which was accreted and stripped by the Milky Way. For that reason Jeans models might not be sufficient to model a cluster with such a violent history. 

The profile of NGC 6266 is very well reproduced by the model which is surprising considering its large error bars. From all the clusters in our sample NGC 6266 is by far the best candidate for hosting an intermediate-mass black hole. With $\sim 2000 \, M_{\odot}$ it is, however, rather small compared to the black-hole masses found in clusters from previous studies. \cite{mcnamara_2012} compared the proper motion data and the surface brightness profile of NGC 6266 with N-body simulations and found that a centrally located IMBH is not required to match these profiles but given the uncertainties in the  position of the center, a black hole with a mass of a few thousand solar masses cannot be excluded. The fact that we use a different center than \cite{mcnamara_2012} might explain the stronger signature in our profile and the different, but still consistent results on the black hole masses. 

Another topic which needs to be discussed is the influence of dark remnants to the velocity-dispersion profile and the probability of this mimicking the effect of a single IMBH at the center. It is known that due to mass-segregation, stars more massive than the mean sink to the center of the cluster, which causes a higher concentration of massive stars/remnants in the cluster's core. This would challenge the detection of an IMBH since the rise in the $M/L_V$ causes a rise in the velocity-dispersion profile which could mimic the influence of an IMBH. The largest influence on the velocity-dispersion profile is produced by the most massive remnants such as black holes and neutron stars. However, the retention fraction of these objects is very poorly understood. Neutron stars are believed to get kicks of order a few hundred km/sec at birth \citep{lyne_1994}, which would imply that only few of them are retained in a globular cluster. Hence, neutron stars are unlikely to dominate the mass budget at any radius in a globular cluster.  In contrast, due to their higher mass, black holes are believed to get smaller  kick velocities and therefore remain in a larger fraction in star clusters after their birth. However, recent population synthesis modeling of Galactic low-mass X-ray binaries by \cite{repetto_2012}  has shown that black holes might get natal kicks that are of the same  order as those of neutron stars. In addition, $N$-body simulations by \cite{mackey_2008} and \cite{merritt_2004} as well as our own unpublished simulations show that star clusters with many stellar mass black holes have very large cores that are best fit by low concentration King models. Since the clusters in our sample all follow high-concentration King profiles, it is unlikely that they contain a significant concentration of  stellar mass black holes in their centres. In the case of white dwarfs, our N-body models assume a retention fraction of 100\% and despite the central rise in $M/L_V$ caused by the white dwarfs, we still need IMBHs to fit the velocity-dispersion profile in some of our clusters. We therefore have reasons to assume that remnants in the cluster center do not influence our results by a large amount and especially that our conclusions concerning the possible presence of IMBHs are not significantly effected.  Nevertheless, this conclusion can not be exclusively proven and our derived IMBH masses have to be treated with care. Detailed N-body simulations should and will be applied in follow-up papers in order to finally answer this question and disentangle kinematic signatures of remnants and IMBHs.

The search for intermediate-mass black holes in globular clusters became a promising new field of research in the last decade. With new observational  techniques such as integral-field spectroscopy and refined analysis methods for deriving proper motions it became possible to resolve the central kinematics where a possible black hole would start influencing the motions of the stars. With our current sample of ten Galactic globular clusters, including $\omega$ Centauri, NGC 6388, NGC 2808 and NGC 5286 (Feldmeier et al. in preparation) we have the largest sample of clusters studied with integral field units up to date. A next step in terms of analysis will be a comparison study of all clusters. We plan to correlate black-hole masses with structural and kinematic properties of the host cluster in order to find possible links to formation scenarios and environments for intermediate-mass black holes. We are also currently running N-body simulations, which will help us understand the dynamical processes in clusters with massive central black holes and to plan future observations for revealing the secrets of intermediate-mass black holes.

\begin{acknowledgements}
This research was supported by the DFG cluster of excellence Origin and Structure of the Universe (www.universe-cluster.de). H.B. acknowledges support from the Australian Research Council through Future Fellowship grant FT0991052.      
\end{acknowledgements}

\bibliographystyle{aa}
\bibliography{ref}

\longtab{3}{
\tiny
\begin{longtable}{ccccc|ccccc}
\caption{\label{tab:moments} Derived surface brightness profiles of all clusters in the V-band. $\Delta V_h$ and $\Delta V_l$ are the high and low values of the errors, respectively} \\
\hline \hline
\noalign{\smallskip}
  
Name &$r$ 	& $V$ & $\Delta V_l$ & $\Delta V_h$ &Name &$r$ 	& $V$ & $\Delta V_l$ & $\Delta V_h$ \\

&[arcsec] & [mag/arcsec$^2$] &[mag/arcsec$^2$] &[mag/arcsec$^2$]& &[arcsec] & [mag/arcsec$^2$] &[mag/arcsec$^2$] &[mag/arcsec$^2$]  \\ 
\noalign{\smallskip}
\hline
\noalign{\smallskip}
\endfirsthead
\caption{continued.}\\
\hline \hline
\noalign{\smallskip}
  
Name &$r$ 	& $V$ & $\Delta V_l$ & $\Delta V_h$ &Name &$r$ 	& $V$ & $\Delta V_l$ & $\Delta V_h$ \\

&[arcsec] & [mag/arcsec$^2$] &[mag/arcsec$^2$] &[mag/arcsec$^2$]& &[arcsec] & [mag/arcsec$^2$] &[mag/arcsec$^2$] &[mag/arcsec$^2$]  \\ 
\noalign{\smallskip}
\hline
\noalign{\smallskip}
\endhead
\noalign{\smallskip}
\hline
\endfoot

 \textbf{NGC 1851}&$-0.40$&$13.48$&$ 0.52$&$ 0.35$&NGC 5824&$ 1.06$&$17.64$&$ 0.02$&$ 0.02$\\
 NGC 1851&$-0.12$&$13.53$&$ 0.31$&$ 0.24$&NGC 5824&$ 1.20$&$18.29$&$ 0.02$&$ 0.02$\\
 NGC 1851&$ 0.00$&$13.49$&$ 0.31$&$ 0.24$&NGC 5824&$ 1.31$&$18.85$&$ 0.02$&$ 0.02$\\
 NGC 1851&$ 0.18$&$13.68$&$ 0.20$&$ 0.17$&NGC 5824&$ 1.39$&$19.33$&$ 0.02$&$ 0.02$\\
 NGC 1851&$ 0.30$&$13.82$&$ 0.15$&$ 0.13$&NGC 5824&$ 1.43$&$19.54$&$ 0.02$&$ 0.02$\\
 NGC 1851&$ 0.48$&$14.15$&$ 0.09$&$ 0.08$&NGC 5824&$ 1.47$&$19.73$&$ 0.02$&$ 0.02$\\
 NGC 1851&$ 0.72$&$14.53$&$ 0.05$&$ 0.05$&NGC 5824&$ 1.50$&$19.91$&$ 0.02$&$ 0.02$\\
 NGC 1851&$ 0.88$&$15.18$&$ 0.04$&$ 0.03$&NGC 5824&$ 1.52$&$20.01$&$ 0.02$&$ 0.02$\\
 NGC 1851&$ 1.00$&$15.71$&$ 0.03$&$ 0.03$&NGC 5824&$ 1.57$&$20.24$&$ 0.02$&$ 0.02$\\
 NGC 1851&$ 1.10$&$16.19$&$ 0.03$&$ 0.03$&NGC 5824&$ 1.61$&$20.49$&$ 0.02$&$ 0.02$\\
 NGC 1851&$ 1.18$&$16.52$&$ 0.03$&$ 0.03$&NGC 5824&$ 1.64$&$20.67$&$ 0.02$&$ 0.02$\\
 NGC 1851&$ 1.30$&$17.01$&$ 0.02$&$ 0.02$&NGC 5824&$ 1.66$&$20.78$&$ 0.02$&$ 0.02$\\
 NGC 1851&$ 1.40$&$17.69$&$ 0.02$&$ 0.02$&NGC 5824&$ 1.71$&$21.05$&$ 0.02$&$ 0.02$\\
 NGC 1851&$ 1.48$&$18.05$&$ 0.02$&$ 0.01$&NGC 5824&$ 1.75$&$21.28$&$ 0.02$&$ 0.02$\\
 NGC 1851&$ 1.60$&$18.72$&$ 0.01$&$ 0.01$&NGC 5824&$ 1.77$&$21.37$&$ 0.02$&$ 0.02$\\
 NGC 1851&$ 1.70$&$19.32$&$ 0.01$&$ 0.01$&NGC 5824&$ 1.77$&$21.38$&$ 0.02$&$ 0.02$\\
 NGC 1851&$ 1.78$&$19.79$&$ 0.01$&$ 0.01$&NGC 5824&$ 1.82$&$21.67$&$ 0.02$&$ 0.02$\\
 NGC 1851&$ 1.85$&$20.18$&$ 0.01$&$ 0.01$&NGC 5824&$ 1.86$&$21.89$&$ 0.02$&$ 0.02$\\
 NGC 1851&$ 1.90$&$20.39$&$ 0.01$&$ 0.01$&NGC 5824&$ 1.86$&$21.90$&$ 0.02$&$ 0.02$\\
 NGC 1851&$ 1.95$&$20.61$&$ 0.01$&$ 0.01$&NGC 5824&$ 1.87$&$21.94$&$ 0.02$&$ 0.02$\\
 NGC 1851&$ 2.00$&$20.78$&$ 0.01$&$ 0.01$&NGC 5824&$ 1.91$&$22.19$&$ 0.02$&$ 0.02$\\
 NGC 1851&$ 2.08$&$21.61$&$ 0.01$&$ 0.01$&NGC 5824&$ 1.93$&$22.32$&$ 0.02$&$ 0.02$\\
 NGC 1851&$ 2.12$&$21.88$&$ 0.01$&$ 0.01$&NGC 5824&$ 1.95$&$22.41$&$ 0.02$&$ 0.02$\\
 NGC 1851&$ 2.16$&$22.19$&$ 0.01$&$ 0.01$&NGC 5824&$ 1.95$&$22.42$&$ 0.02$&$ 0.02$\\
 NGC 1851&$ 2.22$&$22.62$&$ 0.01$&$ 0.01$&NGC 5824&$ 1.99$&$22.69$&$ 0.02$&$ 0.02$\\
 NGC 1851&$ 2.23$&$22.69$&$ 0.01$&$ 0.01$&NGC 5824&$ 2.01$&$22.82$&$ 0.02$&$ 0.02$\\
 NGC 1851&$ 2.30$&$23.14$&$ 0.01$&$ 0.01$&NGC 5824&$ 2.02$&$22.86$&$ 0.02$&$ 0.02$\\
 NGC 1851&$ 2.31$&$23.24$&$ 0.01$&$ 0.01$&NGC 5824&$ 2.03$&$22.91$&$ 0.02$&$ 0.02$\\
 NGC 1851&$ 2.38$&$23.73$&$ 0.01$&$ 0.01$&NGC 5824&$ 2.05$&$23.03$&$ 0.02$&$ 0.02$\\
 NGC 1851&$ 2.46$&$24.38$&$ 0.01$&$ 0.01$&NGC 5824&$ 2.08$&$23.25$&$ 0.02$&$ 0.02$\\
 NGC 1851&$ 2.48$&$24.52$&$ 0.01$&$ 0.01$&NGC 5824&$ 2.09$&$23.33$&$ 0.02$&$ 0.02$\\
 NGC 1851&$ 2.48$&$24.54$&$ 0.01$&$ 0.01$&NGC 5824&$ 2.13$&$23.59$&$ 0.02$&$ 0.02$\\
 NGC 1851&$ 2.54$&$24.93$&$ 0.01$&$ 0.01$&NGC 5824&$ 2.16$&$23.74$&$ 0.02$&$ 0.02$\\
 NGC 1851&$ 2.62$&$25.64$&$ 0.01$&$ 0.01$&NGC 5824&$ 2.18$&$23.90$&$ 0.02$&$ 0.02$\\
 NGC 1851&$ 2.78$&$26.95$&$ 0.01$&$ 0.01$&NGC 5824&$ 2.23$&$24.18$&$ 0.02$&$ 0.02$\\
 \textbf{NGC 1904}&$-0.45$&$15.83$&$ 0.94$&$ 0.49$&NGC 5824&$ 2.23$&$24.21$&$ 0.02$&$ 0.02$\\
 NGC 1904&$-0.14$&$15.83$&$ 0.41$&$ 0.30$&NGC 5824&$ 2.23$&$24.23$&$ 0.02$&$ 0.02$\\
 NGC 1904&$ 0.05$&$15.89$&$ 0.27$&$ 0.22$&NGC 5824&$ 2.27$&$24.44$&$ 0.02$&$ 0.02$\\
 NGC 1904&$ 0.18$&$15.85$&$ 0.22$&$ 0.18$&NGC 5824&$ 2.29$&$24.62$&$ 0.02$&$ 0.02$\\
 NGC 1904&$ 0.28$&$15.82$&$ 0.16$&$ 0.14$&NGC 5824&$ 2.32$&$24.80$&$ 0.02$&$ 0.02$\\
 NGC 1904&$ 0.36$&$15.99$&$ 0.17$&$ 0.14$&NGC 5824&$ 2.35$&$24.98$&$ 0.02$&$ 0.02$\\
 NGC 1904&$ 0.66$&$16.23$&$ 0.05$&$ 0.05$&NGC 5824&$ 2.38$&$25.18$&$ 0.02$&$ 0.02$\\
 NGC 1904&$ 0.83$&$16.44$&$ 0.04$&$ 0.04$&NGC 5824&$ 2.38$&$25.22$&$ 0.02$&$ 0.02$\\
 NGC 1904&$ 0.96$&$16.69$&$ 0.04$&$ 0.04$&NGC 5824&$ 2.43$&$25.59$&$ 0.02$&$ 0.02$\\
 NGC 1904&$ 1.06$&$17.06$&$ 0.04$&$ 0.03$&NGC 5824&$ 2.48$&$25.93$&$ 0.02$&$ 0.02$\\
 NGC 1904&$ 1.20$&$17.43$&$ 0.03$&$ 0.03$&NGC 5824&$ 2.57$&$26.54$&$ 0.02$&$ 0.02$\\
 NGC 1904&$ 1.31$&$17.92$&$ 0.03$&$ 0.03$&NGC 5824&$ 2.64$&$27.06$&$ 0.02$&$ 0.02$\\
 NGC 1904&$ 1.40$&$18.26$&$ 0.03$&$ 0.03$&NGC 5824&$ 2.73$&$27.74$&$ 0.02$&$ 0.02$\\
 NGC 1904&$ 1.47$&$18.76$&$ 0.04$&$ 0.03$&\textbf{NGC 6093}&$-0.22$&$14.80$&$ 0.29$&$ 0.23$\\
 NGC 1904&$ 1.53$&$18.97$&$ 0.03$&$ 0.03$&NGC 6093&$ 0.04$&$14.95$&$ 0.20$&$ 0.17$\\
 NGC 1904&$ 1.56$&$19.10$&$ 0.03$&$ 0.03$&NGC 6093&$ 0.24$&$14.78$&$ 0.15$&$ 0.13$\\
 NGC 1904&$ 1.56$&$19.11$&$ 0.03$&$ 0.03$&NGC 6093&$ 0.40$&$14.98$&$ 0.10$&$ 0.10$\\
 NGC 1904&$ 1.59$&$19.25$&$ 0.03$&$ 0.03$&NGC 6093&$ 0.65$&$15.36$&$ 0.05$&$ 0.05$\\
 NGC 1904&$ 1.64$&$19.52$&$ 0.03$&$ 0.03$&NGC 6093&$ 0.88$&$15.82$&$ 0.03$&$ 0.03$\\
 NGC 1904&$ 1.65$&$19.55$&$ 0.03$&$ 0.03$&NGC 6093&$ 1.00$&$16.33$&$ 0.03$&$ 0.03$\\
 NGC 1904&$ 1.69$&$19.76$&$ 0.03$&$ 0.03$&NGC 6093&$ 1.10$&$16.55$&$ 0.03$&$ 0.03$\\
 NGC 1904&$ 1.73$&$19.98$&$ 0.03$&$ 0.03$&NGC 6093&$ 1.18$&$16.77$&$ 0.02$&$ 0.02$\\
 NGC 1904&$ 1.74$&$20.02$&$ 0.03$&$ 0.03$&NGC 6093&$ 1.30$&$17.37$&$ 0.02$&$ 0.01$\\
 NGC 1904&$ 1.77$&$20.19$&$ 0.03$&$ 0.03$&NGC 6093&$ 1.40$&$17.75$&$ 0.01$&$ 0.01$\\
 NGC 1904&$ 1.80$&$20.40$&$ 0.03$&$ 0.03$&NGC 6093&$ 1.48$&$18.15$&$ 0.01$&$ 0.01$\\
 NGC 1904&$ 1.82$&$20.52$&$ 0.03$&$ 0.03$&NGC 6093&$ 1.60$&$18.69$&$ 0.01$&$ 0.01$\\
 NGC 1904&$ 1.87$&$20.76$&$ 0.03$&$ 0.03$&NGC 6093&$ 1.70$&$19.34$&$ 0.01$&$ 0.01$\\
 NGC 1904&$ 1.91$&$21.04$&$ 0.03$&$ 0.03$&NGC 6093&$ 1.78$&$19.59$&$ 0.01$&$ 0.01$\\
 NGC 1904&$ 1.95$&$21.26$&$ 0.03$&$ 0.03$&NGC 6093&$ 1.85$&$20.23$&$ 0.01$&$ 0.01$\\
 NGC 1904&$ 2.00$&$21.59$&$ 0.03$&$ 0.03$&NGC 6093&$ 1.90$&$20.20$&$ 0.01$&$ 0.01$\\
 NGC 1904&$ 2.00$&$21.60$&$ 0.03$&$ 0.03$&NGC 6093&$ 1.95$&$20.76$&$ 0.01$&$ 0.01$\\
 NGC 1904&$ 2.01$&$21.69$&$ 0.03$&$ 0.03$&NGC 6093&$ 2.00$&$20.99$&$ 0.01$&$ 0.01$\\
 NGC 1904&$ 2.03$&$21.78$&$ 0.03$&$ 0.03$&NGC 6093&$ 2.05$&$21.47$&$ 0.01$&$ 0.01$\\
 NGC 1904&$ 2.07$&$22.08$&$ 0.03$&$ 0.03$&NGC 6093&$ 2.07$&$21.59$&$ 0.01$&$ 0.01$\\
 NGC 1904&$ 2.09$&$22.18$&$ 0.03$&$ 0.03$&NGC 6093&$ 2.09$&$21.76$&$ 0.01$&$ 0.01$\\
 NGC 1904&$ 2.09$&$22.18$&$ 0.03$&$ 0.03$&NGC 6093&$ 2.12$&$21.96$&$ 0.01$&$ 0.01$\\
 NGC 1904&$ 2.12$&$22.43$&$ 0.03$&$ 0.03$&NGC 6093&$ 2.13$&$22.04$&$ 0.01$&$ 0.01$\\
 NGC 1904&$ 2.17$&$22.76$&$ 0.03$&$ 0.03$&NGC 6093&$ 2.16$&$22.24$&$ 0.01$&$ 0.01$\\
 NGC 1904&$ 2.21$&$23.06$&$ 0.03$&$ 0.03$&NGC 6093&$ 2.16$&$22.27$&$ 0.01$&$ 0.01$\\
 NGC 1904&$ 2.23$&$23.24$&$ 0.03$&$ 0.03$&NGC 6093&$ 2.17$&$22.31$&$ 0.01$&$ 0.01$\\
 NGC 1904&$ 2.25$&$23.34$&$ 0.03$&$ 0.03$&NGC 6093&$ 2.21$&$22.63$&$ 0.01$&$ 0.01$\\
 NGC 1904&$ 2.31$&$23.84$&$ 0.03$&$ 0.03$&NGC 6093&$ 2.22$&$22.71$&$ 0.01$&$ 0.01$\\
 NGC 1904&$ 2.38$&$24.33$&$ 0.03$&$ 0.03$&NGC 6093&$ 2.23$&$22.80$&$ 0.01$&$ 0.01$\\
 NGC 1904&$ 2.40$&$24.51$&$ 0.03$&$ 0.03$&NGC 6093&$ 2.23$&$22.83$&$ 0.01$&$ 0.01$\\
 NGC 1904&$ 2.48$&$25.21$&$ 0.03$&$ 0.03$&NGC 6093&$ 2.25$&$22.93$&$ 0.01$&$ 0.01$\\
 NGC 1904&$ 2.57$&$25.94$&$ 0.03$&$ 0.03$&NGC 6093&$ 2.25$&$22.94$&$ 0.01$&$ 0.01$\\
 \textbf{NGC 5694}&$-0.20$&$15.73$&$ 0.29$&$ 0.23$&NGC 6093&$ 2.29$&$23.30$&$ 0.01$&$ 0.01$\\
 NGC 5694&$ 0.06$&$15.83$&$ 0.19$&$ 0.16$&NGC 6093&$ 2.31$&$23.47$&$ 0.01$&$ 0.01$\\
 NGC 5694&$ 0.26$&$16.16$&$ 0.13$&$ 0.12$&NGC 6093&$ 2.32$&$23.53$&$ 0.01$&$ 0.01$\\
 NGC 5694&$ 0.44$&$16.55$&$ 0.09$&$ 0.08$&NGC 6093&$ 2.35$&$23.75$&$ 0.01$&$ 0.01$\\
 NGC 5694&$ 0.66$&$17.22$&$ 0.05$&$ 0.05$&NGC 6093&$ 2.38$&$24.00$&$ 0.01$&$ 0.01$\\
 NGC 5694&$ 0.83$&$17.75$&$ 0.03$&$ 0.03$&NGC 6093&$ 2.38$&$24.05$&$ 0.01$&$ 0.01$\\
 NGC 5694&$ 0.96$&$18.30$&$ 0.03$&$ 0.03$&NGC 6093&$ 2.40$&$24.19$&$ 0.01$&$ 0.01$\\
 NGC 5694&$ 1.03$&$18.73$&$ 0.03$&$ 0.03$&NGC 6093&$ 2.43$&$24.51$&$ 0.01$&$ 0.01$\\
 NGC 5694&$ 1.04$&$18.78$&$ 0.03$&$ 0.03$&NGC 6093&$ 2.47$&$24.81$&$ 0.01$&$ 0.01$\\
 NGC 5694&$ 1.06$&$18.86$&$ 0.03$&$ 0.03$&NGC 6093&$ 2.48$&$24.94$&$ 0.01$&$ 0.01$\\
 NGC 5694&$ 1.09$&$18.98$&$ 0.03$&$ 0.03$&NGC 6093&$ 2.53$&$25.37$&$ 0.01$&$ 0.01$\\
 NGC 5694&$ 1.12$&$19.10$&$ 0.03$&$ 0.03$&NGC 6093&$ 2.55$&$25.53$&$ 0.01$&$ 0.01$\\
 NGC 5694&$ 1.13$&$19.16$&$ 0.03$&$ 0.03$&NGC 6093&$ 2.57$&$25.73$&$ 0.01$&$ 0.01$\\
 NGC 5694&$ 1.17$&$19.34$&$ 0.03$&$ 0.03$&NGC 6093&$ 2.59$&$25.88$&$ 0.01$&$ 0.01$\\
 NGC 5694&$ 1.21$&$19.55$&$ 0.03$&$ 0.03$&NGC 6093&$ 2.64$&$26.41$&$ 0.01$&$ 0.01$\\
 NGC 5694&$ 1.23$&$19.65$&$ 0.03$&$ 0.03$&NGC 6093&$ 2.68$&$26.75$&$ 0.01$&$ 0.01$\\
 NGC 5694&$ 1.29$&$19.93$&$ 0.03$&$ 0.03$&NGC 6093&$ 2.73$&$27.31$&$ 0.01$&$ 0.01$\\
 NGC 5694&$ 1.30$&$19.98$&$ 0.03$&$ 0.03$&\textbf{NGC 6266}&$-0.17$&$14.79$&$ 0.37$&$ 0.28$\\
 NGC 5694&$ 1.34$&$20.18$&$ 0.03$&$ 0.03$&NGC 6266&$ 0.14$&$14.85$&$ 0.16$&$ 0.14$\\
 NGC 5694&$ 1.39$&$20.42$&$ 0.03$&$ 0.03$&NGC 6266&$ 0.44$&$15.04$&$ 0.08$&$ 0.07$\\
 NGC 5694&$ 1.39$&$20.42$&$ 0.03$&$ 0.03$&NGC 6266&$ 0.68$&$15.42$&$ 0.05$&$ 0.05$\\
 NGC 5694&$ 1.43$&$20.64$&$ 0.03$&$ 0.03$&NGC 6266&$ 0.83$&$15.66$&$ 0.04$&$ 0.04$\\
 NGC 5694&$ 1.47$&$20.85$&$ 0.03$&$ 0.03$&NGC 6266&$ 0.96$&$15.78$&$ 0.03$&$ 0.03$\\
 NGC 5694&$ 1.48$&$20.89$&$ 0.03$&$ 0.03$&NGC 6266&$ 1.06$&$15.95$&$ 0.03$&$ 0.03$\\
 NGC 5694&$ 1.50$&$21.04$&$ 0.03$&$ 0.03$&NGC 6266&$ 1.14$&$16.26$&$ 0.03$&$ 0.03$\\
 NGC 5694&$ 1.55$&$21.31$&$ 0.03$&$ 0.03$&NGC 6266&$ 1.26$&$16.61$&$ 0.02$&$ 0.02$\\
 NGC 5694&$ 1.56$&$21.38$&$ 0.03$&$ 0.03$&NGC 6266&$ 1.36$&$16.89$&$ 0.03$&$ 0.03$\\
 NGC 5694&$ 1.57$&$21.39$&$ 0.03$&$ 0.03$&NGC 6266&$ 1.44$&$17.27$&$ 0.05$&$ 0.05$\\
 NGC 5694&$ 1.64$&$21.86$&$ 0.03$&$ 0.03$&NGC 6266&$ 1.52$&$17.46$&$ 0.02$&$ 0.02$\\
 NGC 5694&$ 1.65$&$21.89$&$ 0.03$&$ 0.03$&NGC 6266&$ 1.53$&$17.51$&$ 0.02$&$ 0.02$\\
 NGC 5694&$ 1.71$&$22.27$&$ 0.03$&$ 0.03$&NGC 6266&$ 1.56$&$17.62$&$ 0.02$&$ 0.02$\\
 NGC 5694&$ 1.74$&$22.43$&$ 0.03$&$ 0.03$&NGC 6266&$ 1.59$&$17.70$&$ 0.02$&$ 0.02$\\
 NGC 5694&$ 1.77$&$22.63$&$ 0.03$&$ 0.03$&NGC 6266&$ 1.59$&$17.74$&$ 0.02$&$ 0.02$\\
 NGC 5694&$ 1.82$&$22.96$&$ 0.03$&$ 0.03$&NGC 6266&$ 1.60$&$17.74$&$ 0.02$&$ 0.02$\\
 NGC 5694&$ 1.82$&$22.99$&$ 0.03$&$ 0.03$&NGC 6266&$ 1.61$&$17.78$&$ 0.02$&$ 0.02$\\
 NGC 5694&$ 1.85$&$23.17$&$ 0.03$&$ 0.03$&NGC 6266&$ 1.63$&$17.89$&$ 0.02$&$ 0.02$\\
 NGC 5694&$ 1.87$&$23.27$&$ 0.03$&$ 0.03$&NGC 6266&$ 1.64$&$17.94$&$ 0.02$&$ 0.02$\\
 NGC 5694&$ 1.91$&$23.54$&$ 0.03$&$ 0.03$&NGC 6266&$ 1.65$&$17.97$&$ 0.02$&$ 0.02$\\
 NGC 5694&$ 1.91$&$23.57$&$ 0.03$&$ 0.03$&NGC 6266&$ 1.69$&$18.14$&$ 0.02$&$ 0.02$\\
 NGC 5694&$ 1.95$&$23.80$&$ 0.03$&$ 0.03$&NGC 6266&$ 1.71$&$18.24$&$ 0.02$&$ 0.02$\\
 NGC 5694&$ 2.00$&$24.17$&$ 0.03$&$ 0.03$&NGC 6266&$ 1.73$&$18.32$&$ 0.02$&$ 0.02$\\
 NGC 5694&$ 2.00$&$24.20$&$ 0.03$&$ 0.03$&NGC 6266&$ 1.74$&$18.34$&$ 0.02$&$ 0.02$\\
 NGC 5694&$ 2.01$&$24.27$&$ 0.03$&$ 0.03$&NGC 6266&$ 1.77$&$18.48$&$ 0.02$&$ 0.02$\\
 NGC 5694&$ 2.06$&$24.61$&$ 0.03$&$ 0.03$&NGC 6266&$ 1.77$&$18.49$&$ 0.02$&$ 0.02$\\
 NGC 5694&$ 2.09$&$24.79$&$ 0.03$&$ 0.03$&NGC 6266&$ 1.77$&$18.51$&$ 0.02$&$ 0.02$\\
 NGC 5694&$ 2.17$&$25.44$&$ 0.03$&$ 0.03$&NGC 6266&$ 1.80$&$18.65$&$ 0.02$&$ 0.02$\\
 NGC 5694&$ 2.27$&$26.18$&$ 0.03$&$ 0.03$&NGC 6266&$ 1.82$&$18.75$&$ 0.02$&$ 0.02$\\
 \textbf{NGC 5824}&$-0.23$&$14.45$&$ 0.37$&$ 0.28$&NGC 6266&$ 1.86$&$18.91$&$ 0.02$&$ 0.02$\\
 NGC 5824&$ 0.00$&$14.85$&$ 0.22$&$ 0.18$&NGC 6266&$ 1.87$&$18.96$&$ 0.02$&$ 0.02$\\
 NGC 5824&$ 0.30$&$15.30$&$ 0.11$&$ 0.10$&NGC 6266&$ 1.88$&$19.04$&$ 0.02$&$ 0.02$\\
 NGC 5824&$ 0.44$&$15.72$&$ 0.09$&$ 0.09$&NGC 6266&$ 1.91$&$19.19$&$ 0.02$&$ 0.02$\\
 NGC 5824&$ 0.66$&$16.01$&$ 0.05$&$ 0.05$&NGC 6266&$ 1.91$&$19.19$&$ 0.02$&$ 0.02$\\
 NGC 5824&$ 0.83$&$16.67$&$ 0.03$&$ 0.03$&NGC 6266&$ 1.95$&$19.38$&$ 0.02$&$ 0.02$\\
 NGC 5824&$ 1.06$&$17.64$&$ 0.02$&$ 0.02$&NGC 6266&$ 1.95$&$19.41$&$ 0.02$&$ 0.02$\\
 NGC 5824&$ 1.20$&$18.29$&$ 0.02$&$ 0.02$&NGC 6266&$ 1.97$&$19.50$&$ 0.02$&$ 0.02$\\
 NGC 5824&$ 1.31$&$18.85$&$ 0.02$&$ 0.02$&NGC 6266&$ 2.00$&$19.67$&$ 0.02$&$ 0.02$\\
 NGC 5824&$ 1.39$&$19.33$&$ 0.02$&$ 0.02$&NGC 6266&$ 2.01$&$19.76$&$ 0.02$&$ 0.02$\\
 NGC 5824&$ 1.43$&$19.54$&$ 0.02$&$ 0.02$&NGC 6266&$ 2.04$&$19.92$&$ 0.02$&$ 0.02$\\
 NGC 5824&$ 1.47$&$19.73$&$ 0.02$&$ 0.02$&NGC 6266&$ 2.06$&$20.03$&$ 0.02$&$ 0.02$\\
 NGC 5824&$ 1.50$&$19.91$&$ 0.02$&$ 0.02$&NGC 6266&$ 2.07$&$20.08$&$ 0.02$&$ 0.02$\\
 NGC 5824&$ 1.52$&$20.01$&$ 0.02$&$ 0.02$&NGC 6266&$ 2.07$&$20.11$&$ 0.02$&$ 0.02$\\
 NGC 5824&$ 1.57$&$20.24$&$ 0.02$&$ 0.02$&NGC 6266&$ 2.09$&$20.20$&$ 0.02$&$ 0.02$\\
 NGC 5824&$ 1.61$&$20.49$&$ 0.02$&$ 0.02$&NGC 6266&$ 2.10$&$20.30$&$ 0.02$&$ 0.02$\\
 NGC 5824&$ 1.64$&$20.67$&$ 0.02$&$ 0.02$&NGC 6266&$ 2.12$&$20.43$&$ 0.02$&$ 0.02$\\
 NGC 5824&$ 1.66$&$20.78$&$ 0.02$&$ 0.02$&NGC 6266&$ 2.14$&$20.58$&$ 0.02$&$ 0.02$\\
 NGC 5824&$ 1.71$&$21.05$&$ 0.02$&$ 0.02$&NGC 6266&$ 2.15$&$20.66$&$ 0.02$&$ 0.02$\\
 NGC 5824&$ 1.75$&$21.28$&$ 0.02$&$ 0.02$&NGC 6266&$ 2.17$&$20.74$&$ 0.02$&$ 0.02$\\
 NGC 5824&$ 1.77$&$21.37$&$ 0.02$&$ 0.02$&NGC 6266&$ 2.17$&$20.78$&$ 0.02$&$ 0.02$\\
 NGC 5824&$ 1.77$&$21.38$&$ 0.02$&$ 0.02$&NGC 6266&$ 2.20$&$20.99$&$ 0.02$&$ 0.02$\\
 NGC 5824&$ 1.82$&$21.67$&$ 0.02$&$ 0.02$&NGC 6266&$ 2.21$&$21.02$&$ 0.02$&$ 0.02$\\
 NGC 5824&$ 1.86$&$21.89$&$ 0.02$&$ 0.02$&NGC 6266&$ 2.21$&$21.03$&$ 0.02$&$ 0.02$\\
 NGC 5824&$ 1.86$&$21.90$&$ 0.02$&$ 0.02$&NGC 6266&$ 2.22$&$21.09$&$ 0.02$&$ 0.02$\\
 NGC 5824&$ 1.87$&$21.94$&$ 0.02$&$ 0.02$&NGC 6266&$ 2.25$&$21.31$&$ 0.02$&$ 0.02$\\
 NGC 5824&$ 1.91$&$22.19$&$ 0.02$&$ 0.02$&NGC 6266&$ 2.28$&$21.55$&$ 0.02$&$ 0.02$\\
 NGC 5824&$ 1.93$&$22.32$&$ 0.02$&$ 0.02$&NGC 6266&$ 2.29$&$21.61$&$ 0.02$&$ 0.02$\\
 NGC 5824&$ 1.95$&$22.41$&$ 0.02$&$ 0.02$&NGC 6266&$ 2.31$&$21.82$&$ 0.02$&$ 0.02$\\
 NGC 5824&$ 1.95$&$22.42$&$ 0.02$&$ 0.02$&NGC 6266&$ 2.31$&$21.83$&$ 0.02$&$ 0.02$\\
 NGC 5824&$ 1.99$&$22.69$&$ 0.02$&$ 0.02$&NGC 6266&$ 2.32$&$21.90$&$ 0.02$&$ 0.02$\\
 NGC 5824&$ 2.01$&$22.82$&$ 0.02$&$ 0.02$&NGC 6266&$ 2.33$&$21.97$&$ 0.02$&$ 0.02$\\
 NGC 5824&$ 2.02$&$22.86$&$ 0.02$&$ 0.02$&NGC 6266&$ 2.36$&$22.18$&$ 0.02$&$ 0.02$\\
 NGC 5824&$ 2.03$&$22.91$&$ 0.02$&$ 0.02$&NGC 6266&$ 2.38$&$22.37$&$ 0.02$&$ 0.02$\\
 NGC 5824&$ 2.05$&$23.03$&$ 0.02$&$ 0.02$&NGC 6266&$ 2.39$&$22.44$&$ 0.02$&$ 0.02$\\
 NGC 5824&$ 2.08$&$23.25$&$ 0.02$&$ 0.02$&NGC 6266&$ 2.40$&$22.53$&$ 0.02$&$ 0.02$\\
 NGC 5824&$ 2.09$&$23.33$&$ 0.02$&$ 0.02$&NGC 6266&$ 2.40$&$22.55$&$ 0.02$&$ 0.02$\\
 NGC 5824&$ 2.13$&$23.59$&$ 0.02$&$ 0.02$&NGC 6266&$ 2.42$&$22.74$&$ 0.02$&$ 0.02$\\
 NGC 5824&$ 2.16$&$23.74$&$ 0.02$&$ 0.02$&NGC 6266&$ 2.46$&$23.11$&$ 0.02$&$ 0.02$\\
 NGC 5824&$ 2.18$&$23.90$&$ 0.02$&$ 0.02$&NGC 6266&$ 2.47$&$23.17$&$ 0.02$&$ 0.02$\\
 NGC 5824&$ 2.23$&$24.18$&$ 0.02$&$ 0.02$&NGC 6266&$ 2.47$&$23.20$&$ 0.02$&$ 0.02$\\
 NGC 5824&$ 2.23$&$24.21$&$ 0.02$&$ 0.02$&NGC 6266&$ 2.50$&$23.45$&$ 0.02$&$ 0.02$\\
 NGC 5824&$ 2.23$&$24.23$&$ 0.02$&$ 0.02$&NGC 6266&$ 2.53$&$23.80$&$ 0.02$&$ 0.02$\\
 NGC 5824&$ 2.27$&$24.44$&$ 0.02$&$ 0.02$&NGC 6266&$ 2.62$&$24.63$&$ 0.02$&$ 0.02$\\
 NGC 5824&$ 2.29$&$24.62$&$ 0.02$&$ 0.02$&NGC 6266&$ 2.70$&$25.59$&$ 0.02$&$ 0.02$\\

\end{longtable}
}

\longtab{6}{
\tiny
\begin{longtable}{cccccccccccccccccccc}
\caption{\label{tab:moments} Kinematics measurements of all clusters obtained from the VLT/FLAMES data as well as outer kinematics. Listed are the radius ($r$) in arcseconds, the non-parametric first and second moment of the velocity distribution ($\tilde{\rm V},\tilde{\sigma}$), the four parametric moments of the velocity distribution (${\rm V}, \sigma, \rm{h}_3, \rm{h}_4$ ), and the signal-to-noise of the binned spectra (S/N).} \\
\hline \hline
\noalign{\smallskip}
  
$r$ 	& \multicolumn{3}{c}{$\tilde{\rm V}$} &\multicolumn{3}{c}{$\tilde{\sigma}$} &\multicolumn{3}{c}{${\rm V}$} &\multicolumn{3}{c}{${\sigma}$} &\multicolumn{3}{c}{h$_3$} &\multicolumn{3}{c}{h$_4$}&  S/N  \\ 

[arcsec] 			& \multicolumn{3}{c}{$[\mbox{km} / \mbox{s}]$} 	&\multicolumn{3}{c}{$[\mbox{km} / \mbox{s}]$} & \multicolumn{3}{c}{$[\mbox{km} / \mbox{s}]$} 	&\multicolumn{3}{c}{$[\mbox{km} / \mbox{s}]$}&\multicolumn{3}{c}{}&\multicolumn{3}{c}{} &  \\ 
\noalign{\smallskip}
\hline
\noalign{\smallskip}
\multicolumn{20}{c}{\textbf{NGC 1851}} \\
\multicolumn{20}{c}{\tiny{FLAMES-ARGUS MEASUREMENTS}} \\
\noalign{\smallskip}
\endfirsthead
\caption{continued.}\\
\hline \hline
\noalign{\smallskip}
  
$r$ 	& \multicolumn{3}{c}{$\tilde{\rm V}$} &\multicolumn{3}{c}{$\tilde{\sigma}$} &\multicolumn{3}{c}{${\rm V}$} &\multicolumn{3}{c}{${\sigma}$} &\multicolumn{3}{c}{h$_3$} &\multicolumn{3}{c}{h$_4$}&  S/N  \\ 

[arcsec] 			& \multicolumn{3}{c}{$[\mbox{km} / \mbox{s}]$} 	&\multicolumn{3}{c}{$[\mbox{km} / \mbox{s}]$} & \multicolumn{3}{c}{$[\mbox{km} / \mbox{s}]$} 	&\multicolumn{3}{c}{$[\mbox{km} / \mbox{s}]$}&\multicolumn{3}{c}{}&\multicolumn{3}{c}{} &  \\ 
\noalign{\smallskip}
\hline
\endhead
\noalign{\smallskip}
\hline
\endfoot

$0.52$ & $0.5$& $\pm$ &$0.3$ & $10.6$& $\pm$ &$1.8$ & $-0.0$& $\pm$ &$0.5$ & $11.8$& $\pm$ &$1.8$ & $0.01$& $\pm$ &$0.05$ & $-0.08$& $\pm$ &$0.03$ & $  109$ \\
$2.08$ & $-1.1$& $\pm$ &$0.3$ & $10.0$& $\pm$ &$0.9$ & $-1.9$& $\pm$ &$0.5$ & $11.0$& $\pm$ &$0.9$ & $0.04$& $\pm$ &$0.05$ & $-0.07$& $\pm$ &$0.03$ & $  109$ \\
$4.16$ & $0.9$& $\pm$ &$0.2$ & $8.4$& $\pm$ &$0.7$ & $0.5$& $\pm$ &$0.3$ & $9.3$& $\pm$ &$0.7$ & $-0.00$& $\pm$ &$0.04$ & $-0.06$& $\pm$ &$0.02$ & $   89$ \\
$6.50$ & $0.7$& $\pm$ &$0.2$ & $8.4$& $\pm$ &$0.5$ & $0.3$& $\pm$ &$0.3$ & $9.4$& $\pm$ &$0.5$ & $-0.00$& $\pm$ &$0.03$ & $-0.08$& $\pm$ &$0.01$ & $   90$ \\
$9.10$ & $-0.5$& $\pm$ &$0.3$ & $8.9$& $\pm$ &$0.4$ & $-2.4$& $\pm$ &$0.5$ & $8.9$& $\pm$ &$0.4$ & $0.15$& $\pm$ &$0.04$ & $-0.05$& $\pm$ &$0.02$ & $   80$ \\
$13.26$ & $-0.4$& $\pm$ &$0.3$ & $7.8$& $\pm$ &$1.4$ & $-1.3$& $\pm$ &$0.3$ & $8.5$& $\pm$ &$1.4$ & $0.06$& $\pm$ &$0.03$ & $-0.08$& $\pm$ &$0.01$ & $   76$ \\
\noalign{\smallskip}
\multicolumn{20}{c}{\tiny{FLAMES-MEDUSA MEASUREMENTS}} \\
\noalign{\smallskip}
$69.57$ & $-0.2$& $\pm$ &$0.9$ & $6.2$& $\pm$ &$0.6$&&&&&&&&&&&&& \\
$175.99$ & $1.1$& $\pm$ &$0.6$ & $4.7$& $\pm$ &$0.4$&&&&&&&&&&&&& \\
$253.75$ & $-1.2$& $\pm$ &$0.6$ & $4.0$& $\pm$ &$0.5$&&&&&&&&&&&&& \\
$492.72$ & $-0.1$& $\pm$ &$0.5$ & $3.6$& $\pm$ &$0.4$&&&&&&&&&&&&& \\
\noalign{\smallskip}
\hline
\noalign{\smallskip}
\multicolumn{20}{c}{\textbf{NGC 1904}} \\
\multicolumn{20}{c}{\tiny{FLAMES-ARGUS MEASUREMENTS}} \\
\noalign{\smallskip}
$0.65$ & $2.4$& $\pm$ &$0.4$ & $8.8$& $\pm$ &$1.7$ & $2.4$& $\pm$ &$0.5$ & $9.9$& $\pm$ &$1.7$ & $0.00$& $\pm$ &$0.05$ & $-0.10$& $\pm$ &$0.03$ & $   68$ \\
$1.82$ & $2.0$& $\pm$ &$0.3$ & $8.0$& $\pm$ &$1.1$ & $1.9$& $\pm$ &$0.4$ & $9.0$& $\pm$ &$1.1$ & $0.01$& $\pm$ &$0.03$ & $-0.10$& $\pm$ &$0.02$ & $   86$ \\
$4.03$ & $-0.4$& $\pm$ &$0.4$ & $8.2$& $\pm$ &$0.7$ & $0.3$& $\pm$ &$0.4$ & $8.4$& $\pm$ &$0.7$ & $-0.08$& $\pm$ &$0.03$ & $-0.03$& $\pm$ &$0.01$ & $   93$ \\
$10.66$ & $-0.5$& $\pm$ &$0.6$ & $6.6$& $\pm$ &$0.5$ & $-0.7$& $\pm$ &$0.5$ & $7.2$& $\pm$ &$0.5$ & $0.01$& $\pm$ &$0.03$ & $-0.06$& $\pm$ &$0.01$ & $   77$ \\
\noalign{\smallskip}
\multicolumn{20}{c}{\tiny{FLAMES-MEDUSA MEASUREMENTS}} \\
\noalign{\smallskip}
$57.29$ & $-0.2$& $\pm$ &$0.5$ & $3.8$& $\pm$ &$0.4$&&&&&&&&&&&&& \\
$163.72$ & $0.1$& $\pm$ &$0.4$ & $2.8$& $\pm$ &$0.3$&&&&&&&&&&&&& \\
$253.75$ & $0.7$& $\pm$ &$0.4$ & $2.2$& $\pm$ &$0.3$&&&&&&&&&&&&& \\
$429.04$ & $-0.3$& $\pm$ &$0.4$ & $1.8$& $\pm$ &$0.3$&&&&&&&&&&&&& \\
\noalign{\smallskip}
\hline
\noalign{\smallskip}
\multicolumn{20}{c}{\textbf{NGC 5694}} \\
\multicolumn{20}{c}{\tiny{FLAMES-ARGUS MEASUREMENTS}} \\
\noalign{\smallskip}
$0.52$ & $-2.6$& $\pm$ &$0.6$ & $9.5$& $\pm$ &$2.2$ & $-2.7$& $\pm$ &$0.5$ & $10.6$& $\pm$ &$2.2$ & $-0.02$& $\pm$ &$0.04$ & $-0.07$& $\pm$ &$0.03$ & $   69$ \\
$2.08$ & $-0.4$& $\pm$ &$0.9$ & $8.8$& $\pm$ &$1.0$ & $-0.5$& $\pm$ &$0.4$ & $9.7$& $\pm$ &$1.0$ & $-0.01$& $\pm$ &$0.04$ & $-0.08$& $\pm$ &$0.02$ & $   88$ \\
$4.16$ & $0.1$& $\pm$ &$0.6$ & $9.3$& $\pm$ &$0.8$ & $0.1$& $\pm$ &$0.5$ & $10.4$& $\pm$ &$0.8$ & $-0.02$& $\pm$ &$0.04$ & $-0.08$& $\pm$ &$0.02$ & $   66$ \\
$6.50$ & $0.4$& $\pm$ &$1.1$ & $6.6$& $\pm$ &$0.9$ & $0.5$& $\pm$ &$0.4$ & $7.3$& $\pm$ &$0.9$ & $-0.04$& $\pm$ &$0.02$ & $-0.08$& $\pm$ &$0.01$ & $   39$ \\
\noalign{\smallskip}
\hline
\noalign{\smallskip}
\multicolumn{20}{c}{\textbf{NGC 5824}} \\
\multicolumn{20}{c}{\tiny{FLAMES-ARGUS MEASUREMENTS}} \\
\noalign{\smallskip}
$0.52$ & $3.5$& $\pm$ &$0.3$ & $13.3$& $\pm$ &$1.7$ & $3.8$& $\pm$ &$0.4$ & $14.9$& $\pm$ &$1.7$ & $-0.00$& $\pm$ &$0.04$ & $-0.10$& $\pm$ &$0.03$ & $   92$ \\
$1.82$ & $0.3$& $\pm$ &$0.3$ & $11.2$& $\pm$ &$0.8$ & $0.3$& $\pm$ &$0.4$ & $12.5$& $\pm$ &$0.8$ & $0.02$& $\pm$ &$0.03$ & $-0.09$& $\pm$ &$0.03$ & $  107$ \\
$3.64$ & $0.3$& $\pm$ &$0.3$ & $11.2$& $\pm$ &$0.6$ & $0.3$& $\pm$ &$0.4$ & $12.5$& $\pm$ &$0.6$ & $0.03$& $\pm$ &$0.03$ & $-0.09$& $\pm$ &$0.03$ & $  104$ \\
$7.54$ & $-0.2$& $\pm$ &$0.3$ & $10.9$& $\pm$ &$0.4$ & $-0.3$& $\pm$ &$0.4$ & $12.1$& $\pm$ &$0.4$ & $0.03$& $\pm$ &$0.03$ & $-0.08$& $\pm$ &$0.02$ & $   94$ \\
\noalign{\smallskip}
\multicolumn{20}{c}{\tiny{FABRY-PEROT MEASUREMENTS}} \\
\noalign{\smallskip}
$25.03$ & $0.2$& $\pm$ &$0.8$ & $8.3$& $\pm$ &$0.6$&&&&&&&&&&&&& \\
\noalign{\smallskip}
\multicolumn{20}{c}{\textbf{NGC 6093}} \\
\multicolumn{20}{c}{\tiny{FLAMES-ARGUS MEASUREMENTS}} \\
\noalign{\smallskip}
$0.52$ & $-3.0$& $\pm$ &$0.2$ & $8.3$& $\pm$ &$1.6$ & $-2.9$& $\pm$ &$0.3$ & $9.3$& $\pm$ &$1.6$ & $-0.00$& $\pm$ &$0.02$ & $-0.08$& $\pm$ &$0.01$ & $   95$ \\
$2.08$ & $-2.0$& $\pm$ &$0.2$ & $9.7$& $\pm$ &$0.9$ & $-2.2$& $\pm$ &$0.4$ & $10.7$& $\pm$ &$0.9$ & $0.03$& $\pm$ &$0.04$ & $-0.07$& $\pm$ &$0.02$ & $  102$ \\
$4.16$ & $-1.0$& $\pm$ &$0.2$ & $9.5$& $\pm$ &$0.7$ & $-1.1$& $\pm$ &$0.3$ & $10.5$& $\pm$ &$0.7$ & $0.02$& $\pm$ &$0.03$ & $-0.07$& $\pm$ &$0.01$ & $   94$ \\
$6.50$ & $0.4$& $\pm$ &$0.2$ & $8.8$& $\pm$ &$0.6$ & $0.6$& $\pm$ &$0.3$ & $9.9$& $\pm$ &$0.6$ & $-0.00$& $\pm$ &$0.03$ & $-0.11$& $\pm$ &$0.01$ & $   86$ \\
$9.10$ & $1.3$& $\pm$ &$0.3$ & $8.4$& $\pm$ &$0.8$ & $1.3$& $\pm$ &$0.4$ & $9.3$& $\pm$ &$0.8$ & $0.01$& $\pm$ &$0.03$ & $-0.06$& $\pm$ &$0.01$ & $   86$ \\
$12.74$ & $2.6$& $\pm$ &$0.4$ & $8.4$& $\pm$ &$0.7$ & $2.7$& $\pm$ &$0.5$ & $9.3$& $\pm$ &$0.7$ & $0.01$& $\pm$ &$0.03$ & $-0.06$& $\pm$ &$0.01$ & $   77$ \\
\noalign{\smallskip}
\multicolumn{20}{c}{\tiny{FABRY-PEROT MEASUREMENTS}} \\
\noalign{\smallskip}
$4.94$ & $1.1$& $\pm$ &$0.7$ & $8.7$& $\pm$ &$0.5$&&&&&&&&&&&&& \\
$11.71$ & $1.2$& $\pm$ &$0.7$ & $9.3$& $\pm$ &$0.5$&&&&&&&&&&&&& \\
$21.70$ & $0.3$& $\pm$ &$0.6$ & $8.4$& $\pm$ &$0.5$&&&&&&&&&&&&& \\
$45.16$ & $-1.9$& $\pm$ &$0.6$ & $7.9$& $\pm$ &$0.4$&&&&&&&&&&&&& \\
\newpage
\noalign{\smallskip}
\multicolumn{20}{c}{\textbf{NGC 6266}} \\
\multicolumn{20}{c}{\tiny{FLAMES-ARGUS MEASUREMENTS}} \\
\noalign{\smallskip}
$0.52$ & $4.1$& $\pm$ &$0.2$ & $16.2$& $\pm$ &$1.6$ & $4.0$& $\pm$ &$0.3$ & $17.7$& $\pm$ &$1.6$ & $0.05$& $\pm$ &$0.02$ & $-0.06$& $\pm$ &$0.02$ & $  145$ \\
$2.08$ & $1.2$& $\pm$ &$0.3$ & $15.2$& $\pm$ &$1.0$ & $1.9$& $\pm$ &$0.5$ & $16.5$& $\pm$ &$1.0$ & $0.00$& $\pm$ &$0.03$ & $-0.05$& $\pm$ &$0.02$ & $  172$ \\
$4.16$ & $0.2$& $\pm$ &$0.2$ & $14.8$& $\pm$ &$0.8$ & $0.5$& $\pm$ &$0.3$ & $15.9$& $\pm$ &$0.8$ & $0.01$& $\pm$ &$0.02$ & $-0.04$& $\pm$ &$0.02$ & $  169$ \\
$6.50$ & $-0.6$& $\pm$ &$0.2$ & $14.9$& $\pm$ &$0.6$ & $-1.2$& $\pm$ &$0.4$ & $15.7$& $\pm$ &$0.6$ & $0.06$& $\pm$ &$0.03$ & $-0.04$& $\pm$ &$0.02$ & $  155$ \\
$9.10$ & $-1.2$& $\pm$ &$0.2$ & $15.0$& $\pm$ &$0.6$ & $-2.1$& $\pm$ &$0.4$ & $15.6$& $\pm$ &$0.6$ & $0.07$& $\pm$ &$0.02$ & $-0.03$& $\pm$ &$0.02$ & $  148$ \\
$20.80$ & $1.3$& $\pm$ &$0.2$ & $14.2$& $\pm$ &$0.5$ & $1.2$& $\pm$ &$0.3$ & $15.2$& $\pm$ &$0.5$ & $0.05$& $\pm$ &$0.02$ & $-0.05$& $\pm$ &$0.01$ & $  120$ \\
\noalign{\smallskip}
\multicolumn{20}{c}{\tiny{PROPER-MOTIONS MEASUREMENTS}} \\
\noalign{\smallskip}
$1.56$ & $2.5$& $\pm$ &$2.5$ & $14.6$& $\pm$ &$1.9$&&&&&&&&&&&&& \\
$4.16$ & $0.6$& $\pm$ &$1.9$ & $14.3$& $\pm$ &$1.5$&&&&&&&&&&&&& \\
$6.50$ & $-1.2$& $\pm$ &$1.9$ & $15.5$& $\pm$ &$1.2$&&&&&&&&&&&&& \\
$9.10$ & $-1.2$& $\pm$ &$1.5$ & $14.0$& $\pm$ &$1.2$&&&&&&&&&&&&& \\
$20.80$ & $0.6$& $\pm$ &$1.2$ & $14.9$& $\pm$ &$0.9$&&&&&&&&&&&&& \\
\end{longtable}
}

\end{document}